\documentclass[longauth]{aa}
\usepackage{euclid}
\usepackage{graphicx}
\usepackage{txfonts}

\usepackage{hyperref}
\hypersetup{
colorlinks=true,
linkcolor=blue,
filecolor=magenta,
urlcolor=blue,
citecolor=blue
}

\newcommand{\YJH}{$Y_{\scriptscriptstyle{\rm E}}$, $J_{\scriptscriptstyle{\rm E}}$, $H_{\scriptscriptstyle{\rm E}}$}
\newcommand{\IE}{$I_{\scriptscriptstyle{\rm E}}$}
\newcommand{\YE}{$Y_{\scriptscriptstyle{\rm E}}$}
\newcommand{\JE}{$J_{\scriptscriptstyle{\rm E}}$}
\newcommand{\HE}{$H_{\scriptscriptstyle{\rm E}}$}
\newcommand{\IYJH}{$I_{\scriptscriptstyle{\rm E}}-Y_{\scriptscriptstyle{\rm E}}$, $J_{\scriptscriptstyle{\rm E}}-H_{\scriptscriptstyle{\rm E}}$}
\newcommand{\uIIJ}{$u-I_{\scriptscriptstyle{\rm E}}$, $I_{\scriptscriptstyle{\rm E}}-J_{\scriptscriptstyle{\rm E}}$}

\begin{document} 

   \title{\Euclid\ preparation. XXII. Selection of quiescent galaxies from mock photometry using machine learning}
   \titlerunning{Selection of quiescent galaxies}
   \authorrunning{A. Humphrey et al.}
   
\author{\normalsize Euclid Collaboration: A.~Humphrey$^{1,2}$\thanks{\email{Andrew.Humphrey@astro.up.pt}}, L.~Bisigello$^{3}$, P.~A.~C.~Cunha$^{1,4}$, M.~Bolzonella$^{3}$, S.~Fotopoulou$^{5}$, K.~Caputi$^{6}$, C.~Tortora$^{7}$, G.~Zamorani$^{3}$, P.~Papaderos\orcid{0000-0002-3733-8174}$^{8,9}$
% ...
}

\institute{$^{1}$ Instituto de Astrof\'isica e Ci\^encias do Espa\c{c}o, Universidade do Porto, CAUP, Rua das Estrelas, PT4150-762 Porto, Portugal\\
$^{2}$ DTx -- Digital Transformation CoLAB, Building 1, Azur\'em Campus, University of Minho, 4800-058 Guimar\~aes, Portugal\\
$^{3}$ INAF-Osservatorio di Astrofisica e Scienza dello Spazio di Bologna, Via Piero Gobetti 93/3, I-40129 Bologna, Italy\\
$^{4}$ Faculdade de Ci\^encias da Universidade do Porto, Rua do Campo de Alegre, 4150-007 Porto, Portugal\\
$^{5}$ School of Physics, HH Wills Physics Laboratory, University of Bristol, Tyndall Avenue, Bristol, BS8 1TL, UK\\
$^{6}$ Kapteyn Astronomical Institute, University of Groningen, PO Box 800, 9700 AV Groningen, The Netherlands\\
$^{7}$ INAF-Osservatorio Astronomico di Capodimonte, Via Moiariello 16, I-80131 Napoli, Italy\\
$^{8}$ Departamento de F\'isica, Faculdade de Ci\^encias, Universidade de Lisboa, Edif\'icio C8, Campo Grande, PT1749-016 Lisboa, Portugal\\
$^{9}$ Instituto de Astrof\'isica e Ci\^encias do Espa\c{c}o, Faculdade de Ci\^encias, Universidade de Lisboa, Tapada da Ajuda, PT-1349-018 Lisboa, Portugal\\
$^{10}$ Dipartimento di Fisica e Astronomia "Augusto Righi" - Alma Mater Studiorum Universit\`{a} di Bologna, via Piero Gobetti 93/2, I-40129 Bologna, Italy\\
$^{11}$ Institute of Cosmology and Gravitation, University of Portsmouth, Portsmouth PO1 3FX, UK\\
$^{12}$ INFN-Bologna, Via Irnerio 46, I-40126 Bologna, Italy\\
$^{13}$ Max Planck Institute for Extraterrestrial Physics, Giessenbachstr. 1, D-85748 Garching, Germany\\
$^{14}$ Universit\"ats-Sternwarte M\"unchen, Fakult\"at f\"ur Physik, Ludwig-Maximilians-Universit\"at M\"unchen, Scheinerstrasse 1, 81679 M\"unchen, Germany\\
$^{15}$ INAF-Osservatorio Astrofisico di Torino, Via Osservatorio 20, I-10025 Pino Torinese (TO), Italy\\
$^{16}$ Department of Mathematics and Physics, Roma Tre University, Via della Vasca Navale 84, I-00146 Rome, Italy\\
$^{17}$ INFN-Sezione di Roma Tre, Via della Vasca Navale 84, I-00146, Roma, Italy\\
$^{18}$ Dipartimento di Fisica, Universit\'a degli Studi di Torino, Via P. Giuria 1, I-10125 Torino, Italy\\
$^{19}$ INFN-Sezione di Torino, Via P. Giuria 1, I-10125 Torino, Italy\\
$^{20}$ INAF-IASF Milano, Via Alfonso Corti 12, I-20133 Milano, Italy\\
$^{21}$ Institut de F\'{i}sica d'Altes Energies (IFAE), The Barcelona Institute of Science and Technology, Campus UAB, 08193 Bellaterra (Barcelona), Spain\\
$^{22}$ Port d'Informaci\'{o} Cient\'{i}fica, Campus UAB, C. Albareda s/n, 08193 Bellaterra (Barcelona), Spain\\
$^{23}$ Institut d'Estudis Espacials de Catalunya (IEEC), Carrer Gran Capit\'a 2-4, 08034 Barcelona, Spain\\
$^{24}$ Institute of Space Sciences (ICE, CSIC), Campus UAB, Carrer de Can Magrans, s/n, 08193 Barcelona, Spain\\
$^{25}$ INAF-Osservatorio Astronomico di Roma, Via Frascati 33, I-00078 Monteporzio Catone, Italy\\
$^{26}$ Department of Physics "E. Pancini", University Federico II, Via Cinthia 6, I-80126, Napoli, Italy\\
$^{27}$ INFN section of Naples, Via Cinthia 6, I-80126, Napoli, Italy\\
$^{28}$ Dipartimento di Fisica e Astronomia "Augusto Righi" - Alma Mater Studiorum Universit\'a di Bologna, Viale Berti Pichat 6/2, I-40127 Bologna, Italy\\
$^{29}$ INAF-Osservatorio Astrofisico di Arcetri, Largo E. Fermi 5, I-50125, Firenze, Italy\\
$^{30}$ Centre National d'Etudes Spatiales, Toulouse, France\\
$^{31}$ Institut national de physique nucl\'eaire et de physique des particules, 3 rue Michel-Ange, 75794 Paris C\'edex 16, France\\
$^{32}$ Institute for Astronomy, University of Edinburgh, Royal Observatory, Blackford Hill, Edinburgh EH9 3HJ, UK\\
$^{33}$ Jodrell Bank Centre for Astrophysics, Department of Physics and Astronomy, University of Manchester, Oxford Road, Manchester M13 9PL, UK\\
$^{34}$ European Space Agency/ESRIN, Largo Galileo Galilei 1, 00044 Frascati, Roma, Italy\\
$^{35}$ ESAC/ESA, Camino Bajo del Castillo, s/n., Urb. Villafranca del Castillo, 28692 Villanueva de la Ca\~nada, Madrid, Spain\\
$^{36}$ Univ Lyon, Univ Claude Bernard Lyon 1, CNRS/IN2P3, IP2I Lyon, UMR 5822, F-69622, Villeurbanne, France\\
$^{37}$ Institute of Physics, Laboratory of Astrophysics, Ecole Polytechnique F\'{e}d\'{e}rale de Lausanne (EPFL), Observatoire de Sauverny, 1290 Versoix, Switzerland\\
$^{38}$ Mullard Space Science Laboratory, University College London, Holmbury St Mary, Dorking, Surrey RH5 6NT, UK\\
$^{39}$ Instituto de Astrof\'isica e Ci\^encias do Espa\c{c}o, Faculdade de Ci\^encias, Universidade de Lisboa, Campo Grande, PT-1749-016 Lisboa, Portugal\\
$^{40}$ Department of Astronomy, University of Geneva, ch. d\'Ecogia 16, CH-1290 Versoix, Switzerland\\
$^{41}$ Universit\'e Paris-Saclay, CNRS, Institut d'astrophysique spatiale, 91405, Orsay, France\\
$^{42}$ Department of Physics, Oxford University, Keble Road, Oxford OX1 3RH, UK\\
$^{43}$ INFN-Padova, Via Marzolo 8, I-35131 Padova, Italy\\
$^{44}$ AIM, CEA, CNRS, Universit\'{e} Paris-Saclay, Universit\'{e} de Paris, F-91191 Gif-sur-Yvette, France\\
$^{45}$ INAF-Osservatorio Astronomico di Trieste, Via G. B. Tiepolo 11, I-34143 Trieste, Italy\\
$^{46}$ FRACTAL S.L.N.E., calle Tulip\'an 2, Portal 13 1A, 28231, Las Rozas de Madrid, Spain\\
$^{47}$ Aix-Marseille Univ, CNRS/IN2P3, CPPM, Marseille, France\\
$^{48}$ Istituto Nazionale di Astrofisica (INAF) - Osservatorio di Astrofisica e Scienza dello Spazio (OAS), Via Gobetti 93/3, I-40127 Bologna, Italy\\
$^{49}$ Istituto Nazionale di Fisica Nucleare, Sezione di Bologna, Via Irnerio 46, I-40126 Bologna, Italy\\
$^{50}$ INAF-Osservatorio Astronomico di Padova, Via dell'Osservatorio 5, I-35122 Padova, Italy\\
$^{51}$ Dipartimento di Fisica "Aldo Pontremoli", Universit\'a degli Studi di Milano, Via Celoria 16, I-20133 Milano, Italy\\
$^{52}$ INAF-Osservatorio Astronomico di Brera, Via Brera 28, I-20122 Milano, Italy\\
$^{53}$ INFN-Sezione di Milano, Via Celoria 16, I-20133 Milano, Italy\\
$^{54}$ Institute of Theoretical Astrophysics, University of Oslo, P.O. Box 1029 Blindern, N-0315 Oslo, Norway\\
$^{55}$ Jet Propulsion Laboratory, California Institute of Technology, 4800 Oak Grove Drive, Pasadena, CA, 91109, USA\\
$^{56}$ von Hoerner \& Sulger GmbH, Schlo{\ss}Platz 8, D-68723 Schwetzingen, Germany\\
$^{57}$ Max-Planck-Institut f\"ur Astronomie, K\"onigstuhl 17, D-69117 Heidelberg, Germany\\
$^{58}$ Universit\'e de Gen\`eve, D\'epartement de Physique Th\'eorique and Centre for Astroparticle Physics, 24 quai Ernest-Ansermet, CH-1211 Gen\`eve 4, Switzerland\\
$^{59}$ Department of Physics and Helsinki Institute of Physics, Gustaf H\"allstr\"omin katu 2, 00014 University of Helsinki, Finland\\
$^{60}$ NOVA optical infrared instrumentation group at ASTRON, Oude Hoogeveensedijk 4, 7991PD, Dwingeloo, The Netherlands\\
$^{61}$ Argelander-Institut f\"ur Astronomie, Universit\"at Bonn, Auf dem H\"ugel 71, 53121 Bonn, Germany\\
$^{62}$ INFN-Sezione di Bologna, Viale Berti Pichat 6/2, I-40127 Bologna, Italy\\
$^{63}$ Department of Physics, Institute for Computational Cosmology, Durham University, South Road, DH1 3LE, UK\\
$^{64}$ Universit\'e C\^{o}te d'Azur, Observatoire de la C\^{o}te d'Azur, CNRS, Laboratoire Lagrange, Bd de l'Observatoire, CS 34229, 06304 Nice cedex 4, France\\
$^{65}$ Institut d'Astrophysique de Paris, 98bis Boulevard Arago, F-75014, Paris, France\\
$^{66}$ Sorbonne Universit{\'e}s, UPMC Univ Paris 6 et CNRS, UMR 7095, Institut d'Astrophysique de Paris, 98 bis bd Arago, 75014 Paris, France\\
$^{67}$ University of Applied Sciences and Arts of Northwestern Switzerland, School of Engineering, 5210 Windisch, Switzerland\\
$^{68}$ European Space Agency/ESTEC, Keplerlaan 1, 2201 AZ Noordwijk, The Netherlands\\
$^{69}$ Department of Physics and Astronomy, University of Aarhus, Ny Munkegade 120, DK-8000 Aarhus C, Denmark\\
$^{70}$ Universit\'e Paris-Saclay, Universit\'e Paris Cit\'e, CEA, CNRS, Astrophysique, Instrumentation et Mod\'elisation Paris-Saclay, 91191 Gif-sur-Yvette, France\\
$^{71}$ Institute of Space Science, Bucharest, Ro-077125, Romania\\
$^{72}$ Dipartimento di Fisica e Astronomia "G.Galilei", Universit\'a di Padova, Via Marzolo 8, I-35131 Padova, Italy\\
$^{73}$ Departamento de F\'isica, FCFM, Universidad de Chile, Blanco Encalada 2008, Santiago, Chile\\
$^{74}$ IFPU, Institute for Fundamental Physics of the Universe, via Beirut 2, 34151 Trieste, Italy\\
$^{75}$ INFN-Sezione di Roma, Piazzale Aldo Moro, 2 - c/o Dipartimento di Fisica, Edificio G. Marconi, I-00185 Roma, Italy\\
$^{76}$ Centro de Investigaciones Energ\'eticas, Medioambientales y Tecnol\'ogicas (CIEMAT), Avenida Complutense 40, 28040 Madrid, Spain\\
$^{77}$ Universidad Polit\'ecnica de Cartagena, Departamento de Electr\'onica y Tecnolog\'ia de Computadoras, 30202 Cartagena, Spain\\
$^{78}$ Infrared Processing and Analysis Center, California Institute of Technology, Pasadena, CA 91125, USA\\
$^{79}$  Universit\'e Paris Cit\'e, CNRS, Astroparticule et Cosmologie, F-75013 Paris, France\\
$^{80}$ INAF-IASF Bologna, Via Piero Gobetti 101, I-40129 Bologna, Italy\\
$^{81}$ INFN, Sezione di Trieste, Via Valerio 2, I-34127 Trieste TS, Italy\\
$^{82}$ SISSA, International School for Advanced Studies, Via Bonomea 265, I-34136 Trieste TS, Italy\\
$^{83}$ Departamento de Astrof\'{i}sica, Universidad de La Laguna, E-38206, La Laguna, Tenerife, Spain\\
$^{84}$ Instituto de Astrof\'isica de Canarias (IAC); Departamento de Astrof\'isica, Universidad de La Laguna (ULL), E-38200, La Laguna, Tenerife, Spain\\
$^{85}$ Institut de Recherche en Astrophysique et Plan\'etologie (IRAP), Universit\'e de Toulouse, CNRS, UPS, CNES, 14 Av. Edouard Belin, F-31400 Toulouse, France\\
$^{86}$ Dipartimento di Fisica - Sezione di Astronomia, Universit\'a di Trieste, Via Tiepolo 11, I-34131 Trieste, Italy\\
$^{87}$ INAF, Istituto di Radioastronomia, Via Piero Gobetti 101, I-40129 Bologna, Italy\\
$^{88}$ Dipartimento di Fisica e Scienze della Terra, Universit\'a degli Studi di Ferrara, Via Giuseppe Saragat 1, I-44122 Ferrara, Italy\\
$^{89}$ Institute for Theoretical Particle Physics and Cosmology (TTK), RWTH Aachen University, D-52056 Aachen, Germany\\
$^{90}$ Instituto de Astrof\'isica de Canarias, Calle V\'ia L\'actea s/n, E-38204, San Crist\'obal de La Laguna, Tenerife, Spain\\
$^{91}$ Department of Physics \& Astronomy, University of California Irvine, Irvine CA 92697, USA\\
$^{92}$ University of Lyon, UCB Lyon 1, CNRS/IN2P3, IUF, IP2I Lyon, France\\
$^{93}$ INFN-Sezione di Genova, Via Dodecaneso 33, I-16146, Genova, Italy\\
$^{94}$ INAF-Istituto di Astrofisica e Planetologia Spaziali, via del Fosso del Cavaliere, 100, I-00100 Roma, Italy\\
$^{95}$ Instituto de F\'isica Te\'orica UAM-CSIC, Campus de Cantoblanco, E-28049 Madrid, Spain\\
$^{96}$ Department of Physics, P.O. Box 64, 00014 University of Helsinki, Finland\\
$^{97}$ Department of Physics, Lancaster University, Lancaster, LA1 4YB, UK\\
$^{98}$ Observatoire de Paris, PSL Research University 61, avenue de l'Observatoire, F-75014 Paris, France\\
$^{99}$ Universit\'e de Paris, F-75013, Paris, France, LERMA, Observatoire de Paris, PSL Research University, CNRS, Sorbonne Universit\'e, F-75014 Paris, France\\
$^{100}$ Department of Physics and Astronomy, University College London, Gower Street, London WC1E 6BT, UK\\
$^{101}$ Code 665, NASA Goddard Space Flight Center, Greenbelt, MD 20771 and SSAI, Lanham, MD 20770, USA\\
$^{102}$ Helsinki Institute of Physics, Gustaf H{\"a}llstr{\"o}min katu 2, University of Helsinki, Helsinki, Finland\\
$^{103}$ Centre de Calcul de l'IN2P3, 21 avenue Pierre de Coubertin F-69627 Villeurbanne Cedex, France\\
$^{104}$ Dipartimento di Fisica, Sapienza Universit\`a di Roma, Piazzale Aldo Moro 2, I-00185 Roma, Italy\\
$^{105}$ Aix-Marseille Univ, CNRS, CNES, LAM, Marseille, France\\
$^{106}$ Institut f\"ur Theoretische Physik, University of Heidelberg, Philosophenweg 16, 69120 Heidelberg, Germany\\
$^{107}$ Zentrum f\"ur Astronomie, Universit\"at Heidelberg, Philosophenweg 12, D- 69120 Heidelberg, Germany\\
$^{108}$ Department of Mathematics and Physics E. De Giorgi, University of Salento, Via per Arnesano, CP-I93, I-73100, Lecce, Italy\\
$^{109}$ INFN, Sezione di Lecce, Via per Arnesano, CP-193, I-73100, Lecce, Italy\\
$^{110}$ Institute for Computational Science, University of Zurich, Winterthurerstrasse 190, 8057 Zurich, Switzerland\\
$^{111}$ Junia, EPA department, F 59000 Lille, France\\
$^{112}$ Department of Astrophysical Sciences, Peyton Hall, Princeton University, Princeton, NJ 08544, USA\\
$^{113}$ Department of Physics, P.O.Box 35 (YFL), 40014 University of Jyv\"askyl\"a, Finland\\
$^{114}$ Ruhr University Bochum, Faculty of Physics and Astronomy, Astronomical Institute (AIRUB), German Centre for Cosmological Lensing (GCCL), 44780 Bochum, Germany}

   \date{Received June 20, 2022; accepted September 15, 2022}

  \abstract{The \Euclid\ Space Telescope will provide deep imaging at optical and near-infrared wavelengths, along with slitless near-infrared
  spectroscopy, across $\sim15\,000\,{\rm deg}^2$ of the sky. \Euclid\ is expected to detect $\sim12$ billion astronomical sources, facilitating new insights into cosmology, galaxy evolution, and various other topics. In order to optimally exploit the expected very large dataset, appropriate methods and software tools need to be developed. Here we present a novel machine-learning-based methodology for the selection of quiescent galaxies using broadband \textit{Euclid} \IE, \YE, \JE, and \HE\
  photometry, in combination with multi-wavelength photometry from other large surveys (e.g. the \textit{Rubin} LSST). The \texttt{ARIADNE} pipeline uses 
  meta-learning to fuse decision-tree ensembles, nearest-neighbours, and deep-learning methods into a single classifier that yields significantly 
  higher accuracy than any of the individual learning methods separately. The pipeline has been designed to have `sparsity awareness', such that missing photometry 
  values are informative for the classification. 
  In addition, our pipeline is able to derive photometric redshifts for galaxies selected as quiescent, aided by the `pseudo-labelling' semi-supervised method,
  and using an outlier detection algorithm to identify and reject likely catastrophic outliers. After the application of the outlier filter, our pipeline achieves a
   normalised mean absolute deviation of $\protect\la 0.03$ and a fraction of catastrophic outliers of $\protect\la 0.02$ when measured against the COSMOS2015 photometric redshifts.
  We apply our classification pipeline to mock galaxy photometry catalogues corresponding to three main scenarios: (i) Euclid Deep Survey
  photometry with ancillary $ugriz$, WISE, and radio data; (ii) Euclid Wide Survey photometry with ancillary $ugriz$, WISE, and radio data;
  and (iii) Euclid Wide Survey photometry only, with no foreknowledge of galaxy redshifts. In a like-for-like comparison, our classification
  pipeline outperforms $UVJ$ selection, in addition to the Euclid \IYJH\ and \uIIJ\ colour-colour methods, with improvements in completeness and the F1-score (the harmonic mean of precision and recall) of up to a factor of 2.
  }

   \keywords{Galaxies: photometry -- Galaxies: high-redshift -- Galaxies: evolution -- Galaxies: general}

   \maketitle
%

\section{Introduction}
\label{sec:introduction}
The study of galaxies plays a pivotal role in the effort to understand how the baryonic component of the Universe evolved across cosmic
time; it is on the size scale of galaxies that key processes such as star formation, chemical evolution, black hole growth, and
feedback predominantly take place. Large, systematic imaging and spectroscopic surveys are among the most valuable resources
for investigating galaxy evolution, allowing a wide range of studies, from the identification and study of rare or elusive objects
\citep[e.g.][]{Alexandroff2013} to statistical studies of large samples of galaxies \citep[e.g.][]{Kauffmann2003}. A variety of ground- or space-based surveys have already
provided rich databases for investigating questions surrounding the evolution of galaxies, for example the Sloan Digital Sky Survey \citep[SDSS;][]{York2000,Gunn1998}, with even more data to be generated by ongoing and future campaigns and facilities that will map the extragalactic sky to
even fainter flux levels, and out to even higher redshifts, for example the \textit{Vera C. Rubin} Observatory Legacy Survey of Space and Time (LSST; \citealt{Ivezic2019}),
the \textit{Nancy Grace Roman} Space Telescope (\citealt{Akeson2019}),
the Dark Energy Spectroscopic Instrument survey (\citealt{Dey2019}),
the 4-metre Multi-Object Spectroscopic Telescope (\citealt{Guiglion2019}),
the Multi Object Optical and Near-infrared Spectrograph for the VLT (\citealt{Taylor2018,Cirasuolo2020}), the Square Kilometer Array (\citealt{Dewdney2009}), and
the extended Roentgen Survey with an Imaging Telescope Array (\citealt{Predehl2021}).
In the coming years, the \Euclid\ Space Telescope will make a substantial contribution to our understanding of galaxy evolution. 
\Euclid\ will observe $\sim15\,000\, {\rm deg}^2$ of the extragalactic sky at visible to near-infrared (NIR) wavelengths,
to a $5\,\sigma$ point-source depth of 26.2\,mag\footnote{AB magnitudes are used herein.} in the Euclid VISible Instrument \citep[VIS;][]{Cropper2016} 
\IE\ ($R$+$I$+$Z$) filter
and 24.5\,mag in the Near Infrared Spectrometer and Photometer \citep[NISP;][]{Maciaszek2016} \YE, \JE, and \HE\ filters \citep{Scaramella2021,Schirmer2022}. In addition, three fields with an area totalling 53 deg$^2$ will be the subject of a deeper survey, to a $5\,\sigma$ depth of 28.2\,mag in \IE\ and 26.5\,mag in \YE, \JE, and \HE. \Euclid\ is expected to detect $\sim12$ billion astronomical sources ($3\,\sigma$), providing
multi-colour imaging at $0\farcs1$--$0\farcs4$ resolution, and is also expected to obtain spectroscopic redshifts for $\sim35$ million galaxies 
\citep[e.g.][]{Laureijs2011}. While the primary science drivers of the \Euclid\ mission are baryonic acoustic oscillations, weak lensing cosmology, and redshift-space distortions, the survey is also expected to enable a multitude of high-impact extragalactic science projects, either in stand-alone form or in combination with multi-wavelength data from other surveys (e.g. LSST). 

Automated classification and derivation of physical properties are crucial steps towards the scientific exploitation of data from any large astronomical survey,
and traditionally this would be done using colour-colour methods \citep[e.g.][]{Haro1956,Daddi2004,Leja2019}
or by fitting spectral models or templates \citep[e.g.][]{Bolzonella2000,Fotopoulou2012,Gomes2017}. Machine-learning techniques have
proven to be particularly powerful in this context since they have the ability to detect structure, correlations, and outliers in large, multi-dimensional
datasets, producing models that are typically stronger and more efficient than the traditional methods
\citep[e.g.][]{Baqui2021,Ulmer2019,Logan2020,clarke2020,Cunha2022}. While machine-learning techniques have been applied to 
extragalactic problems for several decades already \citep[e.g.][]{Odewahn1993},
in recent years there has been a rapid growth in their application to this area. Notable examples include detailed morphological
classification via application of deep learning to galaxy images \citep[e.g.][]{Dieleman2015,Huertas-Company2015,DominguezSanchez2018,Tuccillo2018,Nolte2019,Bowles2021,Bretonniere2021}, high-accuracy
estimation of the redshift ($z$) of galaxies from imaging or photometric data 
\citep[e.g.][]{Collister2004,Brescia2013,Cavuoti2017,Pasquet2019,Razim2021,Guarneri2021}, selection and classification of galaxies into various phenomenological
 types \citep[e.g.][]{Cavuoti2014,Steinhardt2020}, and derivation of their physical properties
\citep[e.g.][]{Bonjean2019,Delli2019,Mucesh2021,Simet2021}. A few studies have taken a hybrid or `cooperative' approach, combining results from
traditional methods (e.g. template fitting) with machine-learning methods, resulting in improved accuracy
\citep[e.g.][]{Fotopoulou2018,Cavuoti2017}. 

\begin{figure}
  \includegraphics[width=1.0\columnwidth]{{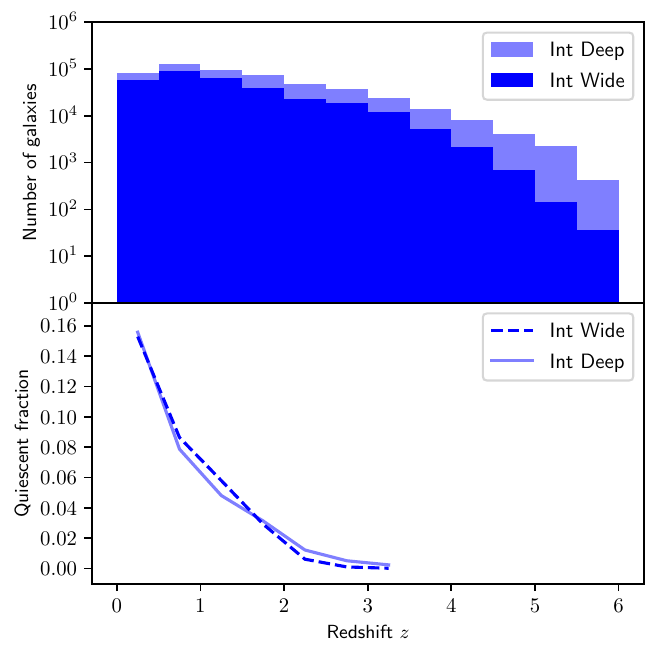}}
  \caption{Redshift distribution of galaxies in the Int Wide and Int Deep catalogues (top panel), 
  and the quiescent galaxy fraction as a function of redshift up to $z=3$ (lower panel).}
  \label{fig:z_distribution}
\end{figure}

The identification of quiescent galaxies\footnote{We adopt the specific star-formation rate (sSFR) of $10^{-10.5} {\rm yr}^{-1}$ as
the boundary between quiescent and star-forming.} is among the most challenging classification problems in extragalactic astronomy and
represents a crucial task in our quest to understand the evolution of galaxies across cosmic time. A particular problem is the fact that there exist
substantial degeneracies between stellar age, metallicity, and reddening by dust \citep[e.g.][]{Worthey1994}, causing the scattering of galaxies across
classification boundaries when using broadband spectral energy distributions (SEDs). These degeneracies can be significantly exacerbated
by the absence of redshift information. 

A frequently applied technique for the selection of quiescent galaxy candidates is $UVJ$ colour-colour selection 
\citep{Strateva2001,Baldry2004,Wuyts2007,Williams2009,Muzzin2013,vanderWel2014,Leja2019,Shahidi2020}. This method selects objects in
rest-frame $U-V$, $V-J_{\scriptscriptstyle{\rm E}}$ colour-colour space that are red because their UV-to-NIR SED is dominated by an old stellar population, as opposed to
star-forming galaxies that are reddened by dust. While the $UVJ$ selection method clearly works \citep[e.g.][]{Fumagalli2014}, there is
a substantial contamination by star-forming galaxies in the region of $\sim10$--30 per cent \citep[e.g.][]{Moresco2013,Schreiber2018,Fang2018}. Various 
observer-frame colour combinations have also been proposed for the selection of quiescent galaxies, such as the $BzK$ method \citep{Daddi2004}, or the
Galaxy Evolution Explorer (GALEX) ${\rm FUV}-V$, $V-J_{\scriptscriptstyle{\rm E}}$, and ${\rm FUV}-V$, $V-W3$ methods proposed by \citet{Leja2019}. Once selected via colour-colour techniques,
SED fitting \citep[e.g.][]{Wiklind2008,Girelli2019} and spectroscopic observations
\citep[e.g.][]{Belli2015,Glazebrook2017,Schreiber2018} can then be employed to confirm their passive nature. Alternative approaches have also been successful at selecting quiescent galaxies, such as template fitting  followed by colour selection \citep[e.g.][]{Laigle2016,Deshmukh2018}.

\begin{figure*}
  \includegraphics[width=1.015\columnwidth]{{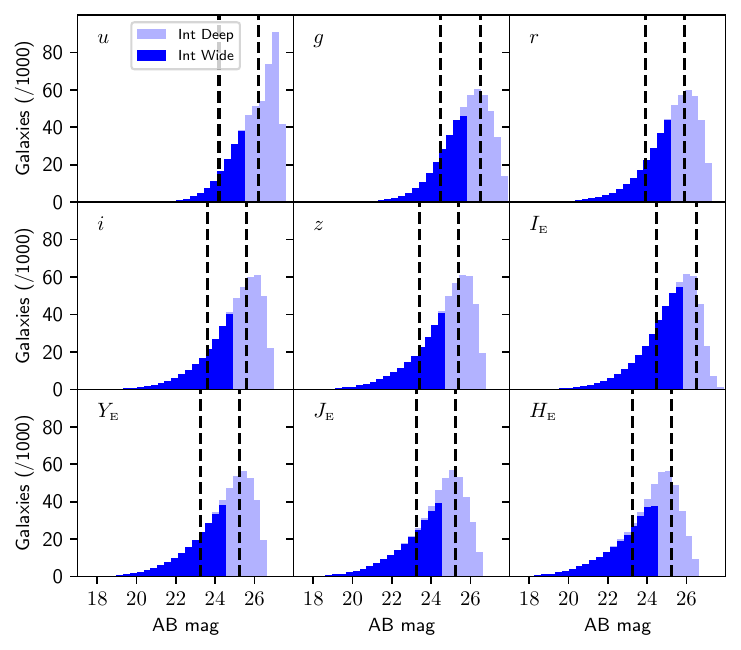}}
  \includegraphics[width=1.03\columnwidth]{{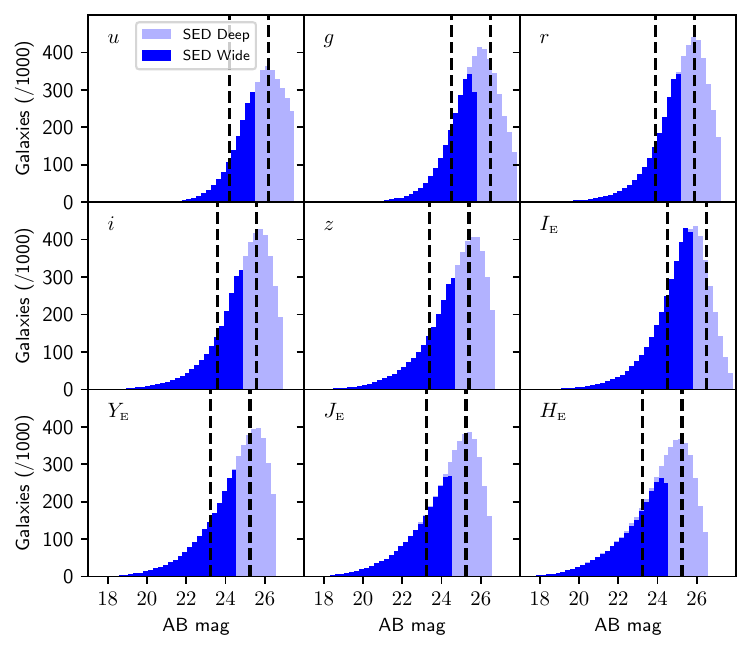}}
  \caption{Distribution of galaxy magnitudes in the Int catalogues (left) and the SED catalogues (right) for the various optical and NIR bands
    used herein. Vertical dashed lines indicate the expected $10\,\sigma$ sensitivity of the Euclid Wide and Deep Surveys in the \IE, \YE, \JE, and \HE\ bands. In all bands, only photometry with a signal-to-noise ratio $\ge 3$ is included.}
  \label{fig:mag_histogram}
\end{figure*}

In the context of preparations for the \textit{Euclid} survey, \citet[][B20 hereinafter]{Bisigello2020} recently developed \IYJH\ and \uIIJ\ 
colour-colour criteria to separate quiescent galaxies from star-forming galaxies up to $z=2.5$, for use in the case where spectroscopic or photometric 
redshifts are available. The proposed colour-colour criteria significantly outperform the traditional $UVJ$ technique that, when derived using only the 
four \textit{Euclid} filters, provides a completeness of only $\sim0.2$ at $z<3$. For example, using their \uIIJ\ criteria, B20 were able to select a sample 
of quiescent galaxies at $0.75<$ $z<1$ with a completeness of $\sim0.7$ and a precision $>0.85$. Their \IYJH\ criteria also allowed the authors to 
select quiescent galaxies at $1<z<2$ with a completeness $>0.65$ and a precision $>0.8$. 

An important limitation of colour-colour selection techniques is that they are, in effect, lossy dimensionality-reduction methods. As such, they are likely
to discard a significant quantity of otherwise useful information that is present in a broadband SED. In this context, machine-learning methods
offer a promising alternative since they are able to perform selection from within highly multi-dimensional datasets, and can be tuned to make a trade-off
between purity and completeness that is appropriate for a desired science application. Indeed, \citet{Steinhardt2020} recently explored the selection of
quiescent galaxies using the unsupervised $t$-distributed stochastic neighbour embedding method \citep{vanderMaaten2008, vanderMaaten2014}, reporting
a significant improvement over the $UVJ$ and template fitting methods. 

In this paper we present a new supervised machine-learning method for the separation of quiescent and star-forming galaxies using \textit{Euclid} and ancillary
photometry, which is designed to handle sparse data and can provide photometric redshift estimates where necessary. The paper is organised
as follows. In Sect.\,\ref{sec:mocks} we describe the mock photometry catalogues used in this study. The metrics of model quality we use are defined
in Sect.\,\ref{sec:metrics}. Full details of the \texttt{ARIADNE} pipeline are given in Sect.\,\ref{sec:pipeline}. We describe in Sect.\,\ref{sec:results} the results of
applying our separation methods to the mock photometric data. Next, in Sect.\,\ref{sec:colour_colour_comparison} we compare our method with colour-colour methods
previously proposed by B20 and others. In Sect.\,\ref{sec:further_analysis} we summarise results from a number of further analyses and tests, 
with full details given in Appendix\,\ref{sec:further_analysis_appendix}. Finally, in Sect.\,\ref{sec:conclusions} we summarise our results and conclusions.

\section{Mock galaxy catalogues}
\label{sec:mocks}

In this work we make use of an updated version of the mock catalogues presented in B20. All catalogues were derived from magnitudes
in the COSMOS2015 multi-wavelength catalogues \citep{Laigle2016}. Objects labelled as stars or X-ray sources, and objects with inadequate\footnote{Sources that 
are `masked in optical broad-bands' in the COSMOS2015 catalogue.} optical photometry,
being removed. After this selection, the COSMOS2015 catalogue contains 518404 objects up to $z=6$. We now briefly introduce the two methods used to derive 
Euclid-like mock catalogues.

In the first method, we interpolated over the observed COSMOS2015 photometry to produce a broken-line template running from ultraviolet to infrared wavelengths,
which is then convolved with the filters of interest to derive mock photometry. As this method is affected directly by the COSMOS2015 photometric errors, which
are similar or larger than those expected in the Euclid Wide Survey, we did not include any additional photometric scatter. The 20\,cm Very Large Array (VLA)
radio continuum flux, where measured, is also included without modification. Using this approach, we derived two separate catalogues, resembling the Euclid Wide Survey \citep{Scaramella2021} and Euclid Deep Survey. We refer to these mock catalogues as `Int Wide' and `Int Deep', respectively.

The question of the potential impact of emission lines on the Int catalogues was discussed previously by B20. In short, the broadband magnitudes in the 
Int and SED catalogues include contributions from emission lines. Because the Int catalogues were constructed using real observed photometry, this means that 
for some bands and some galaxies, a contribution from nebular emission is present. In the case of the SED catalogues, the \texttt{LePhare} code was used with emission 
lines included \citep[see][]{Ilbert2006}, allowing nebular emission to contribute to the mock magnitudes. 

Nevertheless, we do not expect the 
inclusion (or exclusion) of emission lines to have a significant effect on our results. Given the widths of the filters considered herein \citep{Schirmer2022}, 
the effect of emission lines is marginal: observed equivalent widths larger than 350\,\AA, 260\,\AA, 390\,\AA, and 480\,\AA~would be required to produce a boost 
of $\sim0.1$ mag in the VIS, Y, J, and H bands, respectively; such high equivalent widths are rare in the sample of galaxies used for our mock catalogues 
\citep[e.g.,][]{Amorin2015}

\begin{table*}
  \caption{$10\,\sigma$ depth in AB magnitudes of the Wide Survey for the filters included in the mock catalogues.}
        \centering

        \begin{tabular}{cccccccccccc}
            \hline
               \IE\ & \YE\ & \JE\ & \HE\ & CFIS/$u$ & SDSS/$g$ & SDSS/$r$ & SDSS/$i$ & SDSS/$z$ & $W$1    & $W$2 \\
        \hline
            24.5 & 23.25    & 23.25    & 23.25    & 24.20    & 24.50    & 23.90    & 23.60    & 23.40    & 18.39 & 18.04 \\
            \hline
        \end{tabular}
        \label{tab:depth}
	\tablefoot{The Deep Survey is expected to be two magnitudes deeper than the Wide survey in the Euclid, CFIS, and SDSS bands.}
\end{table*}

In the second approach, we used the public code \texttt{LePhare} \citep{Arnouts2007,Ilbert2006} to fit the COSMOS2015 photometry with a large set of
\citet{Bruzual2003} templates, with the redshift fixed at its COSMOS2015 value from \citet{Laigle2016}. In particular, we considered templates 
with two different metallicities ($Z_{\odot}$ and 0.4 $Z_{\odot}$), exponentially declining
star-formation histories with an e-folding timescale $\tau$ varying from 0.1 to 10 Gyr, and ages between 0.1 and 12 Gyr. For the dust extinction, we considered
the reddening law of \citet{Calzetti2000} with 12 values of colour excess from 0 to 1. For each galaxy, we obtained the best SED template applying a $\chi^{2}$
minimisation procedure and we convolved the resulting template with the filters of interest to calculate the desired mock photometry. In effect, the resulting
photometric SED is a synthetic representation of the observed one. For each galaxy, we derived ten mock galaxies by randomising the mock photometry within the
expected photometric errors. As with the first approach, here we derived two different catalogues, one for the Euclid Wide Survey (SED Wide) and one for the 
Euclid Deep Survey (SED Deep). Because the templates do not extend into the radio regime, the SED Wide and SED Deep mock catalogues do not include the 20\,cm radio band. 

We included the VIS \IE\ filter, the NISP \YE, \JE, and \HE\ filters \citep{Schirmer2022}, and the Canada-France Imaging Survey $u$ filter (CFIS/$u$), 
as previously presented in B20. In addition, 
we derived mock fluxes for the SDSS \citep{Gunn1998} $griz$ filters and the Wide-field Infrared Survey Explorer
\citep[WISE;][]{Wright2010} filters at 3.4 and $4.6\,\mu$m ($W$1 and $W$2). For these two WISE filters, we considered the $5\,\sigma$ observational depths of the WISE All-Sky survey: 0.08 and 0.11 mJy, respectively \citep{Wright2010}. For the other filters we instead assumed the observational depths reported in the Euclid Red Book 
\citep{Laureijs2011}, which are expected to be reached with a variety of ground-based telescopes, for which we use the SDSS filters as proxies. The complete list of observational depths
are listed in Table \ref{tab:depth}. Our choice of depths for the Euclid and ground-based photometry is motivated primarily by the need to make a direct comparison with the colour-colour results of B20. Thus, we adopted the depths used in B20\footnote{The \textit{Euclid} photometric depths adopted herein differ slightly from the most recent forecasts: the \IE\ photometry is now forecast to be 0.5\,mag deeper, and the NISP photometry 0.25\,mag deeper \citep{Scaramella2021}, compared to the values
 adopted herein. In the case of ground-based optical photometry overlapping the Euclid survey areas there is no single forecast, since the photometry is expected to come from several different surveys. For instance, at the time the Euclid DR3 release, the UNIONS survey \citep{Chambers2020} is forecast to be 0.6\,mag shallower in $u$ and 0.2\,mag deeper in $r$ compared to the Wide Survey depths adopted herein, with similar depths in the $g$, $i$ and $z$ bands. In the Southern Hemisphere, LSST is expected to provide deeper optical data.}. 

For all bands, only photometry measurements with a signal-to-noise ratio of $\ge3$ are considered. This threshold has been chosen to ensure that only reliable
measurements are used. The use of low signal-to-noise data deserves a detailed and in-depth study, and is under investigation for a future paper.  
For all bands except \IE, non-detections are flagged as missing; objects for which \IE\ has a signal-to-noise below 3 are excluded from the catalogues. 
We refer to B20 for further details on the creation of the mock catalogues.

The specific star-formation rate (sSFR) was derived for each galaxy at $z\le3$ by fitting the 30-band photometric SED of \citet{Laigle2016} using \texttt{LePhare}. Further details of this process are given in B20. The dividing line between `quiescent' and `star-forming' in terms of sSFR is 
somewhat arbitrary, and various different thresholds are in use in the literature, although these are usually in the range 
$< 10^{-10}$ yr$^{-1}$ \citep[e.g.][]{Wu2018} to $< 10^{-11}$ yr$^{-1}$ \citep[e.g.][]{Ilbert2013}. 
For consistency with B20, we define quiescent to mean that a galaxy has sSFR $< 10^{-10.5}$ yr$^{-1}$, and star-forming to mean that
a galaxy has sSFR $\ge 10^{-10.5}$ yr$^{-1}$. In any case, the classification metrics we obtain herein are not significantly dependent on which 
value of sSFR we adopt for the threshold between quiescent and star-forming galaxies, for values of the threshold between $10^{-10}$ yr$^{-1}$ and
 $10^{-11}$ yr$^{-1}$. 

For the redshift labels we adopt the 30-band COSMOS photometric redshifts estimated by \citet{Laigle2016}. For the SED catalogues, the redshift labels 
represent the true redshifts (i.e. with no uncertainty), because the mock photometry is derived directly from templates with known redshifts. 
On the other hand, in the case of the Int catalogues, the redshift labels are merely photometric redshift estimates, and thus have an uncertainty with systematic 
and random components.

\begin{figure*}
  \includegraphics[width=2\columnwidth]{{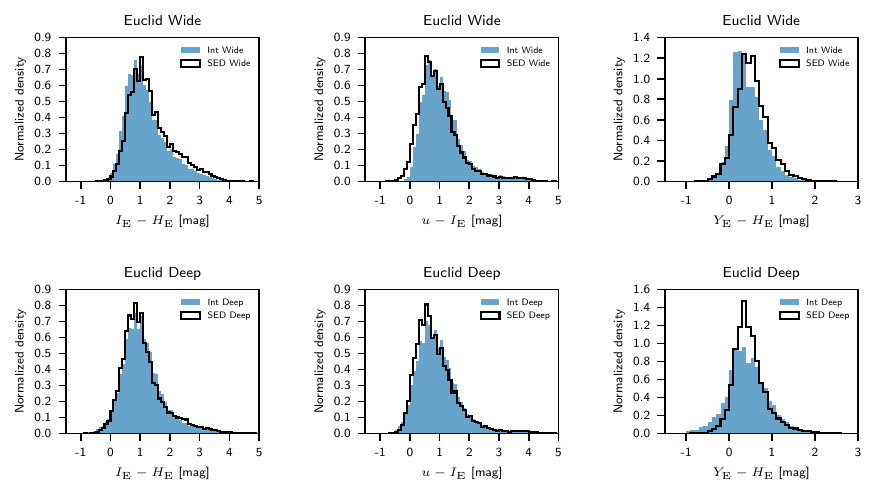}}
  \caption{Comparison between the distribution of the $I_{\scriptscriptstyle{\rm E}}-H_{\scriptscriptstyle{\rm E}}$, $u-I_{\scriptscriptstyle{\rm E}}$, and 
  $Y_{\scriptscriptstyle{\rm E}}-H_{\scriptscriptstyle{\rm E}}$ colours in the Int and SED mock catalogues.
    {\bf Top row:} \textit{Euclid} Wide catalogues. {\bf Bottom row:} \textit{Euclid} Deep catalogues. Significant differences between the Int and SED methods are apparent, 
    most notably with the SED method giving rise to significantly bluer $u-I_{\scriptscriptstyle{\rm E}}$ colours compared to the Int method. These differences arise
    when the galaxy templates are unable to closely match the observed broadband SED; this may be due to the photometric redshift being
     incorrect and/or due to the absence of a template that sufficiently represents the properties of the galaxy.}
  \label{fig:int_vs_sed_colours}
\end{figure*}

\begin{figure*}
   \includegraphics[width=2\columnwidth]{{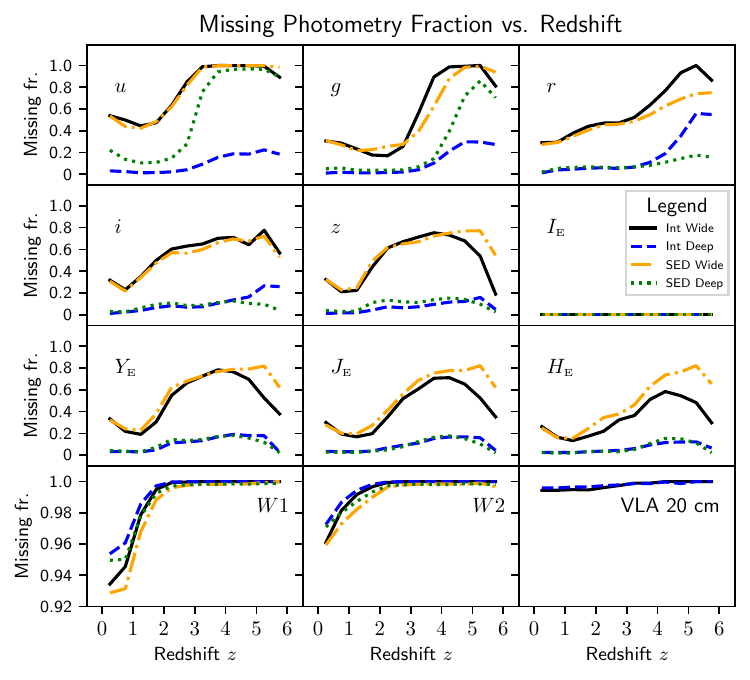}}
  \caption{Fraction of missing photometry measurements vs. redshift for each of the filters used for model training. 
  By construction, the \IE\ band has no missing values. Very few of the galaxies are detected in the $W1$, $W2$, or VLA 20\,cm bands.
    }
  \label{fig:missing}
\end{figure*}

We characterise the properties of the mock catalogues in Figs.\,\ref{fig:z_distribution}--\ref{fig:missing}. In Fig.\,\ref{fig:z_distribution} 
we show the redshift distribution of galaxies. Also shown is the distribution and fraction of quiescent galaxies as a function of redshift, up to $z=3$. 
It can be seen that the quiescent fraction falls rapidly with increasing redshift, starting at $\sim0.16$ at $z\sim0$ and declining to $\le0.05$ by 
$z\sim2.5$, illustrating the highly challenging nature of the search for quiescent (or passive) galaxies at high redshift. 

The distribution of magnitudes is shown in Fig.\,\ref{fig:mag_histogram}, with the expected $10\sigma$ detection limits in the Euclid filters marked. 
This figure shows that the Wide mock catalogues are complete down to the $3\sigma$ detection limit in all optical and NIR bands. However, the Deep catalogues 
are not complete, due to the limits of the COSMOS2015 photometry catalogue. 

Figure\,\ref{fig:int_vs_sed_colours} shows the distribution of $I_{\scriptscriptstyle{\rm E}}-H_{\scriptscriptstyle{\rm E}}$, $u-I_{\scriptscriptstyle{\rm E}}$, and 
$Y_{\scriptscriptstyle{\rm E}}-H_{\scriptscriptstyle{\rm E}}$ colours in the four mock catalogues. There are slight colour differences between the Int and SED catalogues that arise due to their different construction methodologies. For instance, the SED catalogues 
are slightly bluer in terms of their average observer-frame $u-I_{\scriptscriptstyle{\rm E}}$ colours, compared to the Int catalogues. 

In Fig.\,\ref{fig:missing} we show the fraction of missing photometry measurements as a function of redshift for the optical and infrared bands. In the case of our mock catalogues, photometry is flagged as missing when (i) a galaxy or area is unobserved in that band, (ii) a galaxy or area was masked, (iii) a photometry measurement falls below the detection threshold of the catalogue, or (iv) a photometry measurement has a signal-to-noise ratio lower than 3. In general terms, the missing fraction is higher for higher redshifts, although in some cases (notably the NIR bands) there is a turnover and decrease in the missing fraction at very high redshifts ($z \ga 4$). The WISE $W$1 and $W$2 bands and the 20\,cm radio band data are highly sparse, with 
missing fractions of 0.947, 0.976, and 0.992, respectively. Conversely, the \IE\ band has a missing fraction of exactly 0, since detection in this band is 
a requirement for inclusion in our mock catalogues. 

\begin{figure}[!h]
  \centering
  \includegraphics[width=0.95\columnwidth]{{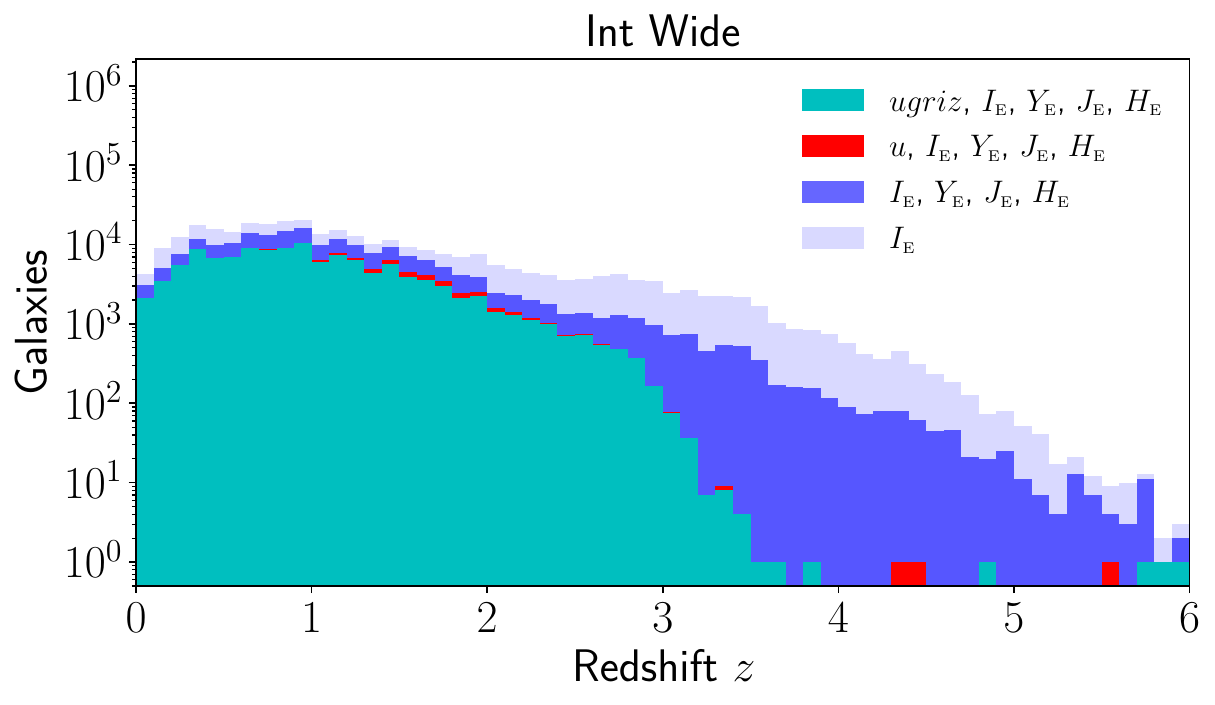}}
  \includegraphics[width=0.95\columnwidth]{{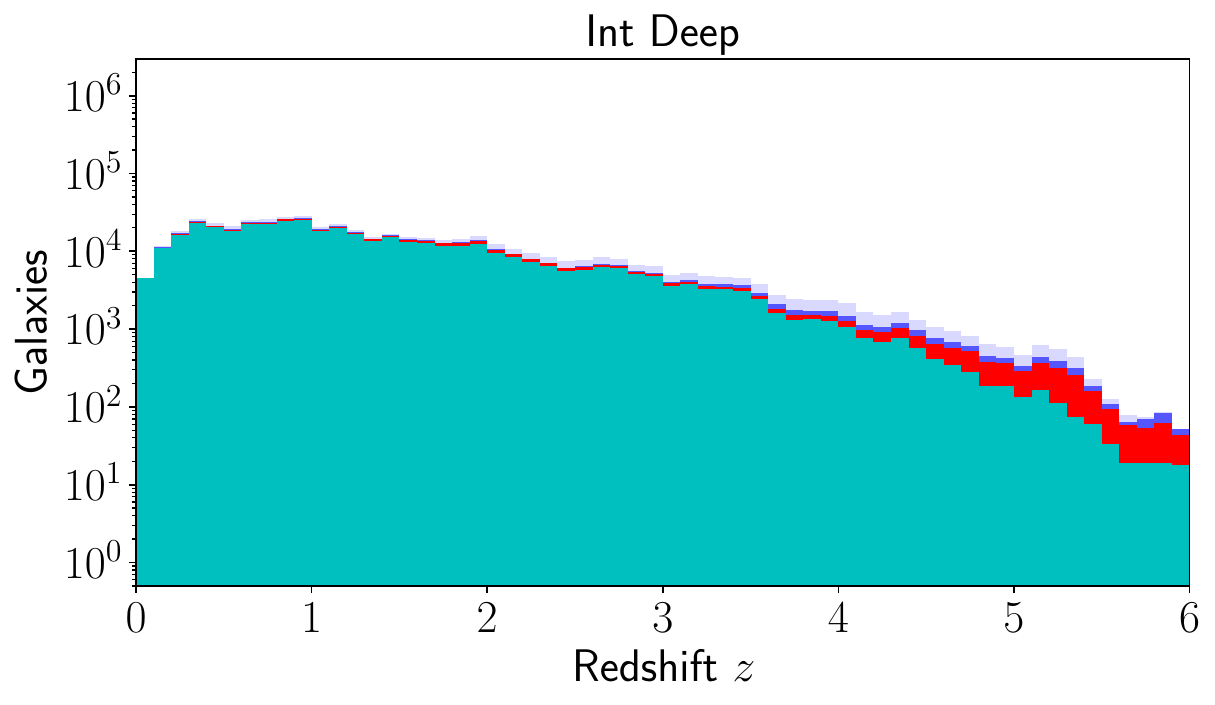}}
  \includegraphics[width=0.95\columnwidth]{{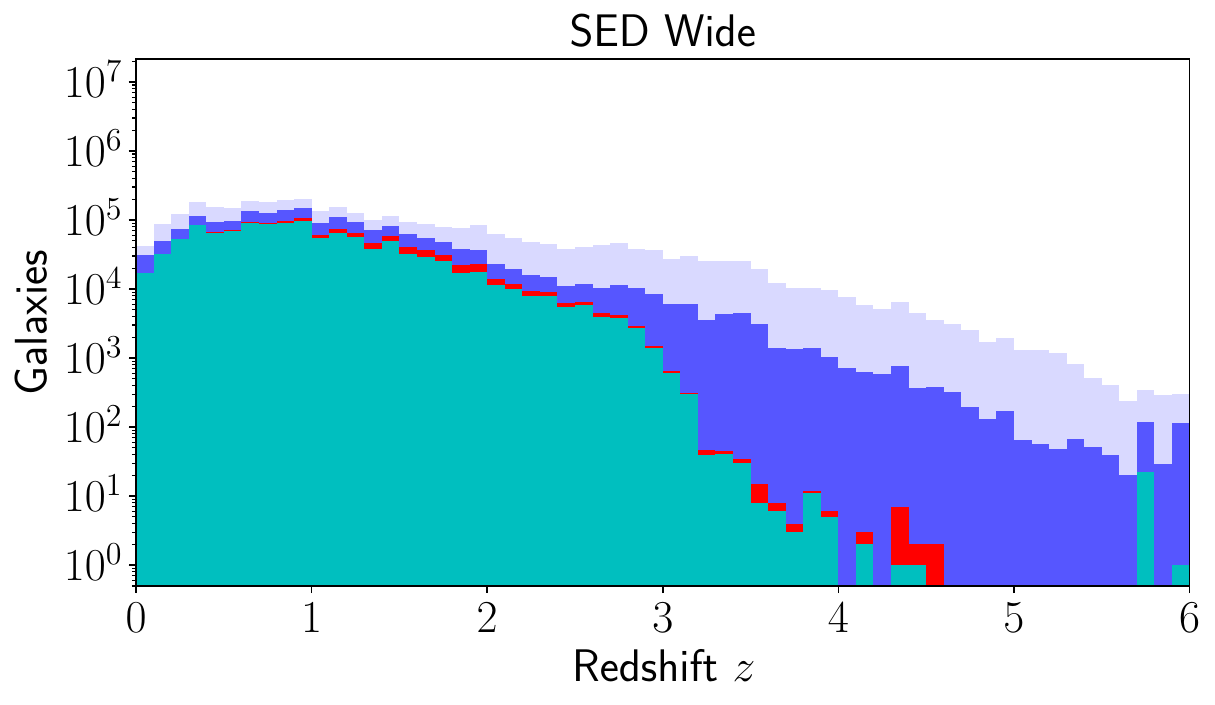}}
  \includegraphics[width=0.95\columnwidth]{{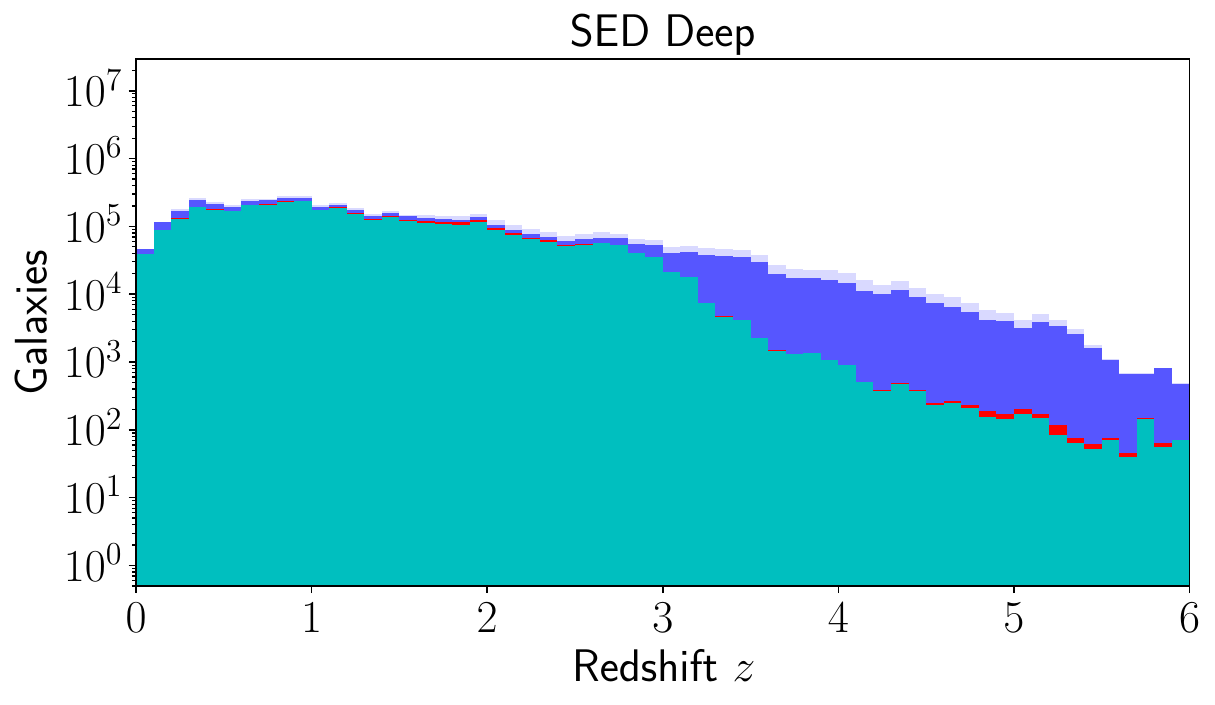}}
  \caption{Effect of various photometric pre-selection criteria on the number and redshift distribution of galaxies in the four mock catalogues.
  We show four cases: (i) only a detection in \IE\ is required; 
  (ii) detection is required in all four \textit{Euclid} bands (\IE, \YJH); 
  (iii) detection is required in $u$ and all four of the \textit{Euclid} bands;  and
  (iv) detection is required in $ugriz$ and all of the \textit{Euclid} bands.
  As described in the main text, for all bands we adopt a signal-to-noise detection threshold $\ge3$.}
  \label{fig:photometry_reqs}
\end{figure}

Finally, Fig.\,\ref{fig:photometry_reqs} shows the impact on the number and redshift distribution of galaxies from application of the main photometric pre-selection criteria used herein: (i) $3\,\sigma$ detections in \IE, \YJH; (ii) $3\,\sigma$ detections in $u$, \IE, \YJH; and (iii) $3\,\sigma$ detections in $ugriz$, \IE, \YJH. As expected, the impact of requiring detection in the $u$ band is to induce a step in the distribution at $z\sim3$, as the Lyman break is redshifted out of the $u$-band filter; this step is much stronger
in the Wide catalogues than in the SED catalogues. In the Int Deep catalogue, the reduction in the number of
sources at $z\ga3$ when detection in $u$ or $ugriz$ is required is surprisingly small; this is likely to be at
least partly caused by the presence of sources with incorrect photometric redshifts in the COSMOS2015 catalogue.  
Table ~\ref{tab:detections} lists the number of galaxies that are detected in each band for each of the mock catalogues.

\section{Metrics of model quality}
\label{sec:metrics}
To evaluate our classification models, we used several metrics that are useful to quantify model quality. Precision, $P$, is the fraction of assignments to a particular class that are correct, calculated as

\begin{equation}
    P = \frac{\rm TP}{\rm TP + \rm FP}\,,
        \label{eq:P}
\end{equation}

\noindent where TP is the number of true positives and FP is the number of false positives. Precision is also sometimes known as `purity' in astronomy, or the `positive predictive value'.

Recall, $R$, is the fraction of galaxies of a particular class within the dataset that are correctly classified as such. It is calculated as

\begin{equation}
    R = \frac{\rm TP}{\rm TP + \rm FN}\,,
        \label{eq:R}
\end{equation} 

\noindent where FN is the number of false negatives. Recall is also sometimes known as `completeness' or `sensitivity'.

The F1-score is the harmonic mean of the precision and the recall and, as such, provides a more general metric for model quality \citep{Dice1945,Sorensen1948}. The
F1-score is calculated as

\begin{equation}
    {\rm F1{\text -}score} = 2\;{P\,R \over P+R}\,,
        \label{eq:F1}
\end{equation}

\noindent where we have opted to give equal weights to $P$ and $R$. The metrics $P$, $R$, and F1-score have values between 0 and 1. Since there is a large imbalance between classes in the datasets used here, we compute the metrics separately for each
class. Unless otherwise stated, the values of $P$, $R$, and the F1-score we quote are computed for the quiescent galaxy class only. 

To assess the quality of photometric redshift estimates produced by our pipeline, we used the following measures. As a measure of accuracy, we
used the normalised median absolute deviation (NMAD), which we calculated as

\begin{equation}
    \mbox{NMAD} = 1.48 \, \mbox{median}\,\left(\frac{\,|z_{\rm phot}-z_{\rm ref}|\,}{1+z_{\rm ref}}\right)\,,
        \label{eq:nmad}
\end{equation}

\noindent where $z_{\rm phot}$ is our photometric redshift and $z_{\rm ref}$ is the reference redshift used as the `ground truth'. In this study, we adopt the 30-band photometric redshifts from COSMOS2015 \citep{Laigle2016} for $z_{\rm ref}$.
 The NMAD can be loosely interpreted as the standard deviation. 
 
 As a further measure of quality, we also calculated the fraction of catastrophic outliers
($f_{\rm out}$). A photometric redshift estimate is considered to be a catastrophic outlier when

\begin{equation}
  \frac{|z_{\rm phot}-z_{\rm ref}|}{1+z_{\rm ref}} > 0.15.
        \label{eq:f_out}
\end{equation}

To test whether (and to which extent) our pipeline systematically over- or underestimates galaxy redshifts, we defined the bias of the photometric redshifts as

\begin{equation}
  \mbox{bias} = \mbox{median}\left(\frac{z_{\rm phot}-z_{\rm ref}}{1+z_{\rm ref}}\right).
        \label{eq:bias}
\end{equation}

It is also useful to have a metric that quantifies the degree to which a false positive classification is in error. We define the `incorrectness' of an
individual false positive classification as

\begin{equation}
    I_{\rm FP} = \logten({\rm sSFR ~yr}) + 10.5.
        \label{eq:I}
\end{equation}

We considered a false positive to be `marginal' when $I_{\rm FP} \le 0.5$, and `catastrophic' when $I_{\rm FP} \ge 1.0$. We also defined an average incorrectness parameter as

\begin{equation}
    \bar{I}_{\rm FP} = \sum\limits_{i=0}^{n}\frac{{\logten({\rm sSFR ~yr}) + 10.5}}{{\rm FP}}\,.
        \label{eq:Ibar}
\end{equation}

\noindent We have not used the commonly used `accuracy' metric, which gives the fraction of 
predictions that are correct, because it can be misleading when the test data are significantly imbalanced, as is the case here.

\section{The ARIADNE pipeline}
\label{sec:pipeline}
\texttt{ARIADNE} is a flexible, modular machine-learning pipeline designed for the purpose of classification and derivation of 
physical properties of astronomical sources on the basis of their photometric SEDs. In a nutshell, \texttt{ARIADNE} takes a table of photometry as input, and uses
algorithms to learn how the input data maps to labels corresponding to galaxy properties. Here we describe the functionality of \texttt{ARIADNE} 
in its classification mode (a flowchart is also shown in Fig.\,\ref{fig:flow_chart}); its application to estimation of physical properties, such as stellar mass, star-formation rate, and extinction, 
will be presented in a future publication (Humphrey et al., in prep.).

\subsection{The learning algorithm}
\label{sec:algo}

After completion of the feature engineering and preprocessing steps described in Sect.\,\ref{sec:feat_eng}, the data are split into a training set and test set with
a 2:1 ratio between the two. This ratio was chosen to give a good balance between having a large number of examples on which to train
the learners, while still having a test set that is representative of the whole dataset\footnote{When applied to the actual Euclid survey data it is expected
  that the ratio used for the train test split may be of order $\sim1:10\,000$}. The split is done randomly, without attempting to balance target labels,
redshift, or any other observational properties, and each execution of the pipeline produces a different train-test split.

\begin{figure*}
\includegraphics[width=2\columnwidth]{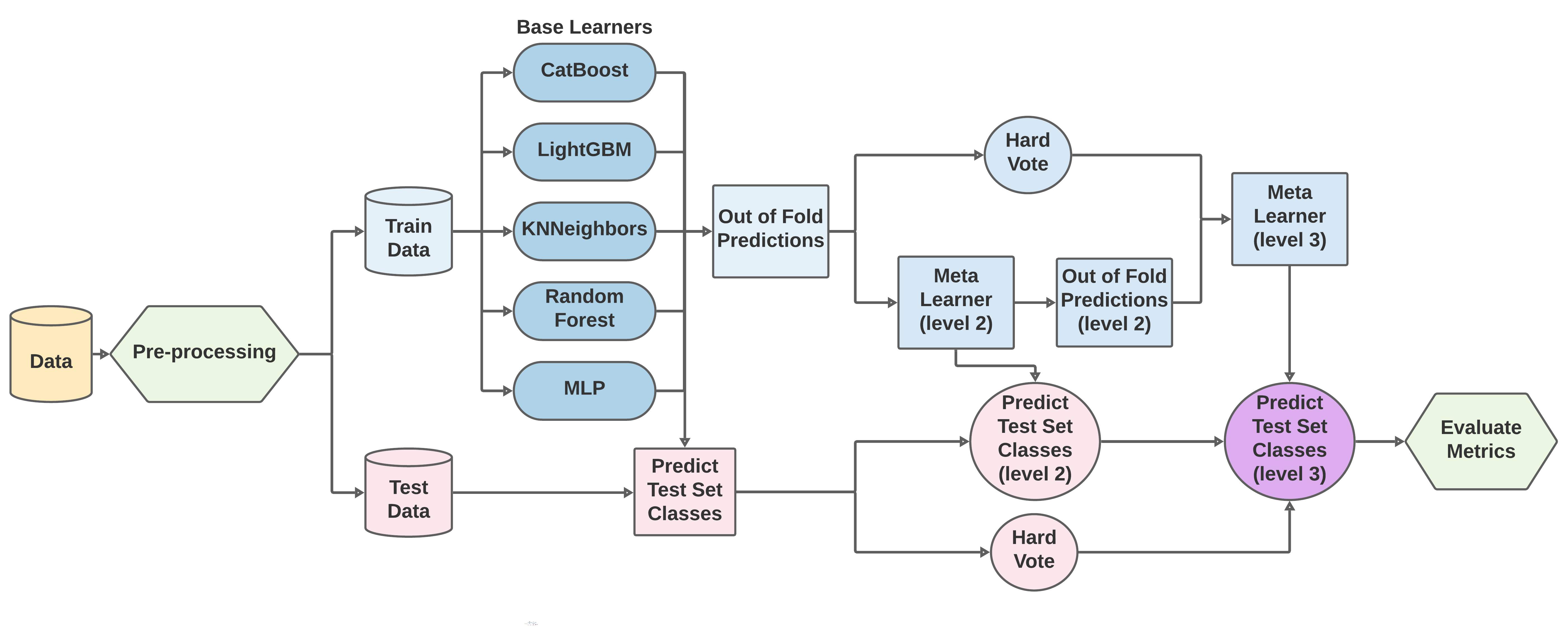}
\caption{Flow chart illustrating the overall learning algorithm used for the \texttt{ARIADNE} classification pipeline.}
\label{fig:flow_chart}
\end{figure*}

The pipeline then trains binary classification models on the training dataset, using five different `base-learners' (see Fig.\,\ref{fig:flow_chart}).
Three of the base-learners are tree-ensemble methods with somewhat different implementations 
(\texttt{CatBoostClassifier\footnote{\href{https://catboost.ai}{https://catboost.ai}} version 0.23.2};
 \texttt{LGBMClassifier\footnote{\href{https://lightgbm.readthedocs.io}{https://lightgbm.readthedocs.io}} version 0.90}; \texttt{RandomForestClassifier}), 
 one is nearest-neighbours-based (\texttt{KNeighborsClassifier}),
  and one is deep-learning-based (\texttt{MLPClassifier}). All are open source and are briefly summarised below.

The Python \texttt{Scikit-Learn\footnote{\href{https://scikit-learn.org}{https://scikit-learn.org}} version 0.22.2} package \citep{Pedregosa2011} offers various machine-learning methods with excellent functionality and
inter-operability with a multitude of in-package preprocessing tools. From \texttt{Scikit-Learn} we make use of the nearest-neighbours-based,
non-parametric \texttt{KNeighborsClassifier} method, the Multi-layer Perceptron-based \texttt{MLPClassifier}, and the randomised decision-tree
classifier \texttt{RandomForestClassifier} \citep{Breiman2001}.

Several other, advanced machine-learning methods are used in our pipeline.
\texttt{CatBoost} \citep{Prokhorenkova2018} is a gradient-boosting, tree-ensemble method that offers high performance classification or
regression, using `ordered-boosting' in place of the classic boosting algorithm to significantly reduce the `prediction shift' commonly
associated with the latter. 

Arguably the most advanced tree-based learning algorithm publicly available at the time of writing is \texttt{LightGBM}, a gradient-boosting,
decision-tree method that deploys a number of key innovations that are especially relevant for the classification problems we approach in this work
\citep{Ke2017}. For instance, \texttt{LightGBM} uses leaf-wise (best-first), rather than level-wise tree-growth, for improved classification
accuracy. Also of relevance is the use of histogram-based algorithms that place continuous feature values into discrete bins, significantly
reducing training time and memory usage.

In addition to the five default base-learners described above, our pipeline also includes the option to use the tree-based method 
\texttt{XGBoost\footnote{\href{https://xgboost.readthedocs.io}{https://xgboost.readthedocs.io}} version 0.25.3} \citep{Chen2016}, for use in situations where this learning algorithm is able to train stronger models than those
 described above. The \texttt{XGBoost} algorithm uses gradient boosting and has several key innovations, including sparsity-aware split-finding, 
 which is of particular relevance for datasets containing a significant fraction of missing values (e.g. photometric non-detections, masked or
 unobserved areas). 

Our application of these learning algorithms to the data at hand reveals that no single algorithm consistently outperforms the others over the full
range of classification problems posed herein. Thus, our pipeline employs model ensembling and interactive algorithm selection. 
 
The base-learners are trained within a stratified $k$-fold procedure: the training data are shuffled and split, with replacement, into five similarly sized folds, ensuring that each fold contains the same proportion of each target class. Each base-learner is trained five times,
each time leaving out a different fold, for which a set of out-of-fold (OOF) class predictions is produced.
This results in an array of OOF predictions for each base-learner. For each iteration of the k-fold procedure, class
predictions are also produced for the test set, averaging the five sets of class predictions for each base-learner.
Unless otherwise stated, we adopt a class probability threshold of 0.5: predictions of $<0.5$ correspond to 
class 0 (star-forming galaxies), and predictions of $\ge0.5$ correspond to class 1 (quiescent galaxies).

The pipeline then performs several iterations of non-linear combination of class predictions from the individual learners. First, a
hard-voting ensemble is produced for the OOF and test predictions. Each base learner contributes one vote towards the class of a galaxy,
and the class with the highest number of votes is chosen. For the data and classification problem considered in this work, the hard-voting
ensemble almost always results in a significantly higher F1-score than any of the individual base-learners, or a simple average of the
class probabilities.

To further improve model quality, the pipeline contains our implementation of the `generalised stacking' method proposed by \citet{Wolpert1992}. A linear
discriminant analysis or \texttt{MLPClassifier} meta-learner is trained to classify galaxies using the OOF predictions from the five base-learners
as features. This is performed within a stratified k-fold procedure as described above, and produces a new set of OOF predictions for the training
dataset, in addition to a new set of predictions for the test dataset. A second iteration of meta-learning is then performed, this time training on
just two features: (i) the results of the hard-voting ensemble and (ii) the OOF predictions produced in the previous iteration of meta-learning. Finally,
the resulting model is used to predict classes for the test data. An alternative implementation of generalised stacking, applied to redshift estimation, 
is presented in \citet{Zitlau2016}.

One of the benefits of generalised stacking is that the optimisation of base-learner hyperparameters, while still necessary to some extent, is not
as crucial for the final performance of the pipeline as it would be when a single learning algorithm is used. This is partly because when the meta-learner
distills the base-classifiers into a single classifier, it performs a process broadly analogous to optimisation and model-selection. Arguably, this
process can be more efficient than traditional optimisation and model selection methods, since it is performed in a single step and is not restricted
to selecting a single model or a single set of hyperparameters, and instead can combine the strengths of several different classifiers that are best able to
model different subsets of the data. In the case where a single base-learner is used within our pipeline, the generalised stacking procedure instead serves
as an `error-correction' algorithm. 

We have not attempted to perform an exhaustive optimisation of the base-learners prior to applying our generalised stacking method. Instead, for each
learner we manually identified a set of default hyperparameters that gives what we judged to be near to the global maximum of the F1-score
for selecting quiescent galaxies from the mock catalogues. This is done to avoid biasing the individual base-learners towards producing classifiers that
all succeed (or fail) in modelling the same subset of the data, and to expand the diversity of classification models available to the meta-learner. 

In the default configuration of our pipeline, all five default base-learners are used. However, the user can instead use a subset of the base learners,
or a single base-learner, when appropriate. For example, we opt to use \texttt{XGBoostClassifier} for the selection of quiescent galaxies at
$2.5<z<3.0$, where its sparsity awareness confers a significant advantage over the other classifiers. In addition, the pipeline has a `fast mode' that uses \texttt{LightGBM} for all classification or regression tasks, at the cost of a small but significant reduction
in $P$, $R$, and F1-score ($\sim 0.01$--0.03). Timed on a mid-range laptop with a quad-core Intel i5-8350U CPU and 16 Gigabytes of RAM, the pipeline used in {fast mode} takes at total of $\sim2$ minutes to train
its classifier on a dataset with $\sim120\,000$ examples and 70 features, compared to $\sim 74$ minutes when using the default (5 base-learners) pipeline configuration. 

Our pipeline makes use of redshift information in one of several ways, depending on the classification problem that is posed. When redshifts are available,
these can be included as an additional feature in the training and test data. In addition, when the objective is to select quiescent galaxies in a specific
redshift interval, pre-binning can be used to discard objects that lie outside the desired redshift interval. 

In the absence of redshifts for the test sample, the pipeline first performs a global selection of quiescent galaxies (without regard to redshift), then
trains a \texttt{KNeighborsRegressor} model to predict photometric redshifts for the selected quiescent galaxies. The photometric redshift point estimates
are refined using our implementation of the semi-supervised `pseudo-labelling' technique \citep{Lee2013}, which aims to use both labelled and unlabelled
data to learn the underlying structure of the data, thereby improving generalisation. Finally, analogous to \citet{Fotopoulou2018,Singal2022}, a
\texttt{KNeighborsRegressor} model is trained to predict whether a galaxy's redshift estimate is a catastrophic outlier, with a tunable probability threshold
to control the strength of the outlier removal.

\begin{table*}
        \centering
        \caption{Impact on the F1-score from using one of several different imputation strategies.}
        \label{tab:impute_tests}
        \begin{tabular}{lcccccr} 
                \hline
                Learning algorithm & Imputed value:& $-99.9$ & mean & $-99.9$ or mean & median & minimum \\
        (1) && (2) & (3) & (4) & (5) & (6) \\
        \hline
        \texttt{CatBoostClassifier}        && 0.633 & 0.632 & 0.633 & 0.644 & 0.621 \\
        \texttt{LightGBMClassifier}        && 0.678 & 0.621 & 0.678 & 0.596 & 0.655 \\
        \texttt{RandomForestClassifier}    && 0.607 & 0.561 & 0.607 & 0.576 & 0.607 \\
        \texttt{MLPClassifier}             && 0.000 & 0.610 & 0.610 & 0.621 & 0.633 \\
        \texttt{KNeighborsClassifier}      && 0.519 & 0.526 & 0.526 & 0.526 & 0.519 \\
        Meta-learner ensemble of the above && 0.667 & 0.600 & 0.656 & 0.623 & 0.610 \\
        \texttt{XGBoostClassifier}        && 0.610 & 0.576 & 0.610  & 0.586 & 0.621 \\
                \hline
        \end{tabular}
	\tablefoot{In this example, we have trained models to select 
        quiescent galaxies from the Int Wide mock catalogue in the redshift bin $2 < z < 2.5$, using $ugriz$, Euclid, $W$1, $W$2, and 20 cm photometry and colours, 
        and with pre-binning by redshift, as described in Sect.\,\ref{sec:euclid_wide}. A single random seed is used for the train/test split and base learners to 
        allow a relatively controlled comparison between methods. 
        The columns are as follows: 
        (1) The learning algorithm; also shown are the final F1-scores after the models produced by the 5 default base-learners have been ensembled using meta-learners;
        (2) F1-score when missing values are imputed with the constant value $-99.9$; 
        (3) F1-score when imputing with the average value of a feature; 
        (4) F1-score when missing values are dynamically imputed with either the constant value $-99.9$ (tree-based learners) or the mean of a feature
        (\texttt{MLPClassifier} and \texttt{KNeighborsClassifier});
        (5) F1-score when imputing with the median value of a feature;
        (6) F1-score when imputing with the minimum value of a feature.
        Some F1-scores differ significantly to those presented in Sect.\,\ref{sec:results}, since here we use only a single random seed instead of averaging results 
        over multiple pipeline runs that use different random seeds. Note that in this test, \texttt{MLPClassifier} was unable to correctly identify any quiescent
        galaxies when missing values were imputed with $-99.9$; nonetheless, this failure did not appear to be detrimental to the final meta-learner ensemble.}
\end{table*}

\begin{figure*}
  \includegraphics[angle=-90,width=0.63\columnwidth]{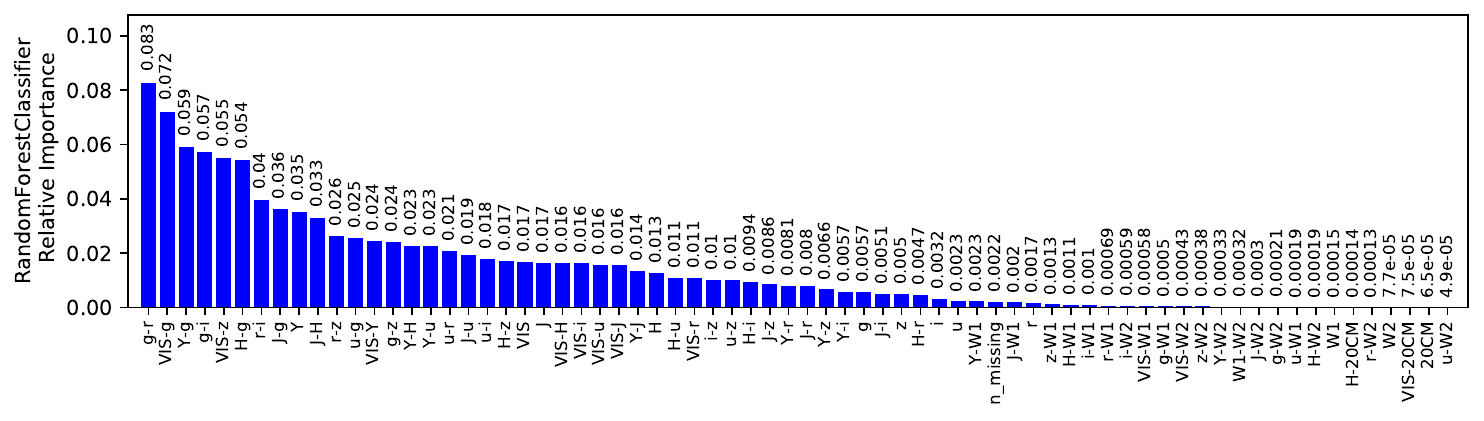}
  \includegraphics[angle=-90,width=0.63\columnwidth]{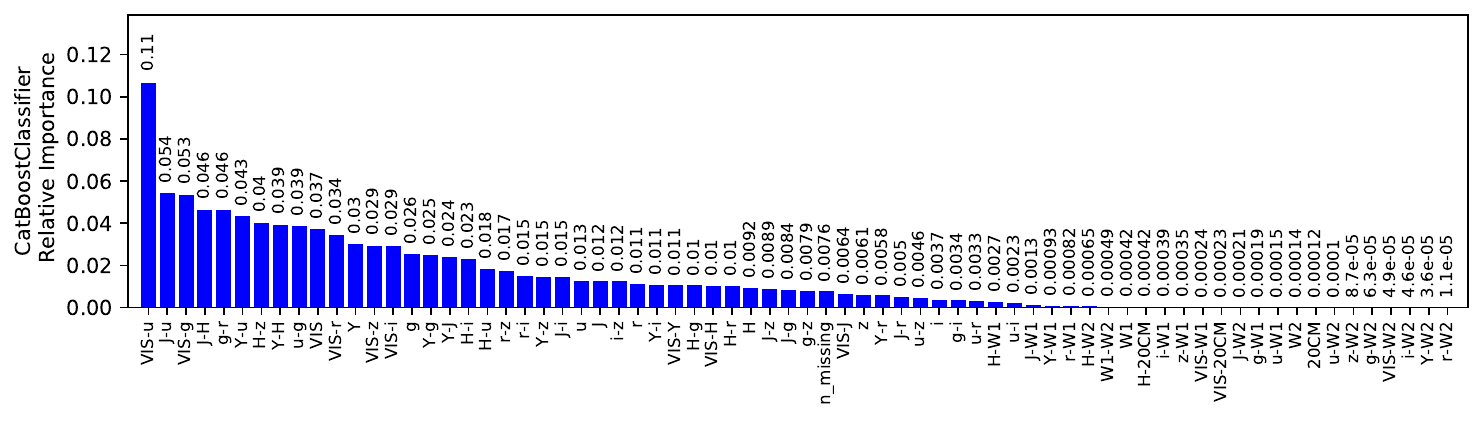}
  \includegraphics[angle=-90,width=0.63\columnwidth]{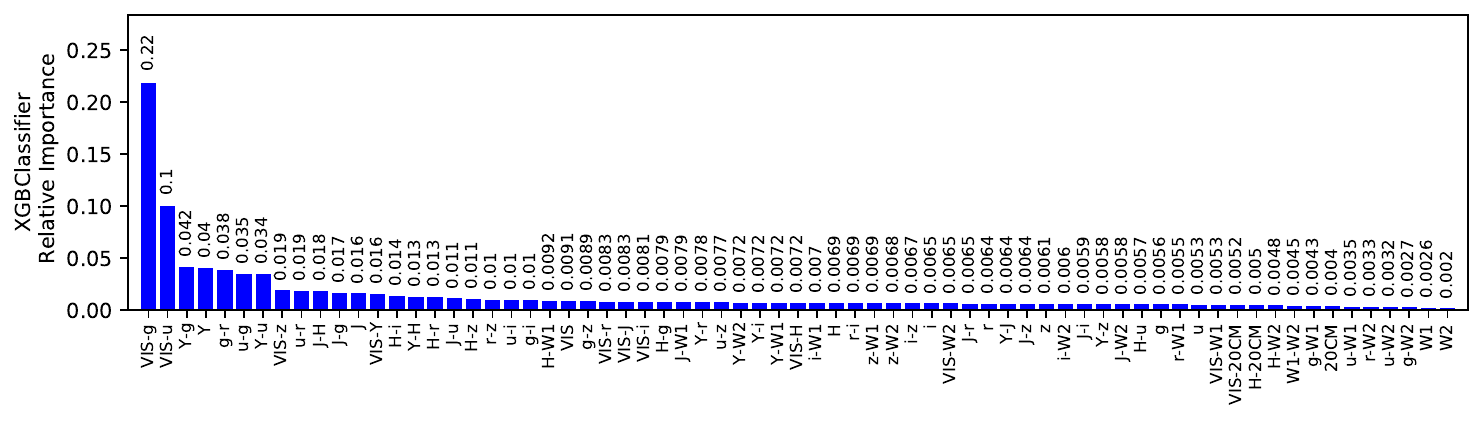}
  \caption{Examples of feature importance derived from the \texttt{RandomForest}, \texttt{CatBoost}, and \texttt{LightGBM} classifiers
    when selecting quiescent galaxies from within the $0<z<3$ interval, without foreknowledge of galaxy redshifts. For each learner, the
    feature importance values are normalised such that their sum is 1.0. The $y$ axis labels correspond to feature names used by the pipeline
    after the pre-processing steps outlined in Sect.\,\ref{sec:feat_eng} have been applied and should be self-explanatory. For example, the feature named `u' 
    is derived from the $u$-band magnitudes (i.e. after our pre-processing steps), the feature `VIS-20CM' is derived from the 
    $I_{\scriptscriptstyle{}\rm E}-20\,{\rm cm}$ colours, etc.
    }
    
  \label{feature_importance}
\end{figure*}

\subsection{Feature engineering}
\label{sec:feat_eng}
\subsubsection{Broadband colours}
Before applying our algorithm to the data, we first performed a pre-processing step known as `feature engineering', whereby the
data are enriched with information to help the algorithm learn more efficiently. We start with the broadband magnitudes and their
$1\,\sigma$ errors as base features. Because colours offer a potentially clearer description of the relative shape of the broadband
SED than magnitudes, we also calculated all possible (unique) broadband colour permutations, and included them as features. In cases where
fluxes are given instead of magnitudes (e.g. VLA radio flux measurements), we converted the values to magnitudes before deriving
the related colours. 

\subsubsection{Missing data imputation strategy}
One of the advantages of our machine-learning approach to galaxy classification is the possibility to efficiently deal with missing
data. A subset of galaxies in the mock catalogues is undetected, or unobserved, in one or more filters; we consider it highly desirable to
include them in our analysis, where possible, for several reasons: (i) such objects enlarge our training dataset; (ii) non-detection in
some bands is likely to carry useful information for our learning algorithm (e.g. $u$-band drop-outs); and (iii) the upcoming large surveys
(e.g. Euclid, LSST, etc.) that motivate this work will produce large datasets where many galaxies have missing data in one or more bands,
and such objects need to be utilised to make the most effective use of the survey data.

Here, we impute values for missing data with a method that is independent of the reason for it to be missing. When using tree-based
learning algorithms (\texttt{CatBoostClassifier}, \texttt{LightGBMClassifier}, \texttt{RandomForestClassifier}, or \texttt{XGBoostClassifier}), 
we impute the missing values with the arbitrarily chosen constant value $-99.9$. All broadband colours that would 
have used the missing value are also set to $-99.9$. Because none of the measured magnitudes or colours have this value (nor do they have 
similar values), information about the presence of missing values is thus preserved such that the tree-based learners can use non-detections to 
aid their classification of sources; this information is essentially lost if the average, median, or minimum value would be used instead for the 
imputation. Conversely, the \texttt{MLPClassifier} and \texttt{KNeighborsClassifier} learning algorithms are generally more sensitive to the 
normalisation of the input features, and imputing data with the value $-99.9$ is likely to create inappropriate and unhelpful artefacts in feature-space; 
therefore, in the cases of \texttt{MLPClassifier} and \texttt{KNeighborsClassifier} we instead impute missing values with the mean of the respective feature, 
computed across the sample of galaxies using all the non-missing values. 

We emphasise that the primary motivation behind our imputation strategy is to flag non-detections such that the learning algorithms can deduce how to 
use them most effectively; an added benefit of this strategy is that it allows the use of objects with photometric SEDs that are missing one or more bands, 
without necessarily having to discard them.

Table~\ref{tab:impute_tests} illustrates the impact on the F1-score from using one of several different imputation strategies, using a fixed random seed for 
the train/test split and base learners. Results for the following strategies are shown: imputation with the mean, the median, the minimum, a constant value of 
$-99.9$, or dynamically switching between $-99.9$ for tree-based learners and the mean for \texttt{MLPClassifier} and \texttt{KNeighborsClassifier}. In each 
case, the F1-scores are shown for the base learners and for the final stacked ensemble classifier. While the results vary significantly depending on the 
choice of random seed, the general outcome is that, as expected, the tree-based learning algorithms (\texttt{CatBoostClassifier}, \texttt{LightGBMClassifier}, 
\texttt{RandomForestClassifier}, or \texttt{XGBoostClassifier}) generally give higher F1-scores when using missing values that are imputed with $-99.9$, 
whereas \texttt{MLPClassifier} and \texttt{KNeighborsClassifier} generally produce higher F1-scores when using the mean, median, or minimum of a 
feature for imputation.

To provide the learning algorithms with additional help to treat missing data, we created a feature (\texttt{n$\_$missing}) that counts the number of 
missing magnitude values for each galaxy. Although not needed in the present study, the \texttt{ARIADNE} pipeline has the capability to make use of
categorical flags that specify the reason for each missing data point in the input data (i.e. non-detection vs. not observed or masked). 

Standardisation was performed using the \texttt{Standard Scaler} from \texttt{Scikit-Learn}, which removes the mean and scales to unit
variance. Missing values flagged with the value $-99.9$ are ignored during the standardisation procedure.

\subsubsection{Target variable}
We generate a target feature representing the binary classification of the galaxies. This feature is dynamically filled, 
depending on the specific subset of galaxies to be selected. At its simplest, the target variable is set to 0 for star-forming
galaxies and 1 for quiescent galaxies. To select quiescent galaxies in a specific redshift range, the target feature is set to 1 for
all quiescent galaxies in that redshift band, and 0 for all other galaxies. 

It is important to note that in the case of the Int catalogues, the sSFR label, and consequently the binary target variable, has 
an intrinsic uncertainty. Depending on the nature of the uncertainties, it is entirely possible that our classification methodology is 
outperforming the initial sSFR evaluation. However, a detailed analysis of this potential effect is beyond the scope of this paper.

\subsubsection{Feature importance and selection}
\label{sec:feat_imp}
It is important to ensure that our models are trained using only features that provide useful information for the prediction of the
target variable. First, we examined the feature importance information provided by three of the tree-based learning algorithms that 
we use here (\texttt{RandomForestClassifier}, \texttt{CatBoostClassifier}, and \texttt{XGBoostClassifier}). Feature importance 
provides a general picture of the relative usefulness of each feature in the construction of the resulting classification models. 
Each of the aforementioned learning algorithms uses a slightly different method to calculate feature importance values. 
\texttt{RandomForestClassifier} calculates the mean decrease in impurity when a feature is used in a split.  
\texttt{CatBoostClassifier} provides several options, from which we select \texttt{PredictionValuesChange}, which indicates the average 
change in the predicted values that result from a change in the feature value. In the case of \texttt{XGBoostClassifier}, we opt to 
use the gain, defined as the improvement in accuracy resulting from the use of a feature in the branches it is on. 

In Fig.\,\ref{feature_importance} we show examples of the feature importances resulting from training each of the three aforementioned 
tree-based learners to select quiescent galaxies in the range $0<z<3$, without 
foreknowledge\footnote{In other words, the input features for the classification models do not contain redshift values, nor are they binned or sorted by redshift. 
In cases where redshift information is included among the input features, this will be stated.} 
of galaxy redshifts. Significant differences are evident among the results in terms of the importance values themselves and feature importance rank, 
reflecting differences among the learning algorithms, the fact that many of the features are strongly correlated, and the somewhat different methods 
used to compute feature importances.

In general, the broadband colours typically show some of the highest feature importance values, indicating they are among the most informative,
as one would expect given the strong correlation between the shape of a galaxy's SED and its activity type. In addition, the broadband magnitudes 
are also clearly useful to some degree. The feature that counts missing values (\texttt{n\_missing}) also appears to be  useful, at least in some circumstances. However, the magnitude errors (not shown) show very low importance values, implying they provide little or no useful information.
An important caveat is that feature importance values do not necessarily give an accurate picture of how the inclusion (or removal) of a
particular feature affects the metric used to quantify model performance. Moreover, our feature analysis applies only to the tree-based
learning algorithms we used, and there is no guarantee that it applies to any of the other learning algorithms present in our pipeline.

Thus, to better understand the general usefulness of each of our features, we tested how removing a feature affects the
F1-score metric that we calculate after execution of the complete pipeline. We conducted split-run tests in which our algorithm is trained and
cross-validated twice, once with a particular feature removed, and once more with this feature reinstated. All other parameters were kept constant
between the two training runs, including the training-test data split and all random seeds; this split-run test process was repeated multiple
times ($>10$) for each feature, each time using a new random seed to ensure the split-run test results are not dependent on which random seed is
used. The resulting difference in the F1-scores between the two split-run test runs then indicates whether the inclusion of a feature is useful
(higher F1-score) or not (lower or unchanged F1-score).

We find that removal of any of the broadband colours or magnitude values results in a noticeable decrease in the F1-score, indicating
they are all useful for training our algorithm. While the broadband magnitudes technically provide the learning algorithms with a full
description of the broadband SED enclosed by the respective wavelength range, it is clear that explicitly providing spectral slopes in
the form of broadband colours allows significantly more accurate model training. 

We also find that inclusion of \texttt{n$\_$missing} results in a significant improvement in the F1-score, although the
size of the improvement appears to depend somewhat on the classification objective, such as the redshift range of galaxies under selection,
and the base learner used. Conversely, removing any (or all) of the magnitude error features results in a small but significant improvement
in the F1-score, indicating they are uninformative and merely add noise to the training data.
Therefore, our machine-learning pipeline trains on the following features: (i) the magnitudes; (ii) 
the broadband colours; and (iii) the \texttt{n$\_$missing} feature.

\begin{figure*}
  \includegraphics[width=0.67\columnwidth]{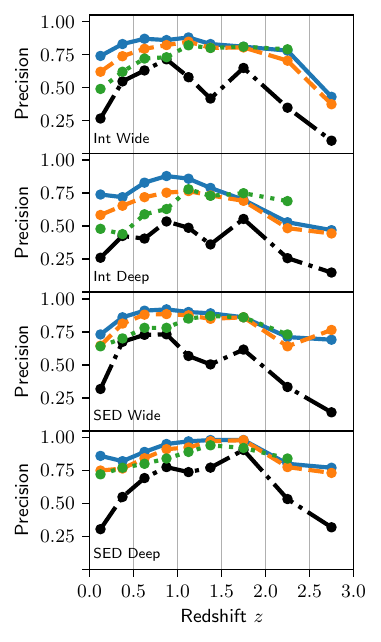}
  \includegraphics[width=0.67\columnwidth]{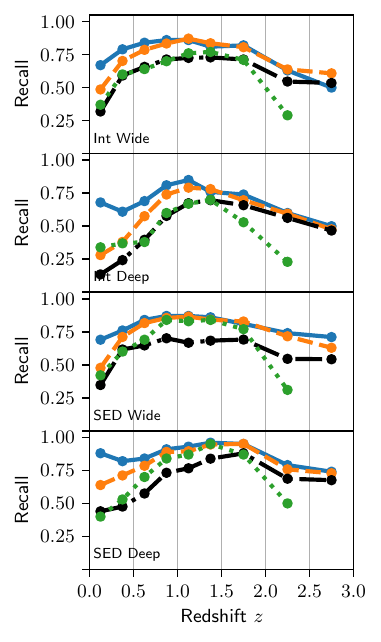}
  \includegraphics[width=0.68\columnwidth]{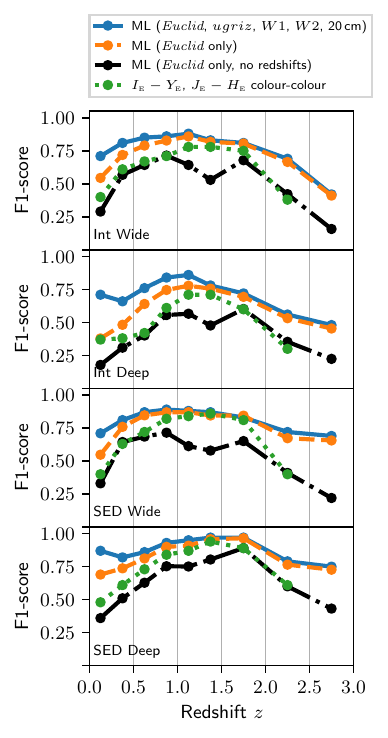}
  
  \caption{Precision, recall, and F1-score for various methods of identifying quiescent galaxies. We show results for the three cases discussed
    in Sect.\,\ref{sec:separation}: (i) Euclid Deep Survey photometry with supporting $ugriz$, $W1$, $W2$, and 20\,cm photometry; (ii) Euclid Wide Survey
    photometry with supporting $ugriz$, $W1$, $W2$, and 20\,cm photometry; and (iii) Euclid Wide Survey photometry only. Two curves are shown for the
    Euclid-only case, corresponding to results obtained either with or without foreknowledge of the galaxy redshifts. All other results shown
    in this figure were obtained assuming foreknowledge of redshifts. 
    In addition, we show the result of applying the \IYJH\ colour-colour selection method developed by B20,
    assuming foreknowledge of (photometric) redshifts.
    In this and subsequent plots showing metrics vs. redshift, the $x$ axis represents the `ground truth' photometric 
    redshifts from COSMOS2015 \citep{Laigle2016}.}
  \label{fig:bisigello_comparison_euclid}
\end{figure*}

\begin{figure*}
  \includegraphics[width=0.675\columnwidth]{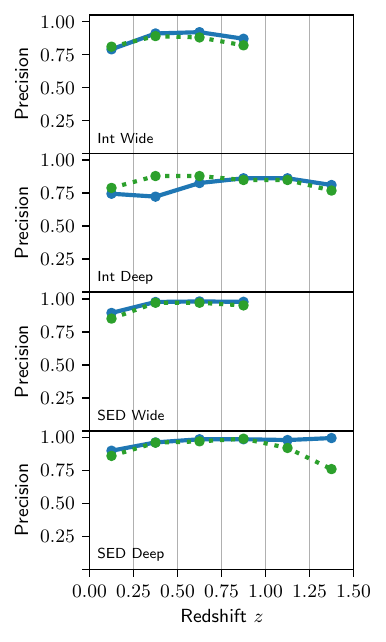}
  \includegraphics[width=0.675\columnwidth]{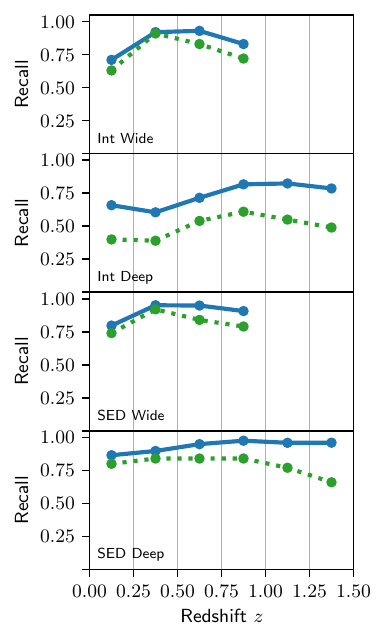}
  \includegraphics[width=0.68\columnwidth]{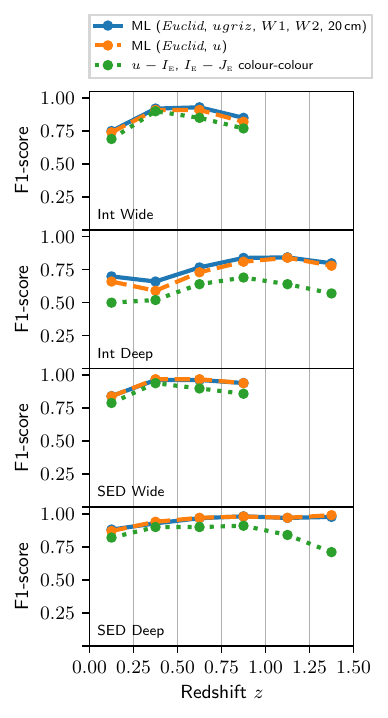}
  
  \caption{Precision, recall, and F1-score for quiescent galaxy identification methods when applied to the subset of galaxies detected
    in $u$, \IE, and \JE, using features derived from the $ugriz$, Euclid, $W$1, $W$2, and (for the Int catalogues) the 20 cm bands.
    For comparison, we show the F1-scores when only features derived from the $u$ and Euclid bands are used.
    We also show curves representing the \uIIJ\ colour-colour selection method proposed by B20.} 
  \label{fig:bisigello_comparison_euclid_u}
\end{figure*}

\begin{table*}
        \centering
        \caption{Results from global selection of quiescent galaxies at $0 \le z \le 3$, and photometric redshift estimation.}
    \label{tab:global_selection}
        \begin{tabular}{llccccccc} 
                \hline
                & & \multicolumn{3}{c}{$\overbrace{\rule{10em}{0em}}^{\text{\small classification statistics}}$} & \multicolumn{4}{c}{$\overbrace{\rule{17em}{0em}}^{\text{\small photometric redshift statistics}}$}\\
                Catalogue & Detections Required & $P$ & $R$ & F1-score & NMAD & $f_{\rm out}$ & Bias & $f_{\rm rej}$\\
                (1) & (2) & (3) & (4) & (5) & (6) & (7) & (8) & (9) \\ 
                \hline
                Int Wide        & \IE, \YJH & 0.85 & 0.75 & 0.80 & 0.027 & 0.022 & 0.0031 & 0.14 \\
                SED Wide        & \IE, \YJH & 0.88 & 0.76 & 0.82 & 0.033 & 0.017 & 0.0000 & 0.22 \\
                Wide (averaged) & \IE, \YJH & 0.87 & 0.76 & 0.81 & 0.030 & 0.019 & 0.0016 & 0.18 \\
                \hline
                Int Deep        & \IE, \YJH & 0.77 & 0.60 & 0.67 & 0.030 & 0.022 & 0.0002 & 0.44 \\
                SED Deep        & \IE, \YJH & 0.88 & 0.79 & 0.83 & 0.022 & 0.015 & $-0.0008$ & 0.29 \\
                Deep (averaged) & \IE, \YJH & 0.83 & 0.70 & 0.75 & 0.026 & 0.018 & $-0.0003$ & 0.37 \\
                \hline
                Int Wide        & $ugriz$, \IE, \YJH & 0.88 & 0.85 & 0.86 & 0.023 & 0.020 & 0.0024 & 0.20 \\                
                SED Wide        & $ugriz$, \IE, \YJH & 0.96 & 0.91 & 0.94 & 0.024 & 0.009 & 0.0008 & 0.01 \\
                Wide (averaged) & $ugriz$, \IE, \YJH & 0.92 & 0.88 & 0.90 & 0.024 & 0.014 & 0.0016 & 0.10 \\
                \hline
                Int Deep        & $ugriz$, \IE, \YJH & 0.80 & 0.63 & 0.70 & 0.030 & 0.020 & 0.0013 & 0.39 \\
                SED Deep        & $ugriz$, \IE, \YJH & 0.96 & 0.92 & 0.94 & 0.020 & 0.010 & 0.0000 & 0.01 \\
                Deep (averaged) & $ugriz$, \IE, \YJH & 0.88 & 0.78 & 0.82 & 0.025 & 0.015 & 0.0007 & 0.20 \\
                \hline
        \end{tabular}
	\tablefoot{The columns are as follows:
          (1) Mock catalogue;
          (2) bands in which galaxies are required to be detected;
          (3) precision $P$ for quiescent galaxy selection;
          (4) recall $R$ for quiescent galaxy selection;
          (5) the F1-score for quiescent galaxy selection;
          (6) normalised median absolute deviation (NMAD) for the photometric redshifts;
          (7) catastrophic outlier fraction ($f_{\rm out}$) for the photometric redshifts;
          (8) bias of the photometric redshifts;
          (9) fraction of quiescent galaxy redshifts rejected as potential catastrophic outliers ($f_{\rm rej}$).
          The typical uncertainties on the $P$, $R$, and F1-score values herein are $\le 0.01$.}
\end{table*}

\section{Results}
\label{sec:results}
\subsection{Selection in redshift bins}
\label{sec:separation}
We now apply our classification pipeline to the problem of selecting quiescent galaxies from the mock Euclid photometry catalogues.
The objective is to examine the suitability of our method for the separation of quiescent and star-forming galaxies, to facilitate
expected Euclid legacy science related to quiescent galaxies. We assume that prior to application of our pipeline, the following steps have been performed: (i) correction for Galactic extinction; (ii) pre-classification into star, active galaxy, and non-active galaxy classes; and (iii) photometric or spectroscopic redshifts have been determined, where applicable. 

To allow a direct comparison with the colour-colour methods of B20, we select quiescent galaxies in redshift bins delimited
by the values $z=0$, 0.25, 0.5, 0.75, 1.0, 1.25, 1.5, 2.0, 2.5. We include an additional bin covering the range $2.5 \le z<3$. The binning is performed using the 30-band photometric redshifts from \citet{Laigle2016}, and assumes that high
quality photometric (or spectroscopic) redshifts will be available for all the galaxies. Hereinafter, all redshift values correspond to the 30-band photometric redshifts from \citet{Laigle2016}, with the obvious exception of those derived using our pipeline in 
Sect.\,\ref{sec:global_selection}. 

A significant fraction of the Euclid survey area is expected to have deep, overlapping ground-based imaging
observations from optical surveys (e.g. LSST), but in some areas these observations may be sparse or non-existent. Therefore, we test the
performance of our pipeline under the three main expected cases in terms of photometric depth and coverage: (i) Euclid Deep Survey photometry
with supporting $ugriz$, $W1$, $W2$, and 20\,cm photometry; (ii) Euclid Wide Survey photometry with supporting $ugriz$, $W1$, $W2$, and 20\,cm photometry;
and (iii) Euclid Wide Survey photometry only. Results for these, and additional cases, are shown in Fig.~\ref{fig:bisigello_comparison_euclid}
and~\ref{fig:bisigello_comparison_euclid_u}. 

\subsubsection{Deep survey: \Euclid, $ugriz$, $W1$, $W2$, 20\,cm}
\label{sec:euclid_deep}

When selecting from the Int Deep catalogue using features derived from the Euclid, $ugriz$, WISE, and VLA photometry (blue points, second row of
Fig.~\ref{fig:bisigello_comparison_euclid}), the F1-score shows a general rise from values of $\sim0.7$ at low-$z$, before peaking at a value
of 0.86 in the $1.0<z<1.25$ bin and declining towards higher redshifts, reaching a value of 0.48 in the $2.5<z<3.0$ bin. 

Selecting from the SED Deep catalogue results in a broadly similar F1-curve (blue points, fourth row of Fig.~\ref{fig:bisigello_comparison_euclid}),
albeit with a broad plateau over the range $0.75\la z \la 2.0$, with maximum and minimum values of 0.97 and 0.75, respectively. The F1-scores
are systematically higher by $\sim0.1$--0.3 compared to values obtained in the same redshift bins using the Int Deep catalogue. 

Interestingly, at $z\ga1$ there is only a marginal reduction in F1-score when the selection is performed without the $ugriz$, WISE, and 20\,cm photometry. 
In other words, provided the redshifts are known beforehand, these bands are largely superfluous for the selection of quiescent galaxies, presumably due to the fortuitous positioning of the 4000-\AA~break within the Euclid broadband SED. 

\subsubsection{Wide survey: \Euclid, $ugriz$, $W1$, $W2$, 20\,cm}
\label{sec:euclid_wide}

The situation is broadly similar when selecting from the Wide catalogue using features derived from the Euclid, $ugriz$, WISE and 20\,cm photometry (blue points in the first and third rows of Fig.~\ref{fig:bisigello_comparison_euclid}). The F1-curve shows a gradual increase from $z=0$ to a broad peak or plateau at $0.75 \la z \la 2.0$, after which there is a gradual decline towards higher redshifts. In the case of the Int Wide selection, the maximum and minimum F1-scores are 0.87 and 0.42, respectively. For the SED Wide selection, these values are 0.89 and 0.69.

As before, the F1-scores are usually higher when selecting from the SED Wide catalogue compared to the Int Wide catalogue, with values that are up to $\sim0.3$ higher. Again, at $z\ga1$ there is only a marginal reduction in F1-score when the selection is performed without the $ugriz$, WISE, and VLA photometry. 

As discussed in Sect.\,\ref{sec:algo}, when selecting from the $2.5 < z < 3.0$ bin we used a single base-learner, \texttt{XGBoost}, together with the generalised stacking algorithm (see also Fig.~\ref{fig:stacking}, right panel). When selecting quiescent galaxies from the Wide catalogues in this redshift bin, this pipeline setup provides significantly higher F1-scores compared to the default pipeline configuration where five base-learners are employed. Although it is not immediately clear why this is the case, we suggest that its ability to understand which values are missing, and its subsequent use of missing values when performing splits, allows the \texttt{XGBoost} algorithm to build stronger classifiers than other learners when there is a high fraction of (informative) missing values in the dataset, as is the case here (see Fig.~\ref{fig:missing}). 

It is also interesting to note that the Int Wide catalogue contains only 16 quiescent galaxies in the range $2.5 \le z \le 3$ with detections in all of
the Euclid bands. As such, the training set contains on average 10.7 quiescent galaxies, and the test set 5.3, making this a `few-shot learning' 
problem. Remarkably, despite the small number of examples in this redshift band, our pipeline is able to obtain $P$, $R$, and, F1-score of $\sim 0.43$.

\subsubsection{Wide survey: \Euclid\ only}
\label{sec:euclid_only}
We also examine the performance of our pipeline when only Euclid observations are available and for which a reliable redshift is not available (black points in Fig.~\ref{fig:bisigello_comparison_euclid}). These conditions are likely to pertain to a small, but potentially significant number of galaxies in the Wide survey. In this case, our classification pipeline must learn to place galaxies simultaneously into the correct activity class and into the correct redshift bin. 

The quality of the classification varies substantially across the redshift range, with a peak F1-score of 0.71 in the $z=0.75$--1 bin, and with minima of $\sim0.2$--0.3 at either end of the range. The results are largely independent of whether the Int Wide or SED Wide catalogue is used.

Without $ugriz$ photometry or redshifts, the classification problem becomes much more challenging, and unsurprisingly the resulting selection is of reduced quality compared to the cases discussed in Sects.\,\ref{sec:euclid_deep} and \ref{sec:euclid_wide}. In particular, the F1-scores are consistently lower than those obtained when also using $ugriz$ photometry and redshifts, with differences of a factor of 2 or more occurring near endpoints of the considered redshift range. Thus, a key result is that redshift information allows for a significantly more accurate selection of quiescent galaxies. 

\subsection{Global selection and redshift estimation}
\label{sec:global_selection}
An alternative to selection in bins (Sect.\,\ref{sec:separation}) is first to perform a global selection of quiescent galaxies, ignoring redshift
information, and subsequently derive photometric redshifts for the selected galaxies. In this approach, we set the Target variable to 1 for all 
quiescent galaxies in the range $0 \le z \le 3$. The Target is set to 0 for star-forming galaxies at $z \le 3$, and also for all galaxies at $z>3$,
regardless of their sSFR. This analysis is performed on two different subsets of the mock data. Casting a relatively wide net, we use all galaxies 
from the subset of each mock catalogue for which there is a detection in all of the Euclid bands. In addition, the analysis is performed for the
subset of galaxies for which there is a detection in each of the $ugriz$ and Euclid bands. 

The results are shown in Table ~\ref{tab:global_selection}. The F1-score, $P$, and $R$ metrics can vary significantly depending on which mock 
catalogue is used and whether galaxies with a non-detection in an optical band are included or rejected. When detections are required in 
Euclid \IE, \YE, \JE, and \HE\ bands only, we obtain $P=0.85$, $R=0.75$, and an F1-score of 0.80 for the Int Wide catalogue, or $P=0.88$, 
$R=0.76$, and an F1-score of 0.82 for the SED Wide catalogue. Using the Int Deep catalogue, we obtain the metric values $P=0.77$, $R=0.60$, and 
an F1-score of 0.67, which are significantly lower than those obtained with Int Wide. On the other hand, when using the SED Deep catalogue, 
the metrics are $P=0.88$, $R=0.79$, and an F1-score of 0.83. 

When detection in Euclid \IE, \YE, \JE, \HE, and all of $ugriz$ is required, the metrics are substantially improved, with
$P=0.88$, $R=0.85$, and an F1-score of 0.86 for the Int Wide catalogue, or $P=0.96$, $R=0.91$, and an F1-score of 0.94 for the SED Wide catalogue. 
These improvements are predominantly due to the fact that requiring a detection in each of the optical and NIR bands reduces the 
input data to a substantially smaller subset with relatively well constrained broadband SEDs 
(see Fig.~\ref{fig:photometry_reqs} and Table ~\ref{tab:detections}). Under these conditions, for the Int Deep catalogue, we obtain 
$P=0.80$, $R=0.63$, and an F1-score of 0.70; for SED Deep, $P=0.96$, $R=0.92$, and an F1-score of 0.94 are obtained. 

To evaluate the quality of the photometric redshift estimates, the 30-band photometric redshifts derived by \citet{Laigle2016} were used as the 
ground truth. In Table.~\ref{tab:global_selection}, we also give the values of NMAD, bias, and the catastrophic outlier fraction ($f_{\rm out}$) 
for the quiescent subset of galaxies whose photometric redshifts were not flagged as outliers by our pipeline. 

Removal of likely catastrophic outliers was performed as described in Sect.\,\ref{sec:algo}, using a relatively stringent probability threshold of 0.15. 
In other words, we rejected photometric redshift estimates that were assigned a probability of $\ge0.15$ of being a catastrophic outlier. 
As with most outlier removal methods, the removal of genuine catastrophic outliers usually comes at the cost of also removing cases that are not
catastrophic outliers. 
The fraction of quiescent galaxies whose photometric redshifts were rejected as catastrophic outliers ($f_{\rm rej}$), and thus were not used to 
calculate NMAD, bias, or $f_{\rm out}$, is also shown. 

Like the classification metrics, the metrics of photometric redshift quality show variation depending on which mock catalogue (or subset thereof) is 
used. We obtain values for NMAD between 0.020 and 0.033, catastrophic outlier fractions between 0.009 and 0.022, and values of bias in the range 
$-0.0008$ to $0.0031$. While these values appear to improve on the results of the recent Euclid Photometric Redshift Challenge \citep{Desprez2020},
it is important to recognise that the mock Euclid photometry catalogue used therein has a significantly different construction to those we have
used herein, making the inter-comparison of photometric redshift metrics potentially unreliable.

\begin{figure}
  \includegraphics[width=1\columnwidth]{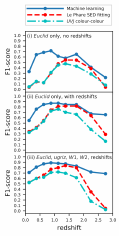}\llap{\makebox[0.58\columnwidth][l]{\raisebox{0.39cm}{\includegraphics[width=0.3\columnwidth]{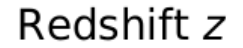}}}}
  \caption{Comparison between our machine-learning method, \texttt{LePhare} SED fitting, and $UVJ$ colour-colour method, for the selection of quiescent galaxies
    from the SED Wide mock Euclid catalogue. The following three configurations of input data were used:
    (i) Euclid photometry only, with no redshift information;
    (ii) Euclid photometry only, this time with photometric redshifts from \citet{Laigle2016};
    and (iii) Euclid, $ugriz$, $W1$ and $W2$ photometry, again using photometric redshifts from \citet{Laigle2016}.} 
  \label{fig:lephare_uvj}
\end{figure}

\section{Comparison with other methods}
\label{sec:colour_colour_comparison}
We test the performance of our quiescent galaxy selection pipeline against four different colour-colour selection methods. We make like-for-like
comparisons, such that our pipeline and the colour-colour method under consideration are applied to photometrically identical subsets of the
mock catalogues. In summary, our machine-learning method outperforms each of the colour-colour methods we tested; full details are given in the 
following subsections.

\subsection{\IYJH\ selection}
To perform a like-for-like comparison with the \IYJH\ selection method, we select from the mock catalogues those galaxies
that have a detection in all of the Euclid bands (i.e. \IE, \YE, \JE, and \HE). We then bin the galaxies by redshift as described in Sect.\,\ref{sec:separation},
and for all bins (except $z=2.5$--3.0 where there are too few galaxies; see B20), we select quiescent galaxies using the \IYJH\ criteria given by B20. 
The results are shown by the green points in Fig.~\ref{fig:bisigello_comparison_euclid}.

Our selection pipeline outperforms the \IYJH\ for almost every combination of redshift and mock catalogue, with the greatest improvements
in the F1-score occurring near each endpoint of the considered redshift range. When our pipeline uses all the available photometry (blue line in
Fig.~\ref{fig:bisigello_comparison_euclid}), the improvement in the F1-score ranges from negligibly small at $z\sim1.5$, to a factor of $\sim2$
in the $z=0$--0.25 and $z=2$--2.5 bins.

While the inclusion of additional photometry bands is clearly part of the reason for the improvements we have obtained, it is not the whole story.
We find that, even when our pipeline has access to exactly the same photometry bands as the \IYJH\ method (i.e. all four Euclid bands; orange
points in Fig.~\ref{fig:bisigello_comparison_euclid}), it still significantly outperforms the colour-colour method. This is because the other Euclid
colours ($I_{\scriptscriptstyle{\rm E}}-J_{\scriptscriptstyle{\rm E}}$, $I_{\scriptscriptstyle{\rm E}}-H_{\scriptscriptstyle{\rm E}}$, 
$Y_{\scriptscriptstyle{\rm E}}-J_{\scriptscriptstyle{\rm E}}$, $Y_{\scriptscriptstyle{\rm E}}-H_{\scriptscriptstyle{\rm E}}$) 
carry information regarding the sSFR that is not present in $I_{\scriptscriptstyle{\rm E}}-Y_{\scriptscriptstyle{\rm E}}$ or 
$J_{\scriptscriptstyle{\rm E}}-H_{\scriptscriptstyle{\rm E}}$. 

When our pipeline is configured to select quiescent galaxies in the same set of redshift bins, using Euclid photometry only, and in the
absence of redshift information (black points in Fig.~\ref{fig:bisigello_comparison_euclid}, we find that, in roughly half of the redshift bins,
the resulting F1-scores are similar to those obtained using the \IYJH\ method (which has the benefit of foreknowledge of galaxy redshifts).
This is generally the case for $z\la1$, while in the range $1 \la z \la 1.75$ our pipeline yields significantly lower F1-scores in this
configuration. Interestingly, the F1-scores obtained using our pipeline to select quiescent galaxies in the $z=2$--2.5 bin are practically identical to,
or slightly above, those obtained using the \IYJH\ method. 

\subsection{\uIIJ\ selection}
For our comparison with the \uIIJ\ method, we use only galaxies that are detected in each of the $u$, \IE, and \JE\ bands. Due to the increasing
sparsity of the $u$-band data at high redshifts, B20 were only able to derive \uIIJ\ selection criteria in redshift bins delimited by the
values $z = 0$, 0.25, 0.5, 0.75, 1.0 in the case of the Wide survey mock catalogues, or $z=0$, 0.25, 0.5, 0.75, 1.0, 1.25, 1.5 for the Deep catalogues. To these bins we applied our quiescent galaxy selection pipeline. The results are shown in Fig.~\ref{fig:bisigello_comparison_euclid_u}, together
with the result of applying the \uIIJ\ method of B20.

Once again, our pipeline (blue or orange points in Fig.~\ref{fig:bisigello_comparison_euclid_u}) outperforms the colour-colour selection method
(green points). The improvement is minimal when using the Wide survey mock catalogues, but is more substantial in the case of the Deep catalogues
(up to a factor of $\sim1.4$), mainly due to improved recall. The largest improvements occur at the upper endpoint of the considered redshift ranges. 
There is little to no reduction in the performance of our pipeline when only the $u$, \IE, \YE, \JE, and \HE\ bands are used (orange points in
Fig.~\ref{fig:bisigello_comparison_euclid_u}), compared to when the full suite of photometry is used (blue points in
Fig.~\ref{fig:bisigello_comparison_euclid_u}). Once again, the improvement obtained from using our machine-learning pipeline is largely due to 
its ability to make use of a larger colour- and magnitude-space. 

\subsection{$BzK$ selection}
In addition, we compare the quality of our quiescent galaxy selection method against the $BzK$ selection method, which is designed to select
quiescent galaxies at $1.4 < z < 2.5$ using the $z-K$ and $B-z$ broadband colours \citep{Daddi2004}. For this comparison, we used 
the observed $B$, $z$ and $K$ magnitudes from the COSMOS2015 catalogue, for galaxies in the Int Wide catalogue, adopting the criteria 
of \citet{Daddi2004} of $(z-K)-(B-z)<-0.2$ and $z-K>2.5$ to select quiescent galaxies. To evaluate the success of the $BzK$ method, we use the sSFR values
from the Int Wide catalogue, adopting the same ${\rm sSFR} < 10^{-10.5} {\rm yr}^{-1}$ threshold to define quiescence.

First, we take the subset of the Int Wide catalogue that lies in the redshift range $1.4 < z < 2.5$, and apply the $BzK$ selection method. We obtain
$P$, $R$, and F1-score values of 0.51, 0.50 and 0.50, respectively. Applying the \texttt{ARIADNE} pipeline to the same subset of galaxies, using features 
derived from the Euclid, $ugriz$, $W1$, $W2$, and 20\,cm bands, we obtain $P$, $R$, and F1-score values of 0.79, 0.80, and 0.79, respectively.
Repeating the same tests over the full redshift range of the Int Wide catalogue, the $BzK$ method results in $P$, $R$, and F1-score of 0.44, 0.07, and 0.13, 
compared to 0.85, 0.75, 0.80 from \texttt{ARIADNE}. In summary, our machine-learning method strongly outperforms the $BzK$ selection method.

\subsection{SED fitting with \texttt{LePhare} and $UVJ$ selection}
\label{sec:le_phare_method}
We also tested, though not exhaustively, our selection pipeline against an SED fitting method wherein the galaxies are separated into 
quiescent and star-forming on the basis of the sSFR estimated from the \texttt{LePhare} best fit to the mock photometry. The fitting procedure is identical to that described in Sect.\,\ref{sec:mocks}, except that the fitting is performed on SED Wide mock photometry, instead of on observational data. Three slightly different approaches are taken, as follows: (i) Only Euclid photometry is used, with redshift being a free parameter; (ii) only Euclid photometry is used, but redshift is fixed to the value given in \citet{Laigle2016}; and (iii) Euclid, $ugriz$, $W1$, and $W2$ photometry is used, including upper limits, with redshift fixed. We derive rest-frame $UVJ$ colours from the best-fitting spectral templates, and use the $UVJ$ selection criteria of \citet{Whitaker2011}. The approach used here may overestimate how well \texttt{LePhare} and the $UVJ$ method can recover the galaxy class,
 since the creation of the SED Wide mock catalogue and our subsequent fitting were in both cases performed using \texttt{LePhare}, with an identical 
 set of base templates. 

The F1-scores as a function of redshift are shown in Fig.~\ref{fig:lephare_uvj} for the \texttt{LePhare} fitting and $UVJ$ methods. Also shown is the F1-score for our machine-learning method when applied to the same subset of data, with identical redshift information. 

Our machine-learning method significantly outperforms both approaches, most noticeably at $z<1$ and $z\ga2.5$. While the \texttt{LePhare} SED fitting method sometimes comes close to reaching the F1-scores obtained by our machine-learning method within the $z\sim 1$--2.5 range, in most redshift bins its F1-scores are dramatically lower; in the case of the $UVJ$ method, the F1-scores are always dramatically lower, typically by $\sim0.2$, than those obtained with our machine-learning method. The 
superior performance of our method is at least partly due its ability to learn how to optimally weight the different bands and colours in different regions of feature-space, unlike the \texttt{LePhare} fitting and $UVJ$ methods.

\section{Further analysis and tests}
\label{sec:further_analysis}
Here we summarise some additional analysis and tests we have conducted. Full details are provided in Appendix~\ref{sec:further_analysis_appendix}. 

\subsection{Stacking versus individual learners}
Our implementation of the generalised stacking method demonstrably improved classification performance. 
With few exceptions, the stacking method consistently outperforms each individual base-learner, as well as outperforming model averaging 
and hard-voting (Fig.~\ref{fig:stacking}). The method is robust against pollution by multiple low-quality classifier models, and can be used as a form of model
selection. Finally, when applied to a single classifier model, the meta-learner often makes a substantial improvement over the original model.

\subsection{The nature of the false positives}
We also examined the distribution of false positives within sSFR space, with the following three main conclusions. First, as expected, the incorrectly classified objects cluster around the class threshold value ($10^{-10.5} {\rm yr}^{-1}$), 
        with a density that is highest in the bins immediately adjacent to the class boundary (Figs.~\ref{fig:errors_int} and~\ref{fig:errors_sed}). Second, the precise distribution of the incorrect classifications differs between the different mock catalogues. Finally, our pipeline offers a significant improvement in $\bar{I}_{\rm FP}$ over the $I_{\scriptscriptstyle{\rm E}}-Y_{\scriptscriptstyle{\rm E}}$, 
        $J_{\scriptscriptstyle{\rm E}}-H_{\scriptscriptstyle{\rm E}}$ and $u-I_{\scriptscriptstyle{\rm E}}$, $I_{\scriptscriptstyle{\rm E}}-J_{\scriptscriptstyle{\rm E}}$ colour-colour methods, 
        reducing the degeneracy between quiescent galaxies and dusty, star-forming galaxies as shown in Fig.~\ref{fig:errors_VIS_Y_J_H} (in addition to improving
        on the P, R, and the F1-score metrics as described above).

\begin{table*}
        \centering
        \caption{Global selection of quiescent galaxies at $0 \le z \le 3$ for different pipeline and data configurations.}

        \resizebox{\textwidth}{!}{%
        \begin{tabular}{lccccccl} 
                \hline
                Catalogue & Bands Used & Detections Required & Redshifts & $P$ & $R$ & F1-score \\
                (1) & (2) & (3) & (4) & (5) & (6) & (7) \\
                \hline
                Int Wide        & $ugriz$, \IE, \YJH, $W1$, $W2$, 20\,cm & \IE, \YJH          & none    & 0.85 & 0.75 & 0.80 \\
                Int Wide (fast mode) & $ugriz$, \IE, \YJH, $W1$, $W2$, 20\,cm & \IE, \YJH     & none    & 0.84 & 0.74 & 0.79 \\
                SED Wide        & $ugriz$, \IE, \YJH, $W1$, $W2$         & \IE, \YJH          & none    & 0.88 & 0.76 & 0.82 \\
                SED Wide (fast mode) & $ugriz$, \IE, \YJH, $W1$, $W2$         & \IE, \YJH     & none    & 0.88 & 0.75 & 0.81 \\
                Wide (averaged) & $ugriz$, \IE, \YJH, $W1$, $W2$, 20\,cm & $ugriz$, \IE, \YJH & none    & 0.92 & 0.88 & 0.90 \\
                Wide (averaged) & $ugriz$, \IE, \YJH, $W1$, $W2$, 20\,cm & $ugriz$, \IE, \YJH & $100\%$ & 0.92 & 0.91 & 0.92 \\
                Wide (averaged) & \IE, \YJH                              & \IE, \YJH          & none    & 0.81 & 0.68 & 0.74 \\
                Wide (averaged) & $ugriz$, \IE, \YJH, $W1$, $W2$, 20\,cm & \IE, \YJH          & none    & 0.87 & 0.76 & 0.81 \\
               Wide (averaged) & $ugriz$, \IE, \YJH, $W1$, $W2$, 20\,cm & \IE, \YJH          & $50\%$  & 0.86 & 0.79 & 0.82 \\
                Wide (averaged) & $ugriz$, \IE, \YJH, $W1$, $W2$, 20\,cm & \IE, \YJH          & $100\%$ & 0.86 & 0.83 & 0.85 \\
                Wide (averaged, $\sigma_z=0.025$) & $ugriz$, \IE, \YJH, $W1$, $W2$, 20\,cm & \IE, \YJH  & $100\%$ & 0.86 & 0.83 & 0.84 \\
                Wide (averaged, $\sigma_z=0.05$)  & $ugriz$, \IE, \YJH, $W1$, $W2$, 20\,cm & \IE, \YJH  & $100\%$ & 0.86 & 0.82 & 0.84 \\
                Wide (averaged, $\sigma_z=0.075$) & $ugriz$, \IE, \YJH, $W1$, $W2$, 20\,cm & \IE, \YJH  & $100\%$ & 0.86 & 0.81 & 0.83 \\
                \hline
                Int Deep        & $ugriz$, \IE, \YJH, $W1$, $W2$, 20\,cm & \IE, \YJH          & none    & 0.77 & 0.60 & 0.67 \\
                Int Deep (fast mode) & $ugriz$, \IE, \YJH, $W1$, $W2$, 20\,cm & \IE, \YJH     & none    & 0.77 & 0.59 & 0.67 \\
                SED Deep        & $ugriz$, \IE, \YJH, $W1$, $W2$         & \IE, \YJH          & none    & 0.88 & 0.79 & 0.83 \\
                SED Deep (fast mode) & $ugriz$, \IE, \YJH, $W1$, $W2$    & \IE, \YJH          & none    & 0.88 & 0.78 & 0.83 \\
                Deep (averaged) & $ugriz$, \IE, \YJH, $W1$, $W2$, 20\,cm & $ugriz$, \IE, \YJH & none    & 0.88 & 0.78 & 0.82 \\ 
                Deep (averaged) & $ugriz$, \IE, \YJH, $W1$, $W2$, 20\,cm & $ugriz$, \IE, \YJH & $100\%$ & 0.89 & 0.85 & 0.87 \\
                Deep (averaged) & \IE, \YJH                              & \IE, \YJH          & none    & 0.73 & 0.54 & 0.62 \\
                Deep (averaged) & $ugriz$, \IE, \YJH, $W1$, $W2$, 20\,cm & \IE, \YJH          & none    & 0.83 & 0.70 & 0.75 \\
                Deep (averaged) & $ugriz$, \IE, \YJH, $W1$, $W2$, 20\,cm & \IE, \YJH          & $50\%$  & 0.83 & 0.74 & 0.78 \\
                Deep (averaged) & $ugriz$, \IE, \YJH, $W1$, $W2$, 20\,cm & \IE, \YJH          & $100\%$ & 0.85 & 0.81 & 0.83 \\               
\hline
SED Wide (train), Int Wide (test) & $ugriz$, \IE, \YJH, $W1$, $W2$ & \IE, \YJH & $100\%$ & 0.76 & 0.86 & 0.81 \\
Int Wide (train), SED Wide (test) & $ugriz$, \IE, \YJH, $W1$, $W2$ & \IE, \YJH & $100\%$ & 0.89 & 0.76 & 0.82 \\
Int Deep (train), Int Wide (test) & $ugriz$, \IE, \YJH, $W1$, $W2$, 20\,cm & \IE, \YJH & none & 0.76 & 0.72 & 0.74 \\
SED Deep (train), SED Wide (test) & $ugriz$, \IE, \YJH, $W1$, $W2$ & \IE, \YJH & none & 0.71 & 0.83 & 0.76 \\
\hline
        \end{tabular}}
        \label{tab:global_selection2}
	\tablefoot{No pre-binning of the data by redshift was performed. The columns are as follows:
          (1) Mock catalogue; where the metrics from using the SED and Int catalogues have been averaged, this is indicated in parentheses;
          where the pipeline has been used in its fast mode, this is also indicated; 
          in the case of transfer-learning, we specify separately the catalogues from which the training and test data are drawn;  
          (2) bands used for galaxy selection, which includes missing values;
          (3) bands in which galaxies are required to be detected;
          (4) redshift information included in the input data; 
          (none, or a percentage of redshifts included in a feature);
          (5) precision $P$ for quiescent galaxy selection;
          (6) recall $R$ for quiescent galaxy selection;
          (7) the F1-score for quiescent galaxy selection.
          The typical uncertainties on the $P$, $R$, and F1-score values herein are $\le 0.01$.}
\end{table*}

\subsection{Reconciling the Int and SED results}
Generally speaking, our classification results differ depending on whether the mock catalogue was constructed using the Int or SED method. 
Firstly, we argue that results obtained using the Int catalogues are likely to be pessimistic with regard to the performance
of our pipeline when applied to real Euclid data; this is because the Euclid photometry is expected to have significantly higher signal-to-noise
ratios, and because the method of interpolating photometry to simulate the Euclid bands is also likely to introduce errors. 
Conversely, we expect that results obtained with the SED catalogues are likely to be somewhat optimistic, because the construction of the SED 
catalogues involves forcing the photometry to conform to one of a limited range of galaxy templates. 

Thus, we argue that results obtained with the Int and SED catalogues will bracket
the real-Universe performance of our pipeline. As a result, we show performance metrics averaged over the Int catalogue and
the corresponding SED catalogue (i.e. Int Wide and SED Wide; Int Deep and SED Deep) in Tables~\ref{tab:global_selection} and~\ref{tab:global_selection2}, 
where appropriate.

\subsection{Tuning the probability threshold}
We investigated the impact on the precision and recall of tuning the value of the class probability threshold, instead
of adopting the default threshold value of 0.5 (see Fig.~\ref{fig:purity_completeness}). Our two main findings are as follows.\ First, there exists a trade-off between precision and recall such
        that one may be increased, but at the cost of reducing the
        other; tuning the probability threshold allows a balance to be
        struck between $P$ and $R$ that is suitable for different scientific
        needs.

Second, using the case shown in Fig.~\ref{fig:purity_completeness} (bottom panel) as an example, adopting a probability 
        threshold of 0.85 yields a very pure sample of quiescent galaxies (P = 0.98) but with moderate incompleteness (R = 0.56); 
        conversely, a probability threshold of 0.05 results in high completeness (R = 0.97) but moderate purity (P = 0.61).

\subsection{Impact of including redshift as a feature}
We examined the usefulness of including the \citet{Laigle2016} COSMOS2015 photometric redshifts as a feature in the data used for model 
training \citep[e.g.][]{Simet2021}. Selected results are included in Table~\ref{tab:global_selection2}, and are shown in Fig.~\ref{fig:redshift_feature}. 
Our two main findings are as follows.\ First, the inclusion of redshifts as a feature in the data significantly improves the classification metrics by reducing the degeneracy between redshift and sSFR,
        particularly for galaxies at $z \le 0.5$ or $z \ge 2.5$. 

        Second, even when only a subset of the redshifts are included, the classification metrics are still improved compared to the case where no redshift are included. In other words, it is beneficial to include any available redshift information, even if it is somewhat sparse, when training classification models. 

\subsection{The impact of noise}
The impact of different types of noise has been explored. The main results are summarised as follows.\ As expected, adding extra noise to the photometric data results in a reduction in the F1-score when selecting quiescent galaxies. However, even when the data becomes extremely noisy (${\rm S/N} \sim 3$), our classification pipeline remains nominally functional with an F1-score of 
        $\sim 0.67$ (see Fig.~\ref{fig:noise_tests}). 

Our classification pipeline is robustly resistant to labelling errors when these are random. However, systematic labelling errors tend to propagate into the final classification results.

Finally, when photometric redshifts are included as a feature, adding Gaussian noise to their values generally results in little or no reduction in the 
        classification metrics; we find that including noisy redshift values still gives generally better classification results compared to the case where
        no redshifts are included.

\subsection{Transfer learning experiments}
\subsubsection{Training on templates and predicting on real SEDs}
We also experimented with using classification models trained on spectral templates to select quiescent galaxies from
catalogues of observed photometry (see Fig.~\ref{fig:transfer_learning}). We find that classifiers trained on the SED Wide 
catalogue are indeed able to  select quiescent galaxies from the Int Wide catalogue, albeit with marginally lower F1-scores 
compared to models trained on the Int Wide catalogue itself. Thus, machine-learning models trained on synthetic 
galaxy SEDs are a potential alternative to traditional methods used selecting quiescent galaxies at redshifts 
where there are few (or no) known examples \citep[e.g.][]{Girelli2019,Cecchi2019}.

\subsection{Which observables are useful to select quiescent galaxies?}
We investigated the impact of adding one of $u$,$g$,$r$,$i$,$z$, $W$1, $W2$, and 20\,cm to the Euclid bands, when 
selecting quiescent galaxies. The main results are summarised as follows. 

Generally speaking, this results in a significant improvement in the F1-score when
        selecting quiescent galaxies. The improvement varies depending on which band is added, and the redshift interval 
        in which the selection is performed (Figs.~\ref{fig:fimportance_images}, \ref{fig:f1_increase}, \ref{fig:fimportance_images_long_wavelength}, and
        \ref{fig:f1_increase_long_wavelength}). 

As expected, the addition of a longer-wavelength optical band is typically more useful for selecting quiescent galaxies at higher redshift. Conversely, shorter wavelength optical bands are more useful for low-redshift selection. 

Interestingly, the 20\,cm radio band provides a significant improvement in F1-score at $1.75 \la z \la 2.5$, despite these data being 
        very sparse. \section{Summary and concluding remarks}
\label{sec:conclusions}
We have introduced the \texttt{ARIADNE} machine-learning pipeline for the classification of galaxies. It uses a novel architecture
with meta-learning to combine the strengths of tree, nearest-neighbours, and deep-learning methods, resulting in significantly higher classification accuracy
compared to the individual learning algorithms. The most relevant technical conclusions from this study are as follows:

\begin{itemize}

\item We have applied the tree-ensemble methods \texttt{CatBoostClassifier} and \texttt{LightGBM} to the selection of quiescent galaxies,
and find that both offer significant performance improvements over the commonly used \texttt{Scikit-Learn RandomForestClassifier} method, in terms of
model quality and training efficiency. 

\item Providing our pipeline with sparsity awareness, by quantifying the sparsity of the photometry for each galaxy, improves the
classification performance. In addition, the sparsity-aware method \texttt{XGBoostClassifier} was found to be well suited for selecting quiescent
galaxies at high redshifts ($z>2.5$), which tend to have many missing photometry values.

\item We have shown that our implementation of the `generalised stacking' method can be used to perform error correction on individual
machine-learning-based galaxy classification models, sometimes turning a mediocre model into a significantly better one.

\item We have used the pseudo-labelling technique \citep{Lee2013} to improve the quality of our photometric redshift estimates, with improvements in NMAD and the
catastrophic outlier fraction of $\sim 1$--3 per cent. When applied to the high-volume of data that will come from future very large surveys, such as Euclid or LSST, 
we expect pseudo-labelling to have a much greater effect. Further exploration of the application of pseudo-labelling to the estimation of galaxy physical 
properties is presented in \citet{Humphrey2022}.\\

\end{itemize}

We have applied our pipeline to the selection of quiescent galaxies from mock Euclid photometric catalogues, using simulated Euclid \IE, \YE, \JE, and \HE\ photometry,
optionally using ancillary optical, infrared, or radio measurements. The main results are as follows:

\begin{itemize}

\item We have shown that our classification pipeline is able to efficiently select quiescent galaxies from within the
redshift range $0 < z < 3$, using mock Euclid \IE, \YE, \JE, and \HE\ photometry and somewhat sparse supporting data at other wavelengths.
The precision (purity), recall (completeness), and F1-scores vary substantially with redshift, and between the various mock catalogues and subsets thereof.

\item We find that including ancillary $ugriz$, mid-infrared (WISE), and radio (20\,cm) photometry yields substantial improvement in the selection of quiescent galaxies
at $z\la1$. Smaller, but nonetheless significant, improvements were found at $z\ga1$.

\item In like-for-like comparisons, our machine-learning pipeline strongly outperforms the $UVJ$ method \citep{Whitaker2011} when derived from Euclid (and ancillary)
survey mock photometry, and usually outperforms the Euclid-specific \IYJH, and \uIIJ\ colour-colour selection methods (B20). The improvement we
obtain over the colour-colour methods can exceed a factor of 2 in terms of completeness and F1-score, with the greatest improvements occurring at $z\la1$ and $z\ga2$.

\item In addition to being fewer in number, the false positives resulting from our classification pipeline are less extreme than those resulting
from the \IYJH, and \uIIJ\ methods, in that their actual sSFR values are typically closer to the boundary between quiescent and star-forming galaxies.

\item The significantly improved classification compared to the colour-colour, $UVJ$, and template-fitting methods is likely the result 
of the more efficient use of the available data by our machine-learning methodology. Compared to the colour-colour methods, the ability to perform the classification 
in a higher-dimensional colour-magnitude space clearly helps. More generally, machine-learning methodologies have the ability to automatically weight the different
 colours and filters according to their usefulness for the classification task at hand, whereas traditional methods often take a somewhat `blind' approach to data
  weighting.

\item Our pipeline is able to derive photometric redshifts for galaxies selected as quiescent, aided by the pseudo-labelling semi-supervised method,
also using an outlier detection algorithm to identify and reject likely catastrophic outliers. Our pipeline achieves a normalised mean absolute deviation 
of $\la0.03$ and a fraction of catastrophic outliers of $\la0.02$ when measured against the COSMOS2015 photometric redshifts of \citet{Laigle2016}. These 
values appear to improve on the results of the Euclid Photometric Redshift Challenge \citep{Desprez2020}, but we emphasise that the mock photometry catalogue 
used therein is of a significantly different construction to those we have used herein, making any inter-comparison of photometric redshift metrics potentially
unreliable.

\item The inclusion of galaxy redshifts among the train and test datasets offers 
significant improvement in the quality of our classification models, even when the redshifts are relatively 
noisy or incomplete.

\item We have investigated the potential impact of various systematics. Most notably, we find that
our pipeline results are robust against the presence of random errors in the class labels of the training data, 
for label error rates of up to $\sim 33$ per cent.

\end{itemize}

This work has added to the growing body of evidence supporting the importance of machine-learning techniques 
(or artificial intelligence) in astronomy and astrophysics. In particular, we have demonstrated that machine 
learning usually outperforms colour-colour methods for the selection of quiescent galaxies; while part of this
improvement is due to the ability to make use of a larger number of bands and colours, we have also shown that 
machine-learning methods still perform a superior selection even when the training data contains only the bands
used by the respective colour-colour method. In future publications the methods presented herein will be 
further developed and applied to other related problems in extragalactic astrophysics. 

\section*{Acknowledgments}
This work was supported by Funda\c{c}\~ao para a Ci\^encia e a Tecnologia (FCT) through grants UID/FIS/04434/2019, UIDB/04434/2020, 
UIDP/04434/2020 and PTDC/FIS-AST/29245/2017, and an FCT-CAPES Transnational Cooperation Project. 
LB acknowledges financial support by the Agenzia Spaziale Italiana (ASI) under the research contract 2018-31-HH.0. 
KIC acknowledges funding from the European Research Council through the award of the Consolidator Grant ID 681627-BUILDUP.
AH acknowledges support from the NVIDIA Academic Hardware Grant Program.
AH also thanks colleagues Jean Gomes, Jo\~ao Pedroso, Catarina Lobo, Tom Scott, Ana Afonso, Patricio Lagos, Israel Matute, 
Stergios Amarantidis, Jose Afonso, Rodrigo Carvajal, and Ciro Pappalardo for useful discussions or comments. 
\AckEC
In the development of our pipeline, we have made use of the Pandas \citep{McKinney2010}, Numpy \citep{Harris2020}, Scipy \citep{Virtanen2020} and
Dask \citep{Rocklin2015} packages for Python.

\begin{appendix}

\section{Impact of redundant features on feature importance}
\label{appendix_redundancy}
Machine-learning methods that build models using decision-tree ensembles have the potential to 
provide insights into the structure of the training data via analyses of the feature-importances. 
When all features in the dataset are fully independent of each other, the feature-importance can 
provide a relatively straightforward indication of how useful each feature is for the model to 
predict the target labels. However, when there is significant co-linearity between features, the 
importance may be shared between co-linear features, resulting in potentially misleading 
feature-importance information. Indeed, it is likely that significant co-linearity exists among the
various broadband magnitudes and colours used in this work.  
Therefore, here we examine how the feature importance calculations used by the \texttt{RandomForestClassifier}, 
\texttt{CatBoostClassifier}, and \texttt{XGBoostClassifier} tree-based algorithms are affected by the presence of 
multiple co-linear features. 

\begin{figure*}
\includegraphics[width=2\columnwidth]{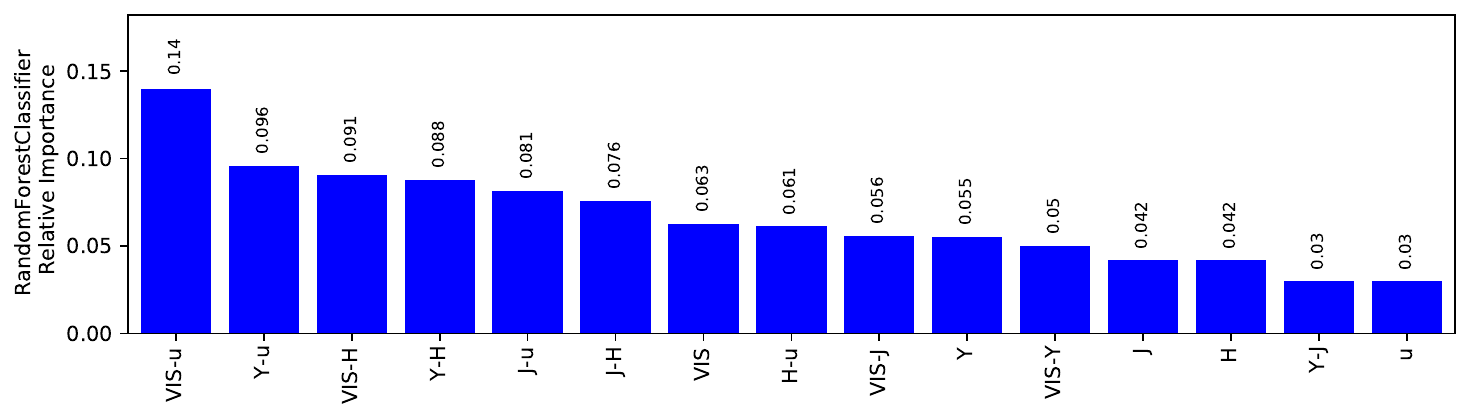}

\includegraphics[width=2\columnwidth]{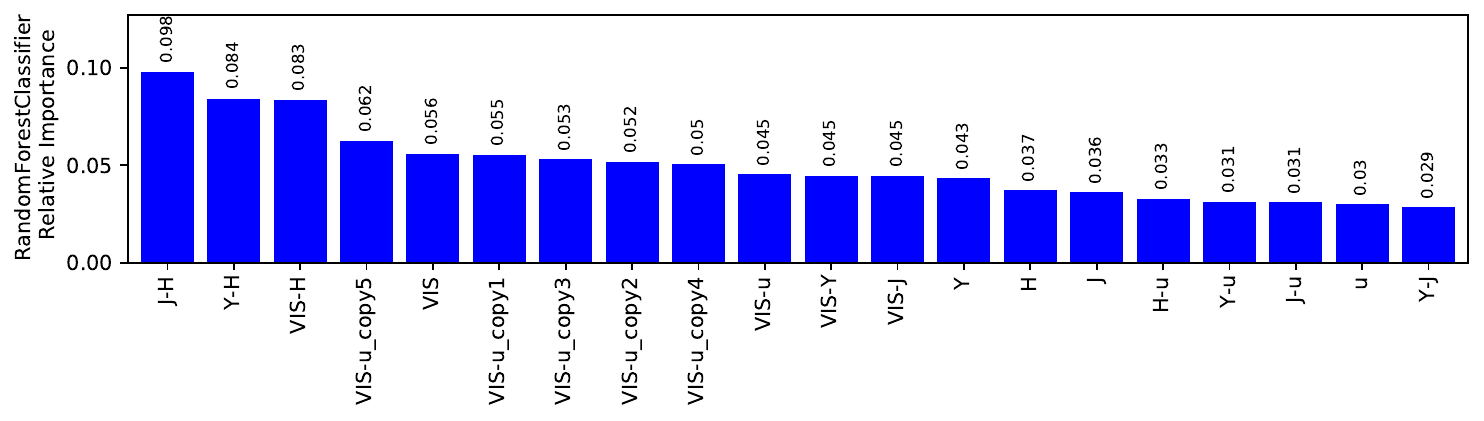}
\caption{Feature importance values when using \texttt{RandomForestClassifier} to select quiescent galaxies 
in the redshift range $0 < z < 0.25$. The results in the upper panel are for the case where none of the 
features have been duplicated. The lower panel shows the feature importance values for the case where
five copies of $I_{\scriptscriptstyle{\rm E}}-u$ have been injected into the dataset prior to model training.
The $x$ axis labels correspond to feature names used by the pipeline after the pre-processing steps outlined 
in Sect.\,\ref{sec:feat_eng} have been applied, and should be self-explanatory (see the caption of Fig.~\ref{feature_importance}.
For example, the feature named `VIS-u\_copy1' is a copy of the feature named 'VIS-u', etc.}
\label{fig:duplicated_features_rf}
\end{figure*}

\begin{figure*}
\includegraphics[width=2\columnwidth]{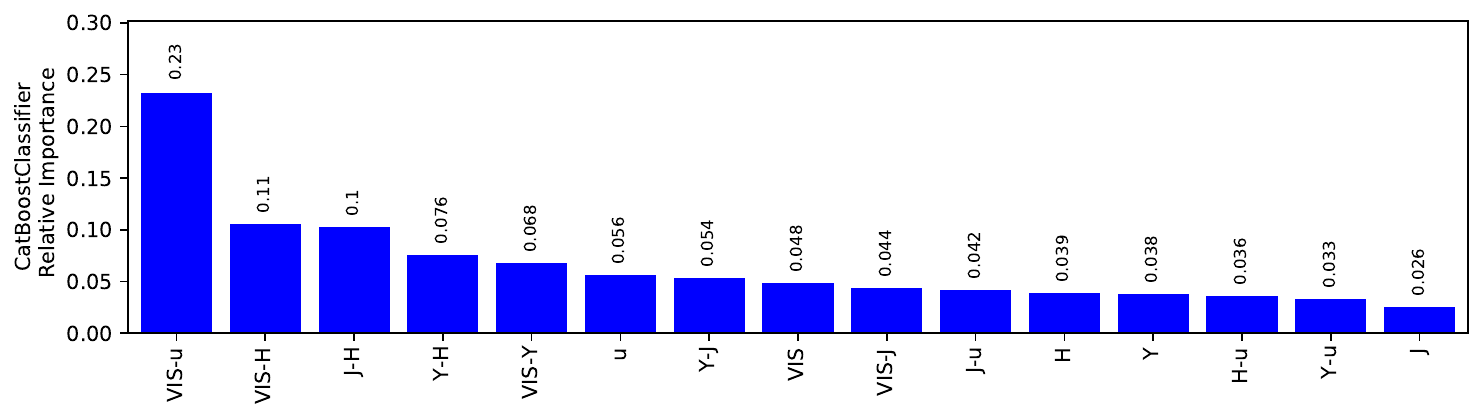}

\includegraphics[width=2\columnwidth]{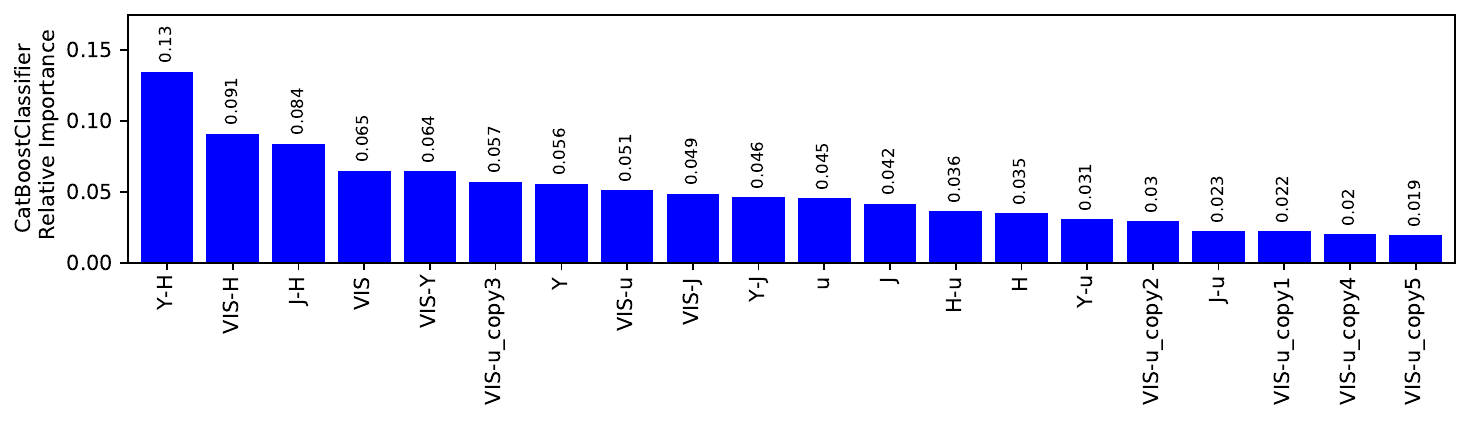}
\caption{Similar to Fig.~\ref{fig:duplicated_features_rf}, but for models trained using \texttt{CatBoostClassifier}.}
\label{fig:duplicated_features_catboost}
\end{figure*}

\begin{figure*}
\includegraphics[width=2\columnwidth]{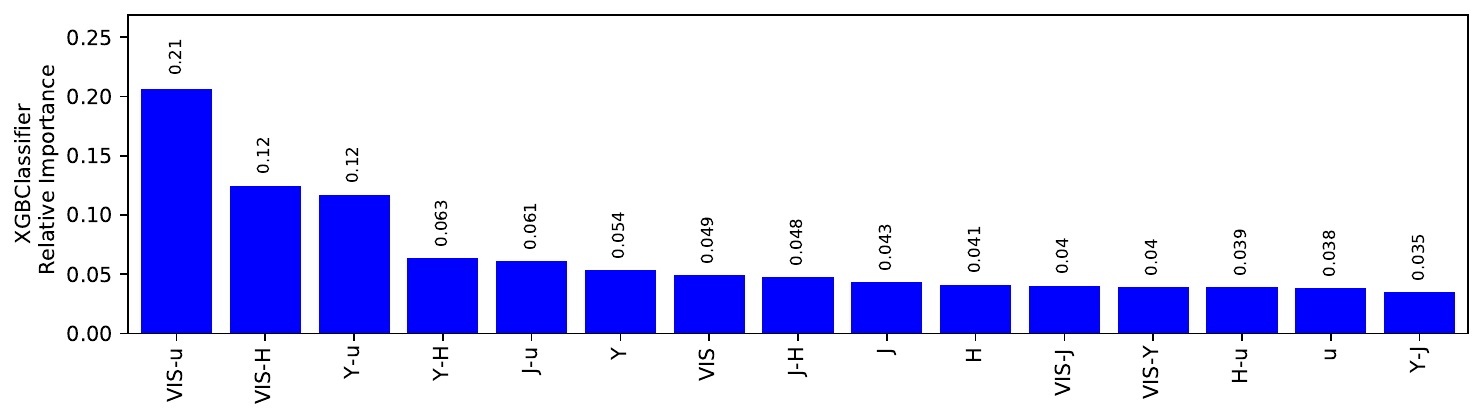}

\includegraphics[width=2\columnwidth]{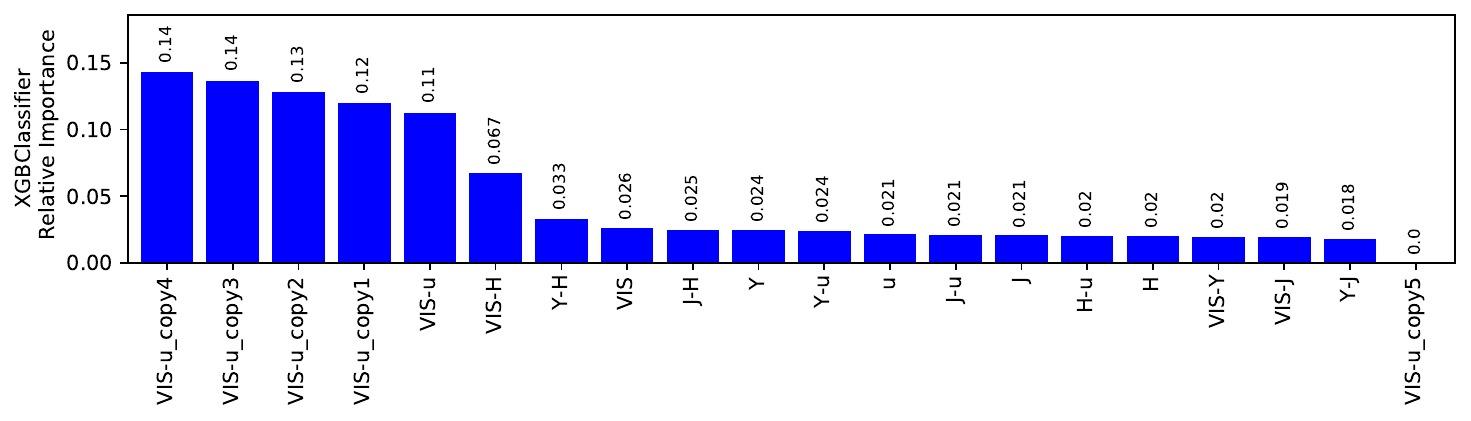}
\caption{Similar to Fig.~\ref{fig:duplicated_features_rf}, but for models trained using \texttt{XGBoostClassifier}.}
\label{fig:duplicated_features_xgb}
\end{figure*}

We first take the Int Wide catalogue, select features derived from $u$, \IE, \YE, \JE, or \HE\ photometry, 
and add five identical copies of the $u-I_{\scriptscriptstyle{\rm E}}$ feature to the dataset. We then
train \texttt{RandomForestClassifier}, \texttt{CatBoostClassifier}, and \texttt{XGBoostClassifier} models to 
select quiescent galaxies in the $0 < z < 0.25$ range. We selected the $u-I_{\scriptscriptstyle{\rm E}}$ colour for duplication because
\texttt{RandomForestClassifier}, \texttt{CatBoostClassifier}, and \texttt{XGBoostClassifier} all find this feature
to be the most important for this particular classification problem. For these model runs, no information was
provided regarding the redshifts of the galaxies. 
The results are shown in Figs.~\ref{fig:duplicated_features_rf}--~\ref{fig:duplicated_features_xgb}.

The impact on the feature importances of duplicating the most important single feature ($I_{\scriptscriptstyle{\rm E}}-u$ in this example) 
is somewhat different for each of the learning algorithms, presumably due to: 
(i) differences in the way the algorithms select features for constructing individual decision trees; 
(ii) how they deal with co-linearity among the input features; 
and (iii) the different methods used to calculate feature importance values.

In the case of \texttt{RandomForestClassifier}, the impact of including duplicates of $I_{\scriptscriptstyle{\rm E}}-u$ is for this feature 
and its duplicates to be demoted to a significantly lower rank of importance (Fig.~\ref{fig:duplicated_features_rf}),
with all six features occupying a similar position within the feature importance ranking. Clearly, it is risky to 
rely on the feature importances produced by \texttt{RandomForestClassifier}, since they are a function of how 
informative a features is and the uniqueness of the information it provides.
When using \texttt{CatBoostClassifier}, the inclusion of duplicates of $I_{\scriptscriptstyle{\rm E}}-u$ also results in the demotion of $I_{\scriptscriptstyle{\rm E}}-u$ 
and its duplicates to significantly lower ranks of importance (Fig.~\ref{fig:duplicated_features_catboost}), with several
being placed at the very end of the importance ranking. 

In contrast, the feature importance values provided by \texttt{XGBoostClassifier} are much more robust against 
the presence of co-linearity among features. As illustrated in Fig.~\ref{fig:duplicated_features_xgb}, $I_{\scriptscriptstyle{\rm E}}-u$, and/or 
several copies thereof are consistently assigned the highest values of feature importance, or else are simply ignored
(feature importance = 0). As such, the feature importances provided by \texttt{XGBoostClassifier} are likely to offer 
a relatively robust method to determine the most relevant observables for the selection of particular galaxy 
types, even when there is significant co-linearity between the observables. 
 
\section{Further analysis and tests}
\label{sec:further_analysis_appendix}

In this appendix we provide full details of the analyses and tests that were summarised in Sect.\,\ref{sec:further_analysis}. 

\begin{figure*}
  \includegraphics[width=0.8\columnwidth]{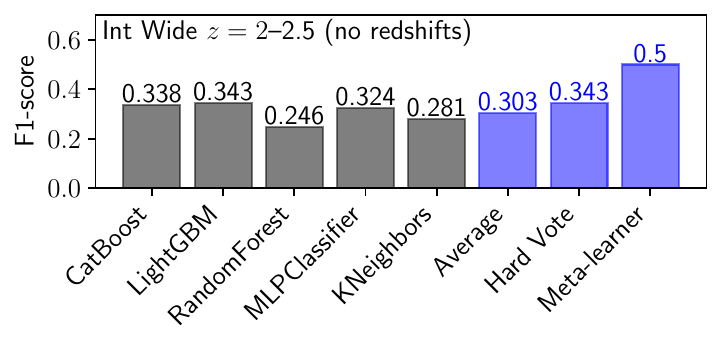}
  \includegraphics[width=0.8\columnwidth]{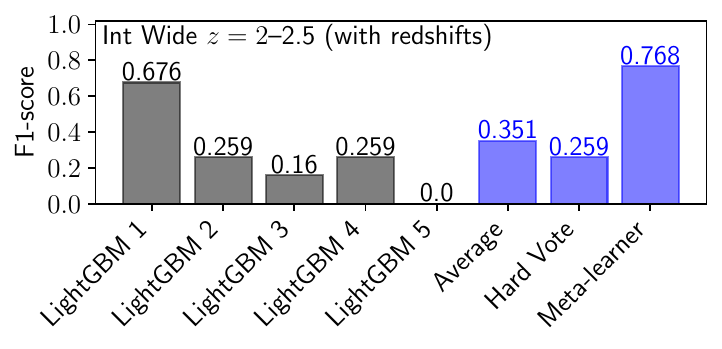}
  \includegraphics[width=0.4\columnwidth]{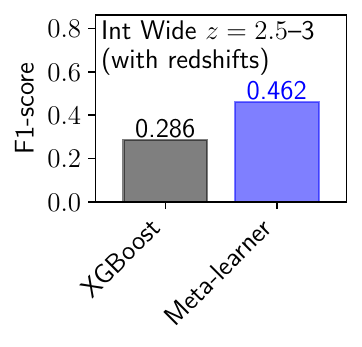}
  \caption{Examples of the F1-scores from individual base-learners and the model ensembling methods.
    {\bf Left:} Selection of quiescent galaxies at $z=2$--2.5 from the Int Wide catalogue using Euclid photometry, without foreknowledge of galaxy redshifts.
    As described in the text, the meta-learner performs a non-linear fusion of the individual classifiers, resulting in a significantly higher F1-score than
    obtained by any of the individual base learners or the two other ensemble methods (averaging and hard-voting).
    {\bf Centre:} Impact of ensembling a \texttt{LightGBMClassifier} model, the hyperparameters of which are well-tuned for this problem (LightGBM 1), with four other
    \texttt{LightGBM} models that have poorly tuned hyperparameters (LightGBM 2,3,4,5). In this case, averaging the model predictions and hard-voting both produce poor results,
    but the meta-learner is able to identify and weight accordingly the low quality class predictions.
    {\bf Right:} Application of a meta-learner to a classification model produced by a single base-learner, in this case \texttt{XGBoostClassifier}. In this circumstance,
    the meta-learner performs `error correction', resulting in a significant improvement in the quality of the classifier.}
  \label{fig:stacking}
\end{figure*}

\begin{figure*}
  \includegraphics[width=2\columnwidth]{{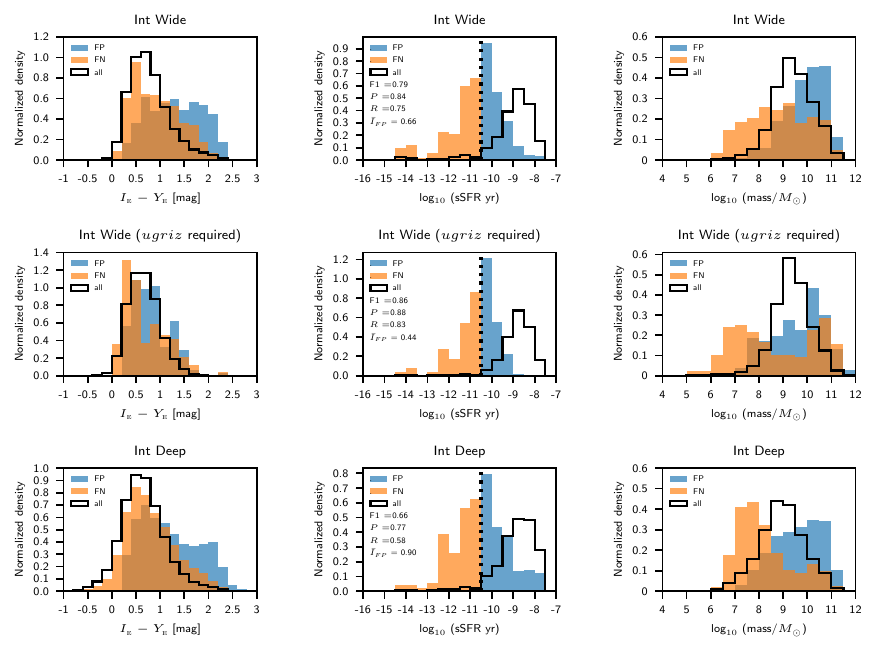}}
  \caption{Distribution of incorrect classifications (FP and FN) with respect to $I_{\scriptscriptstyle{\rm E}}-Y_{\scriptscriptstyle{\rm E}}$, sSFR, and stellar mass, when selecting quiescent galaxies from the Int catalogues.
    Also shown is the overall distribution of galaxies in the catalogue (or catalogue subset) on which the selection was performed.
    {\bf Top row:} Using the $ugriz$, Euclid, $W1$, $W2$, and 20\,cm photometry from the Int Wide catalogue, excluding only those galaxies with a non-detection in any Euclid band.
    {\bf Centre row:} As above, but excluding galaxies without a non-detection in any of the $ugriz$ or Euclid bands.
    {\bf Bottom row:} Using the $ugriz$, Euclid, $W1$, $W2$, and 20\,cm photometry from the Int Deep catalogue, excluding only those galaxies with a non-detection in any Euclid band.
  In each case, we include the values of the F1-score, precision, recall, and $\bar{I}_{\rm FP}$ metrics}
  \label{fig:errors_int}
\end{figure*}

\begin{figure*}
  \includegraphics[width=2\columnwidth]{{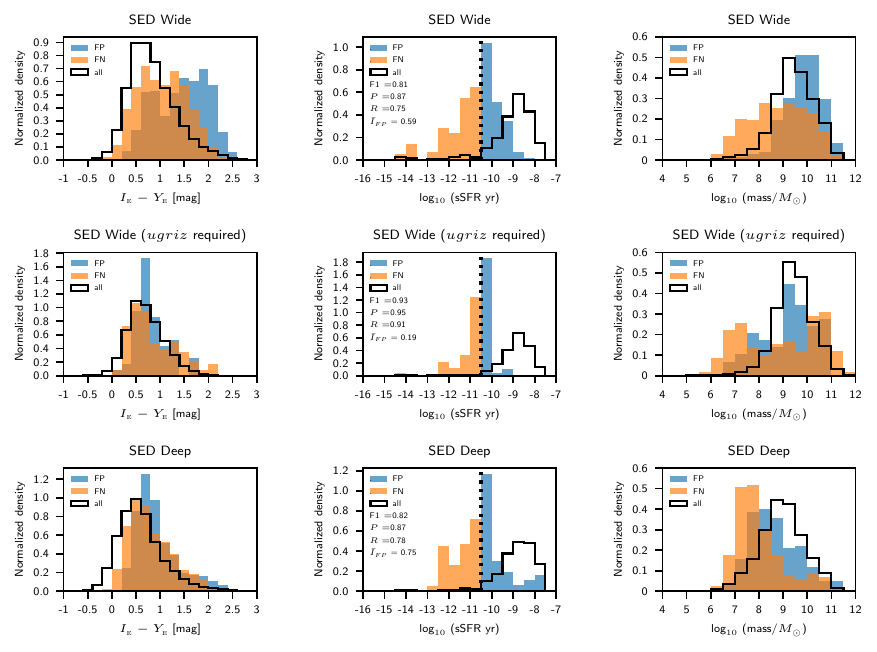}} 
  \caption{Similar to Fig.~\ref{fig:errors_int}, but instead using the SED catalogues.}
  \label{fig:errors_sed}
\end{figure*}

\begin{figure*}
  \centering
  \includegraphics[width=2\columnwidth]{{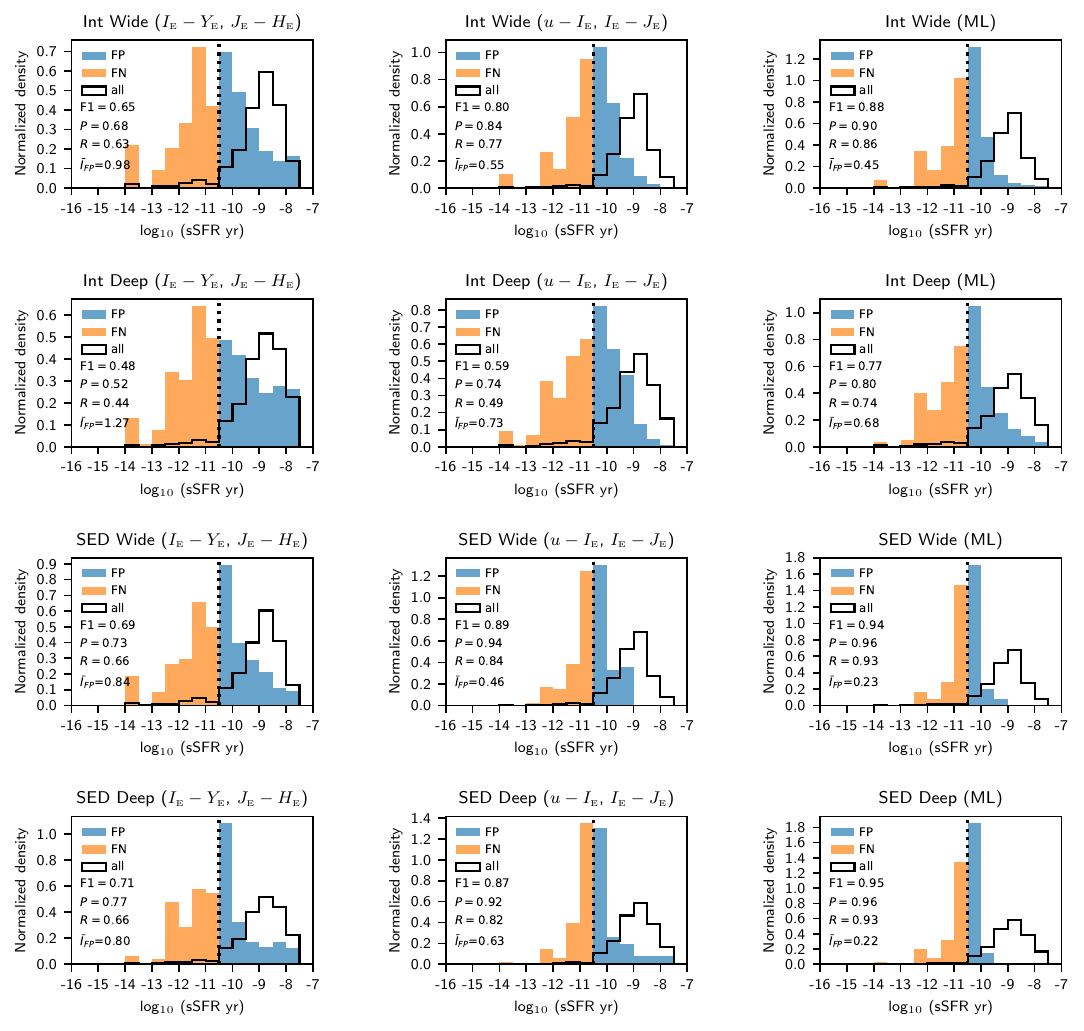}}
  \caption{Distribution of incorrect classifications for the \IYJH\ method of B20 ((left column) and the \uIIJ\ method of B20 (centre column)
    To allow a direct comparison between the \uIIJ\ and our machine-learning selection method, we show results from our pipeline under conditions equivalent
    to those used for the B20 \uIIJ\ method: Galaxy photometric redshifts are included as a feature to be trained on;
    only galaxies detected in $u$, \IE, and \JE\ are used; only galaxies in the redshift ranges $0<z<1$ and $0<z<1.5$ are used for the Wide and Deep catalogues, respectively.
  In each panel, we include the values of the F1-score, precision, recall, and $\bar{I}_{\rm FP}$ metrics. }
  \label{fig:errors_VIS_Y_J_H}
\end{figure*}

\begin{figure}
  \includegraphics[width=1\columnwidth]{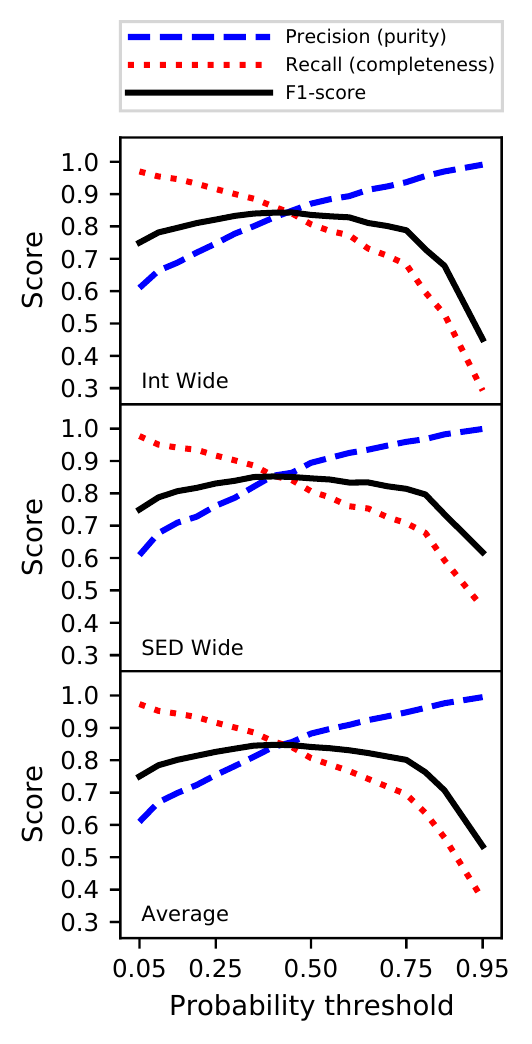}
  \caption{Precision (purity), recall (completeness), and the F1-score as a function of the probability threshold used to separate quiescent
    and star-forming galaxies, for Int Wide (top), SED Wide (centre), and the average result (bottom).} 
  \label{fig:purity_completeness}
\end{figure}

\begin{figure*}
  \includegraphics[width=2.07\columnwidth]{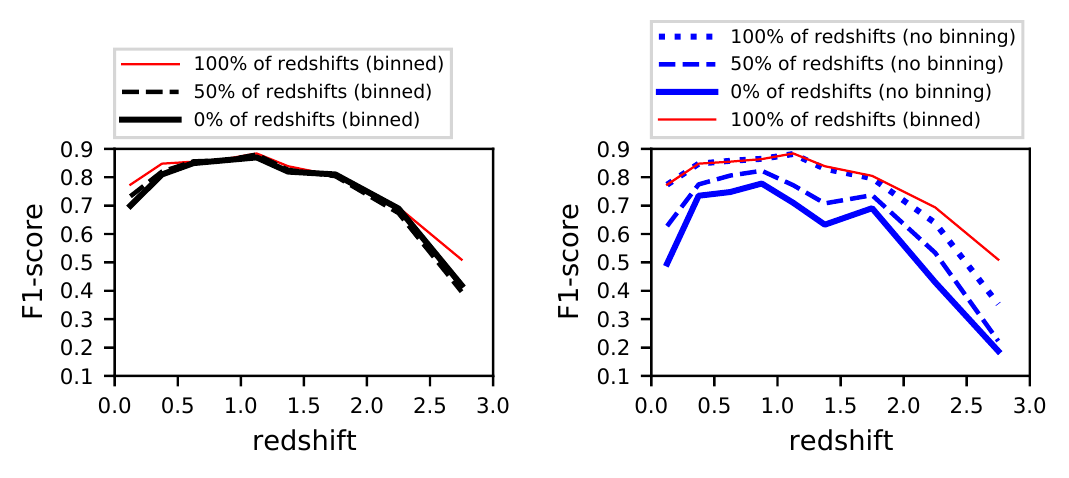}\llap{\makebox[1.63\columnwidth][l]{\raisebox{0.39cm}{\includegraphics[width=0.3\columnwidth]{redshift_z_label_crop.pdf}}}}\llap{\makebox[0.59\columnwidth][l]{\raisebox{0.39cm}{\includegraphics[width=0.3\columnwidth]{redshift_z_label_crop.pdf}}}}
  \caption{Impact of including source redshifts as an additional feature in the input data. {\bf Left:} Case where the dataset is binned by 
  photometric redshift prior to model training, with 100 per cent, 50 per cent, or none of the redshifts included as a feature in the data. 
  {\bf Right:} Illustrating the case where no redshift binning is performed, with classifiers being trained to identify quiescent galaxies at specific
   redshift intervals. For this test, the Int Wide mock catalogue was used.} 
  \label{fig:redshift_feature}
\end{figure*}

\begin{figure}
  \includegraphics[width=1.03\columnwidth]{{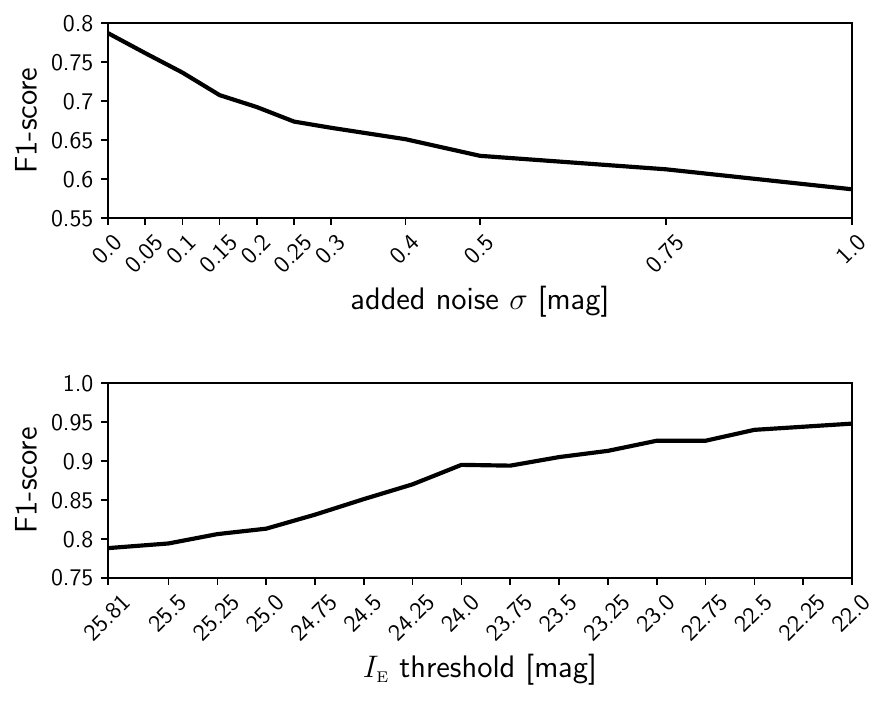}}
  \caption{Testing the impact of photometry measurement uncertainties on the results from our classification pipeline.
    {\bf Top:} Example of how the F1-score is reduced when Gaussian noise is added to the optical and NIR photometry in the
    Int Wide catalogue. {\bf Bottom:} Example of how the F1-score is increased when galaxies fainter than an arbitrary \IE\ 
    threshold are excluded from the Int Wide catalogue.}
  \label{fig:noise_tests}
\end{figure}

\subsection{Stacking versus individual learners}
We discuss the benefits of our implementation of the generalised stacking method, in which meta-learners are trained to fuse the output from several
base-learners into a single classifier. We find that, with very few exceptions, our stacking method consistently outperforms each of the individual base-learners,
in addition to outperforming the traditional ensembling methods of model averaging and hard-voting. This is illustrated in Fig.~\ref{fig:stacking} (left panel), where
we show results from a single run of our pipeline, in this case applied to the selection of quiescent galaxies at $z=2$--2.5 from the Int Wide catalogue using 
Euclid
photometry, and without foreknowledge of galaxy redshifts. In this example, averaging the predictions across the base-learners results in an `averaging-down', while the
hard-vote ensemble method results in an F1-score that matches that of the best individual base-learner (in this case \texttt{LightGBMClassifier}). In contrast, the
meta-learner yields an F1-score that is substantially higher ($\sim46$ per cent in this case) than any of the individual base-learners or other ensemble methods. 

In Fig.~\ref{fig:stacking} (centre panel) we illustrate the robustness of the generalised stacking method against pollution by multiple low-quality classifier models.
We ensembled a LightGBM model with hyperparameters that are well tuned for this problem (LightGBM 1), with four other LightGBM models that have
purposefully poorly tuned hyperparameters (LightGBM 2--5). Whereas the average and hard-voting ensembles give poor results, the meta-learner is able to
discard low-quality class predictions and also make a significant improvement over the single high-quality model (LightGBM 1). 

In addition, we illustrate the usefulness of generalised stacking when applied to a single classifier model. In Fig.~\ref{fig:stacking} (right panel), we show the result
of selecting quiescent galaxies in the $z=2.5$--3 redshift range, using $ugriz$, \Euclid, $W1$, and $W2$ photometry from the Int Wide catalogue, with foreknowledge of
redshifts. The meta-learner is able to substantially improve on the F1-score of the \texttt{XGBoostClassifier}, increasing the score by 61 per cent, 
effectively turning a rather poor classifier into a potentially much more useful one.

\subsection{The nature of the false positives}
\label{sec:errors}
Although the $P$, $R$, and F1-score classification metrics are informative about whether galaxies are correctly or incorrectly classified, they do not provide information
about the {incorrectness} of the incorrect classifications. For instance, the F1-score is insensitive to whether false positives are marginally non-quiescent
(e.g. ${\rm sSFR}\sim 10^{-10.4}\,{\rm yr}^{-1}$) or are in fact powerful starburst galaxies (e.g. $10^{-8} {\rm yr}^{-1}$); when selecting samples of quiescent galaxies, the former case is clearly less baneful
than the latter. Therefore, we now examine the nature of the non-quiescent contaminants in samples selected as quiescent by our classification pipeline, utilising our metrics of
incorrectness $I_{\rm FP}$ and $\bar{I}_{\rm FP}$ (see Eqs.~\ref{eq:I} and \ref{eq:Ibar}).

In Figs.~\ref{fig:errors_int} and~\ref{fig:errors_sed} we show the distribution of incorrect classifications with respect to $I_{\scriptscriptstyle{\rm E}}-H_{\scriptscriptstyle{\rm E}}$, sSFR, and stellar mass, for the Int and SED catalogues.
As is often the case when performing a binary classification on a continuously distributed sample, the incorrectly classified objects cluster around the class threshold value ($10^{-10.5} {\rm yr}^{-1}$), with
a density that is highest in the bins immediately adjacent to the class boundary. However, the precise distribution of classification errors varies between the different catalogues (and
subsets thereof).

When using the Wide survey mock catalogues, $\ga50$ per cent of false positives are what we consider to be marginal classification errors, that is, their sSFR is within 0.5 dex of
the class boundary. Furthermore, there are very few false positives with high values of sSFR: fewer than 25 per cent of the false positives are at sSFR $\ge10^{-9.5} {\rm yr}^{-1}$,
while false positives with sSFR $\ge10^{-9} {\rm yr}^{-1}$ are negligible ($\la5$ per cent). On the other hand, when the Deep survey mock data are used, the fraction of false positives at
relatively high values of sSFR ($\ge10^{-9} {\rm yr}^{-1}$) is non-negligible ($\sim25$ per cent). This is true regardless of whether the Int Deep or SED Deep catalogue is used. 
The distribution of the incorrect classifications with respect to the stellar mass shows a significant diversity and depends strongly on which mock catalogue is used. In general,
the false positives are biased towards relatively high stellar mass (i.e. $\ge10^{9} M_{\odot}$). 

For comparison, in Fig.~\ref{fig:errors_VIS_Y_J_H} we also show the distribution of errors with respect to sSFR for the \IYJH\ (left column) and \uIIJ\ (middle column)
colour-colour methods developed by B20. To allow a relevant comparison between the \uIIJ\ colour-colour method and our machine-learning pipeline, Fig.~\ref{fig:errors_VIS_Y_J_H}
(right column) also shows results from applying our pipeline with conditions equivalent to those used for the \uIIJ\ method:
(i) Galaxy redshifts are included as a feature to be trained on;
(ii) only galaxies detected in $u$, \IE, and \JE\ are used;
(iii) only galaxies in the redshift ranges $0<z<1$ and $0<z<1.5$ are used for the Wide and Deep catalogues, respectively.

Our classification pipeline results in significantly lower $\bar{I}_{\rm FP}$ than the colour-colour methods. This is due to a reduction in the fraction of false positives that are
located at high sSFR (e.g. $\ga 10^{-9} {\rm yr}^{-1}$). In other words, our pipeline not only offers a significant improvement over colour-colour methods, 
in terms of $P$, $R$, and the F1-score, but also significantly reduces the degeneracy between quiescent galaxies and dusty, star-forming galaxies. 

There is another potential source of classification errors that should be mentioned. Throughout this work we have tacitly assumed that the target variable accurately represents the
ground truth. This assumption is clearly valid for the SED catalogues, since they are derived from templates corresponding to known physical properties. However, for the Int catalogues
there is the possibility that some instances flagged as classification errors are, in actual fact, instances where our pipeline provides the correct classification and the target variable
is incorrect.

\subsection{Reconciling the Int and SED results}
\label{sec:reconciling_int_sed}
As discussed above, our machine-learning models were trained and evaluated using one of the four mock catalogues, but there are often significant differences
between the results obtained using the Int or SED catalogues for a given \Euclid\ survey (e.g. Fig.~\ref{fig:bisigello_comparison_euclid}). In particular, the
precision, recall, and F1-scores for quiescent galaxy selection tend to be higher when using an SED mock catalogue, compared to its corresponding
Int catalogue (i.e. SED Wide vs. Int Wide; SED Deep vs. Int Deep).

This is due to the different methods used in the construction of the catalogues (see Sect.\,\ref{sec:mocks} and B20). The Int catalogues have an observationally more realistic starting point since they are constructed with real photometry, albeit twice convolved with a filter, 
but the signal-to-noise ratio of the data is significantly lower than will be the case for the actual \Euclid\ photometry, likely increasing 
the difficulty of the classification problem compared to when the real \Euclid\ (and ancillary) survey data are used. 
Conversely, the mock photometry in the SED catalogues has noise properties that match those expected for \Euclid\ observations, but the SEDs 
themselves are forced to conform to a restricted set of simplified templates, which could potentially simplify the classification problem.

Therefore, results obtained using the Int catalogues are likely to be pessimistic, and results obtained with the SED catalogues are likely to be
 optimistic with regard to the performance of our pipeline when applied to real \Euclid\ (and ancillary) survey data. Thus, to estimate the 
 performance of our pipeline when selecting quiescent galaxies from \Euclid\ (and ancillary) survey data, we use performance metrics averaged 
 over the Int catalogue and the corresponding SED catalogue (i.e. Int Wide and SED Wide; Int Deep and SED Deep). Therefore,
 Tables~\ref{tab:global_selection} and~\ref{tab:global_selection2} include averaged metrics, where appropriate. 
 We consider results obtained with the Int and SED catalogues to bracket the likely range of performance of our pipeline when it is
 applied to real \Euclid\ (plus LSST, etc.) photometry.

\subsection{Tuning the probability threshold}
In some circumstances, it is desirable to maximise either the precision (purity) or the recall (completeness) of the selected quiescent galaxy samples, according to the nature of one's scientific objectives. 
Thus, we investigate and illustrate the impact on the precision and recall of tuning the value of the 
class probability threshold, instead of adopting the default threshold value of 0.5. 

In Fig.~\ref{fig:purity_completeness} we show how precision ($P$), recall ($R$), and the $F1$-score vary with the class probability threshold when selecting
quiescent galaxies from the Int Wide mock catalogue (top) or the SED Wide catalogue (middle). Also shown are the scores when averaged between the two catalogues (bottom).
The redshifts are included as a feature in the input data. There exists a trade-off between precision and recall such that one may be increased, but at the cost of reducing
the other. For example, from the averaged scores (bottom panel), we find that adopting a probability threshold of 0.85 yields a sample of quiescent galaxies that is very
pure ($P=0.98$) but somewhat incomplete ($R=0.56$). Conversely, using a probability threshold of 0.05 results in a sample with moderate purity ($P=0.61$) but
high completeness ($R=0.97$).

Tuning the probability threshold allows a balance to be struck between $P$ and $R$ that is suitable for different scientific needs. For instance, using a probability
threshold of 0.3 gives a sample of quiescent galaxies that is both reasonably pure ($P=0.8$) and reasonably complete ($R=0.9$). While the examples given here
pertain to selection in the redshift range $0 \le z \le 3$, this exercise can, of course, also be performed for the selection of quiescent galaxies in narrower
redshift bands. As an example of this, when selecting quiescent galaxies in the redshift range $1 \le z \le 2$ using a probability threshold of 0.9, we obtain
$P=0.98$ and $R=0.51$, while using a threshold of 0.1 yields $P=0.70$ and $R=0.95$.

\subsection{Impact of including redshift as a feature}
\label{sec:z_as_feature}
Here we investigate the impact of several different methods for the treatment of redshift information in our pipeline. Thus far, we have included redshift information
by pre-binning the mock catalogues using the \citet{Laigle2016} COSMOS2015 photometric redshifts (Sect.\,\ref{sec:separation}). Alternatively, we ignored redshift
information and instead performed a global selection of quiescent galaxies, deriving photometric redshifts subsequently (Sect.\,\ref{sec:global_selection}). A further
possibility that we now also examine is the inclusion of redshifts as a feature in the input data for model training \citep[e.g.][]{Simet2021}.

In Fig.~\ref{fig:redshift_feature} we show F1-score versus redshift when selecting quiescent galaxies from the Int Wide catalogue, considering several different
configurations for the included redshift information. As before, we exclude galaxies that have a non-detection in one or more \Euclid\ bands. The left panel of
Fig.~\ref{fig:redshift_feature} shows results from our pipeline when the mock catalogue data are pre-binned by redshift as described in Sect.\,\ref{sec:separation},
and all available photometry is used (i.e. $ugriz$, \Euclid, $W1$, $W2$, 20\,cm). We find that including the \citet{Laigle2016} photometric redshifts in the input data
significantly increases the F1-scores in bins at $z \le 0.5$ or $z \ge 2.5$, by reducing the degeneracy between redshift and sSFR. However, the F1-score is not significantly
changed for bins in the range $0.5 < z < 2.5$.

We also experimented with the inclusion of redshifts for a fraction of galaxies only, by randomly replacing redshift values with $-99.9$. The left panel of
Fig.~\ref{fig:redshift_feature} also shows the result of including only 50 per cent of the redshifts. At the low end of the redshift range, we find that the 
inclusion of 50 per cent of the redshifts provides a small but significant improvement in the F1-score. However, at the high endpoint of the redshift range, there is 
no noticeable improvement compared to the F1-scores obtained when no redshifts are included. 

We repeated the same experiment, but without pre-binning the galaxies by redshift (see Sect.\,\ref{sec:euclid_only}). As before, we excluded objects with a non-detection in any
of the \Euclid\ bands and used all available photometry from the Int Wide catalogue. The results are shown in the right panel of Fig.~\ref{fig:redshift_feature}.
Within the $0 < z < 2$ range, there is essentially no penalty in terms of F1-score associated with not pre-binning by redshift, provided all redshifts are included as a feature
in the input data (nevertheless, pre-binning does allow considerably faster training of models, since the training set is now much smaller). However, at $z=2$ and above, 
significantly lower F1-scores are obtained compared to the case where the data are pre-binned by redshift. When only 50 per cent of redshift values are included, we find
a significant decrease in F1-scores compared to the case where 100 per cent of redshift values are included, but the F1-score are still usually significantly above those
obtained when none of the redshift values are included. Classification metrics for the global selection of quiescent galaxies, including 100 per cent, 50 per cent, or none of
 the galaxy redshifts are included in Table~\ref{tab:global_selection2}.

\begin{figure}
  \includegraphics[width=1\columnwidth]{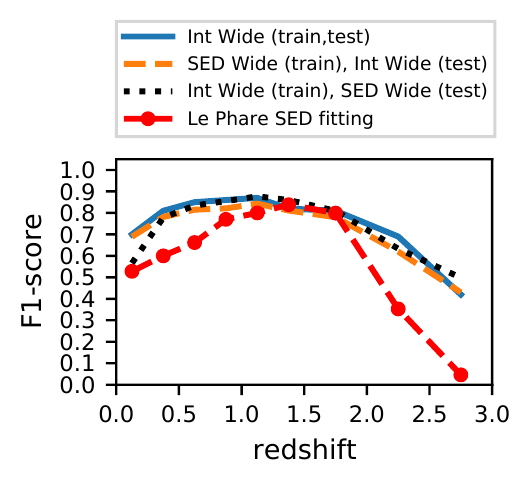}\llap{\makebox[0.57\columnwidth][l]{\raisebox{0.39cm}{\includegraphics[width=0.3\columnwidth]{redshift_z_label_crop.pdf}}}}
  \caption{Results from our transfer learning method. When we train models on SED Wide data, and use them to select quiescent galaxies from Int Wide data 
  (dashed orange curve), the F1-scores are only slightly lower than those we obtain from models trained on Int Wide data (solid blue curve). A generally
   similar result is obtained when we train models on Int Wide data and use them to select quiescent galaxies from SED Wide data (dotted black curve). For
    comparison we also show the results from the \texttt{LePhare} template fitting method applied to the SED Wide mock catalogue 
    (see Sect.\, \ref{sec:le_phare_method}), using \Euclid, $ugriz$, $W1$, $W2$, and 20\,cm photometry, and with redshifts fixed at the \citet{Laigle2016} values.} 
  \label{fig:transfer_learning}
\end{figure}

\subsection{The impact of noise}
\label{sec:noise_tests}
In this subsection we explore the impact of different types of noise that are expected to be present within the data. The experiments 
presented here are intended to be informative and illustrative, but not necessarily exhaustive.

\subsubsection{Adding noise to the photometry}

In Fig.~\ref{fig:noise_tests} we show results from modelling the impact of the addition or the reduction of noise in the data. For this experiment,
we select quiescent galaxies at $0 \le z \le 3$ from the Int Wide mock catalogue, using our pipeline in {fast mode}. The upper panel of
Fig.~\ref{fig:noise_tests} shows how the F1-score decreases when each magnitude measurement has a random offset, drawn from a Gaussian 
distribution of $\sigma$, applied. Interestingly, even when the data are extremely noisy, the pipeline remains nominally functional. For instance, 
even when the photometry has been degraded to a signal-to-noise ratio of $\sim3$, it is nonetheless still able to perform a global selection of 
quiescent galaxies, albeit with somewhat reduced precision, recall, and F1-scores of $\sim0.67$.

To simulate a reduction in noise, we perform a cut to remove galaxies fainter than an arbitrary \IE\ threshold, where 
lower values of this 
threshold result in a higher average signal-to-noise ratio for the mock catalogue (Fig.~\ref{fig:noise_tests}, lower panel). Despite the crudeness 
of these tests, it is clear that reducing (increasing) the signal-to-noise ratio of the data results in lower (higher) quality classification models, 
as evaluated by the F1-scores. While the different noise characteristics probably play a significant role in the differences in pipeline performance
between the Int and SED catalogues, additional effects may also be important. 

\subsubsection{Label noise}
Heretofore, we have tacitly assumed that the quiescent and star-forming labels used in the training and evaluation of our classification models 
give an accurate representation of the ground truth. However, there is the possibility that some labels are incorrect, that is, quiescent galaxies labelled
as star-forming or vice versa. We examine the potential impact of incorrect labels using two slightly different approaches.

In the first approach, we assess the impact on model quality from introducing incorrect labels at random. To do this, we select a subset of galaxies in the mock 
photometry catalogue, and for these galaxies we replace all occurrences of the value 0 with the value 1, and all occurrences of the value 1 with 0. 
We then perform our standard preprocessing and model training steps as described in Sect.\,\ref{sec:pipeline}. For these tests, the \texttt{ARIADNE} 
pipeline is used in {fast mode}. The classification metrics are evaluated for two different cases, where (i) the test set class predictions are
 compared with the original unaltered test set labels, or (ii) the test set class predictions are compared with the altered (noisy) test set labels.

In the context of this experiment we point out that the `random errors' 
discussed here can either be truly random errors arising from the methodology used to generate the labels (e.g. \texttt{LePhare} template fitting)
 or else can be systematic errors that the machine-learning algorithms are unable to model and reproduce. An example of the latter type of error might be
 a systematic error driven by photometric bands that are present in the multi-wavelength dataset used to generate the labels, but which do not appear in the data that are seen by our pipeline.

Under these conditions, the presence of noisy labels varies depending on the particular classification problem that is being addressed. For instance,
when selecting quiescent galaxies at $0 \le z \le 3$ from the Int Wide catalogue, without foreknowledge of redshifts, and with 33 per cent of the 
labels having been flipped, the metrics obtained when the test set labels contain errors are relatively poor, at $P=0.62$, $R=0.12$, and ${\rm F1{\text -} score} = 0.21$. 
This is to be expected, because the classification results are being evaluated against a ground truth that contains many incorrect labels, leading to
artificially poor metrics. 

Conversely, the metrics we obtain are substantially better if we instead evaluate the same classification results against the original ground truth, 
yielding $P=0.86$, $R=0.68$, and ${\rm F1{\text -} score} = 0.76$; these values are only slightly different compared to the case where label errors are not introduced 
at all ($P=0.84$, $R=74$, ${\rm F1{\text -} score}=0.79$). In other words, for this case the presence of randomly incorrect labels in the training data did not have 
a substantial impact on the classification of sources in the test set.

In another example, selecting quiescent galaxies at $1 \le z \le 1.25$ with no foreknowledge of redshifts, and with 10 per cent of labels in error, 
results in an F1-score of 0.13 when evaluated using the test set ground truth with label errors injected. Remarkably, when evaluating the same 
classification results using the ground truth test set  without label errors, the F1-score is 0.69, effectively unchanged from the case where no 
label errors are injected at all (${\rm F1{\text -} score} = 0.69$). In this case, the presence of random errors in the training set labels had no detrimental effect 
on the final classifications of sources in the test set. 

One of the interesting implications of these results is that our pipeline is generally able to ignore random label errors. 
Furthermore, when random label errors are present, our pipeline may even produce predictions that are of higher quality than the labels in the input
data, in terms of how close the predictions are to the real ground truth. Further research (beyond the scope of this paper) is needed to test whether
machine-learning methods such as those discussed herein may be able to improve on traditional SED fitting methods, rather than simply emulating them.

The second approach is similar to the first, but aims to simulate systematic errors in the class labels. For this, we used the \texttt{KMeans} clustering 
algorithm from \texttt{Scikit-Learn} to separate galaxies from the Int Wide catalogue into an arbitrary number of clusters, with an arbitrary number
of the clusters being selected to have an arbitrary fraction of their class labels inverted. For illustrative purposes, we separated the data into 
100 clusters, and inverted 99 per cent of the labels in three randomly selected clusters; the inverted labels represent 11 per cent of all labels. We then
selected quiescent galaxies at $0 \le z \le 3$, again using features derived from the $ugriz$, \Euclid, $W$1, $W$2, and 20 cm bands, and without
foreknowledge of galaxy redshifts. Evaluating the metrics when using the modified labels for the test set, we obtain $P=0.89$, $R=85$, and an F1-score of 0.87. 
In contrast, evaluating
the metrics using the original test set labels, we obtain the substantially lower values $P=0.36$, $R=0.69$, and an F1-score of 0.47. In this case,
our pipeline is able to emulate the systematic label errors present in the training set, resulting in test set class predictions that contain similar
systematic errors. 

\subsubsection{Redshift noise}
As we have demonstrated heretofore, the inclusion of redshift information in the training data often results in classification models that are better able 
to correctly identify quiescent galaxies (e.g. Appendix~\ref{sec:z_as_feature}). A key point to be considered is whether our results are significantly 
affected by the accuracy of the photometric redshifts used, especially since the 30-band COSMOS2015 redshifts \citep{Laigle2016} we have used could potentially be
more accurate than the redshifts that will be estimated from \Euclid\ photometry and the anticipated ancillary data \citep[see e.g.][]{Desprez2020}. 
Therefore, we explore the impact of adding Gaussian noise to the COSMOS2015 redshifts prior to model training. While a full treatment of this issue is 
beyond the scope of the present work, we consider several different cases in order to obtain indicative results. 

Random samples were drawn from a Gaussian distribution with standard deviation $\sigma_z$; these values were multiplied by $1+z$ and then added to the 
photometric redshift values after the Target variable was set, simulating the addition of Gaussian noise. The global selection of quiescent galaxies in 
the range $0 < z < 3$ was then repeated with the (now noisier) redshifts included as a feature, along with the $ugriz$, \Euclid, and Wise photometry, 
and colours derived therefrom. This test was performed for the Int Wide and SED Wide catalogues, for the values $\sigma_z = 0.025, 0.05, 0.075$. 

The results from this test, averaged over an equal number of pipeline runs on the Int Wide and SED Wide catalogues, are shown in 
Table ~\ref{tab:global_selection2}. While $P$ is essentially unchanged, $R$, and the F1-score are slightly reduced by $\sim 0.01$--0.02.

Next, we consider the impact of redshift noise on the selection of quiescent galaxies in the narrower redshift bins used in 
Sect.\,\ref{sec:separation}. As one might expect, it is more difficult to select quiescent galaxies inside these narrower redshift bins 
when the photometric redshifts are noisier. This is because the models, in addition to separating quiescent and star-forming galaxies, must now
also separate quiescent galaxies by their redshift. For example, in the case where $\sigma_z = 0.05$, the F1-score is typically reduced by 
$\sim0.07$ compared to the case where no noise is added. Nevertheless, the F1-scores are still significantly higher than when the redshifts are
not included at all, with the improvement ranging from $\sim 0.02$ at $1 < z < 1.5$, to $\sim 0.2$ at $z < 0.5$ and $z > 2$. 
Interestingly, even in the case where $\sigma_z = 0.1$, representing rather noisy redshifts (NMAD $\sim 0.1$), the F1-scores at $z < 0.5$ and $z > 2$
are still $\sim 0.1$ higher than when redshifts are not included. In other words, even when photometric redshifts are somewhat noisy, 
their inclusion in the data can nevertheless result in significantly stronger classification models, compared to when photometric redshifts are not used.

\subsection{Transfer learning experiments}
\subsubsection{Training on templates and predicting on real SEDs}
\label{transfer_learning}
We also experimented with the possibility of using classification models trained on spectral templates to select quiescent galaxies from
catalogues of observed photometry. To explore this, our pipeline trained its classification models on the SED Wide mock catalogue, and selected
quiescent galaxies from the Int Wide catalogue using the resulting classifier. Essentially, we gave our pipeline the task of identifying and weighting 
the defining characteristics of quiescent and star-forming galaxies from a set of simplifying abstractions (galaxy SED templates), rather than from 
observed SEDs. The train-test split was performed as described in Sect.\,\ref{sec:algo}, ensuring that each galaxy is present in either the training set 
or the test set, but not both. For this experiment, the data are pre-binned by redshift as described in Sect.\,\ref{sec:separation}, but we did not include 
the redshift values as a feature in the input data for the classification pipeline.

The results are shown in Fig.~\ref{fig:transfer_learning}, where it can be seen that classifiers trained on the SED Wide catalogue are indeed able to 
select quiescent galaxies from the Int Wide catalogue, albeit with marginally lower F1-scores compared to models trained on the Int Wide catalogue itself. 
This opens up the interesting possibility of using machine-learning models trained on synthetic galaxy SEDs as a potential alternative to traditional 
colour-colour or template fitting methods \citep[e.g.][]{Girelli2019,Cecchi2019} of selecting quiescent galaxies that have redshifts (or other properties) 
for which there are no (or few) known examples. 

Figure~\ref{fig:transfer_learning} also reveals that the transfer learning method can also function in reverse. When the training and test datasets are 
swapped with each other, such that classification models are trained on the Int Wide catalogue and are then used to select quiescent galaxies from 
the SED Wide catalogue, very similar results are obtained. The exception is in the lowest redshift bin ($0 < z <0.25$), where the F1-score is now 
reduced by $\sim0.12$ compared to the previous case. Classification metrics for the global selection of quiescent galaxies using transfer learning are 
also included in Table~\ref{tab:global_selection2}.

We also show in Fig.~\ref{fig:transfer_learning} (red dashed line) the F1-scores obtained when using \texttt{LePhare} to fit templates to SED Wide mock 
data using the same photometry, and with redshifts fixed to the values from \citet[][see also Sect.\,\ref{sec:le_phare_method}]{Laigle2016}. This template 
fitting method clearly provides results of similar quality to our pipeline in the $1.0 \la z \la 2.0$ redshift range, but at $z\la1.0$ or at $z\ga2$ 
the method significantly under-performs our transfer learning method. The under-performance of our \texttt{LePhare} template fitting method is likely 
caused, at least partly, by the absence of priors concerning (i) the known (or suspected) distribution of the galaxy classes within the redshift and 
colour-spaces, and (ii) the relative importance (or weighting) that should be given to each data point in the broadband SED,
aside from their signal-to-noise ratio. 

\subsubsection{Training on Deep and predicting on Wide survey data}
\label{deep_to_wide}
The Euclid Deep Survey will provide photometry (and spectra) in several fields for which there are pre-existing multi-wavelength observations, 
allowing the construction of source catalogues with high-quality labels. In turn, this is expected to facilitate the training of classifiers, which
can then be used to predict labels for the enormous number of sources that will be detected in the Euclid Wide Survey. An important question,
 however, is whether models trained using the deep field photometry are suitable for use in selecting quiescent galaxies in the wide field survey. 
 While the Deep and Wide Surveys are expected to be somewhat similar (but clearly not identical) within the magnitude-space covered by the Wide Survey, 
 at least 60 per cent of the galaxies in the Deep Survey are below (or close to) the \IE\ $3\sigma$ detection threshold of the Wide Survey. It is not clear
 {\it a priori} how the presence of these faint galaxies will affect the quality of the classification models. 
 
To test this, we train our pipeline using the Int Deep catalogue, and use the resulting classification model to predict classes for the galaxies in the 
Int Wide catalogue. No galaxy was permitted to be present in both the training and the test set. In addition, no foreknowledge of redshifts was assumed, 
all available photometry bands were used ($ugriz$, \Euclid, $W$1, $W$2, and 20 cm), and only galaxies detected in all four \Euclid\ bands were included. 
Under these conditions, the resulting F1-score for selection of quiescent galaxies from the Int Wide catalogue is 0.74, moderately lower than the F1-score 
of 0.80 obtained using models trained using Int Wide (see Table~\ref{tab:global_selection2}). Conducting this test instead using the SED catalogues 
resulted in a similarly reduced F1-score of 0.76, compared to 0.82 when using only SED Wide catalogue. 

Thus, although it is possible to separate quiescent and star-forming galaxies in the Wide Survey mocks, the F1-score suffers a significant penalty and is 
reduced by at least $\sim0.06$. Clearly, the presence of a large number of additional, faint galaxies in the training set exacerbates at least some of the 
degeneracies described in Sect.\,\ref{sec:introduction}, and induces the learning algorithms to place undue weight on galaxies near the faint end of the 
magnitude distribution. 

In this study we are, of course, only able to make use of mock \Euclid\ photometry catalogues, but we must also consider how the training set for 
this task will be constructed from real \Euclid\ data. Based on the analyses presented heretofore, we propose constructing the training 
set(s) from Deep Survey photometry from fields that have high-quality multi-wavelength observations (and spectroscopy), using the deepest available
data in order to generate high-quality ground truth labels, but then training classification models on Wide Survey-depth observations.

\subsection{Alternative targets}
\label{sec:alternative targets}
In this work, we have used target labels that were generated using an sSFR threshold to differentiate between quiescent and star forming galaxies. 
However, it is also interesting to consider other methods for generating the labels. Thus, here we explore the impact of using labels derived from the 
${\rm NUV}-r$ versus $r-J_{\scriptscriptstyle{\rm E}}$ \citep{Ilbert2010,Ilbert2013} or ${\rm NUV}-r$ versus $r-K$ \citep{Arnouts2013} colour-colour methods, instead of the sSFR-based labels used throughout this work. The mock catalogue used for these tests is Int Wide, and features derived from the $ugriz$, \IE, \YJH, $W$1, $W$2 and 20 cm bands were used, 
with detections required in all of the \Euclid\ bands. Our pipeline was used in {fast mode}. The results are summarised in Table \ref{tab:alternative_targets}. 

In the global selection of quiescent galaxies ($0 \le z \le 3$), the impact of using labels the alternative labels is limited. In the case of the labels derived
from the ${\rm NUV}-r$ versus $r-J_{\scriptscriptstyle{\rm E}}$ method, the F1-score is improved by $\sim$0.03 compared to the original (sSFR) labels. On the other hand, the F1-score when 
using the labels derived from the ${\rm NUV}-r$ versus $r-K$ method is $\sim$0.02 lower compared to when the original labels are used. 

When selected quiescent galaxies in narrower redshift ranges, having first removed sources whose photometric redshifts place them outside the range of interest, we usually obtain slightly higher F1-scores when using labels from the ${\rm NUV}-r$ versus $r-J_{\scriptscriptstyle{\rm E}}$ method, compared to when the original labels are used. in the $z = 2.5-3$
bin, there is a particularly large improvement in the F1-score (from $\sim$0.4 to $\sim$0.6). Conversely, in the $z=0-0.25$ bin we obtain a significantly lower 
F1-score when using when using labels from the ${\rm NUV}-r$ versus $r-J_{\scriptscriptstyle{\rm E}}$ method (0.53 compared to 0.69). When labels derived from the ${\rm NUV}-r$ versus $r-K$ method 
are used, the resulting F1-scores are consistently lower compared to the case where the original labels (sSFR) are used. 

\begin{table*}
        \centering
        \caption{Selection of quiescent galaxies in the redshift range $0 \ge z \ge 3$, using the Int Wide catalogue with its labels replaced by
        labels derived from the ${\rm NUV}-r$ vs. $r-J_{\scriptscriptstyle{\rm E}}$ or ${\rm NUV}-r$ vs. $r-K$ colour-colour methods.}
        \resizebox{\textwidth}{!}{%
        \begin{tabular}{ccccccccccc} 
                \hline
                                & & \multicolumn{3}{c}{$\overbrace{\rule{10em}{0pt}}^{\text{\small ${\rm NUV}-r$ vs. $r-J_{\scriptscriptstyle{\rm E}}$ labels}}$} & \multicolumn{3}{c}{$\overbrace{\rule{10em}{0pt}}^{\text{\small ${\rm NUV}-r$ vs. $r-K$ labels}}$} & \multicolumn{3}{c}{$\overbrace{\rule{10em}{0pt}}^{\text{\small sSFR labels}}$} \\
                                Redshift range & Redshift binning & $P$ & $R$ & F1-score & $P$ & $R$ & F1-score & $P$ & $R$ & F1-score \\
                (1) & (2) & (3) & (4) & (5) & (6) & (7) & (8) & (9) & (10) & (11) \\
                \hline
                $0-3$      & No & 0.86 & 0.79 & 0.82 & 0.79 & 0.74 & 0.77 & 0.84 & 0.74 & 0.79 \\
                \hline
                $0-0.25$   & Yes & 0.51 & 0.56 & 0.53 & 0.41 & 0.46 & 0.43 & 0.72 & 0.67 & 0.69 \\
                $0.25-0.5$ & Yes & 0.82 & 0.81 & 0.82 & 0.78 & 0.74 & 0.76 & 0.81 & 0.80 & 0.80 \\
                $0.5-0.75$ & Yes & 0.86 & 0.85 & 0.86 & 0.84 & 0.81 & 0.83 & 0.86 & 0.85 & 0.85 \\
                $0.75-1$   & Yes & 0.90 & 0.90 & 0.90 & 0.80 & 0.85 & 0.82 & 0.85 & 0.87 & 0.86 \\
                $1-1.25$   & Yes & 0.90 & 0.90 & 0.90 & 0.79 & 0.86 & 0.82 & 0.88 & 0.88 & 0.88 \\
                $1.25-1.5$ & Yes & 0.81 & 0.85 & 0.83 & 0.69 & 0.80 & 0.74 & 0.82 & 0.84 & 0.83 \\
                $1.5-2$    & Yes & 0.77 & 0.79 & 0.78 & 0.61 & 0.72 & 0.66 & 0.80 & 0.82 & 0.81 \\
                $2-2.5$    & Yes & 0.70 & 0.69 & 0.69 & --   & --   & --   & 0.62 & 0.69 & 0.65 \\
                $2.5-3$    & Yes & 0.62 & 0.65 & 0.62 & --   & --   & --   & 0.40 & 0.57 & 0.37 \\
                \hline
        \end{tabular}%
}
        \label{tab:alternative_targets}
	\tablefoot{For this test the $ugriz$, \IE, \YJH, $W$1, $W$2 and 20 cm 
        bands were used, with detections required in all of the \Euclid\ bands. Here, the \texttt{ARIADNE} pipeline was used in its fast mode. No results are 
        shown for the ${\rm NUV}-r$ vs. $r-K$ case in the redshift ranges $2-2.5$ or $2.5-3$, due to the very low number of sources labelled as quiescent by this method.\protect\\
        The columns are as follows:\protect\\
          (1) Redshift range in which the test was conducted;  \protect\\
          (2) information on whether photometric redshifts were used to restrict the dataset to source in the specified redshift range (`Yes' or `No'); \protect\\
          (3) precision $P$ for quiescent galaxy selection when using labels derived from the ${\rm NUV}-r$ vs. $r-J_{\scriptscriptstyle{\rm E}}$ method;\protect\\
          (4) recall $R$ for quiescent galaxy selection when using labels derived from the ${\rm NUV}-r$ vs. $r-J_{\scriptscriptstyle{\rm E}}$ method;\protect\\
          (5) the F1-score for quiescent galaxy selection when using labels derived from the ${\rm NUV}-r$ vs. $r-J_{\scriptscriptstyle{\rm E}}$ method;\protect\\
          (6) precision $P$ for quiescent galaxy selection when using labels derived from the ${\rm NUV}-r$ vs. $r-K$ method;\protect\\
          (7) recall $R$ for quiescent galaxy selection when using labels derived from the ${\rm NUV}-r$ vs. $r-K$ method;\protect\\
          (8) the F1-score for quiescent galaxy selection when using labels derived from the ${\rm NUV}-r$ vs. $r-K$ method;\protect\\
          (9) precision $P$ for quiescent galaxy selection when using labels derived from the sSFR;\protect\\
          (10) recall $R$ for quiescent galaxy selection when using labels derived from the sSFR;\protect\\
          (11) the F1-score for quiescent galaxy selection when using labels derived from the sSFR.}
\end{table*}

\subsection{Which observables are useful to select quiescent galaxies?}
\label{sec:xgb_feature_analysis}

\begin{figure*}
\centering
\includegraphics[width=0.785\columnwidth,trim=10 -15 0 0]{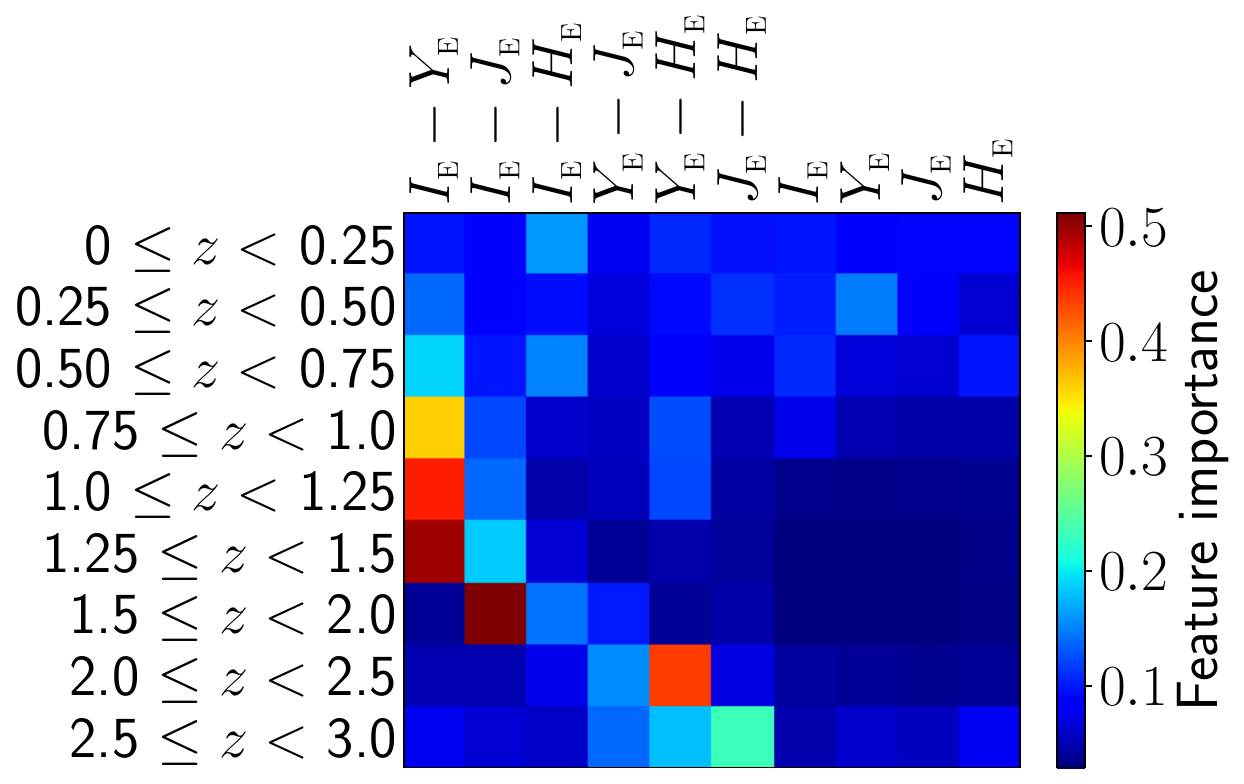}
\includegraphics[width=1.2\columnwidth,trim = -115 0 0 0]{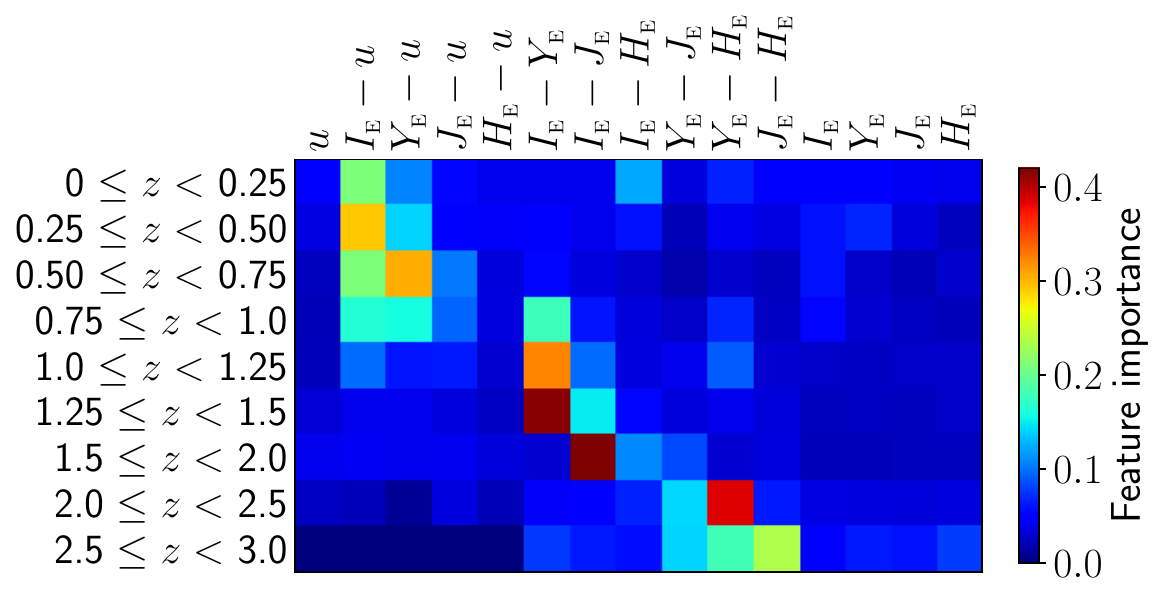}

\includegraphics[width=1\columnwidth]{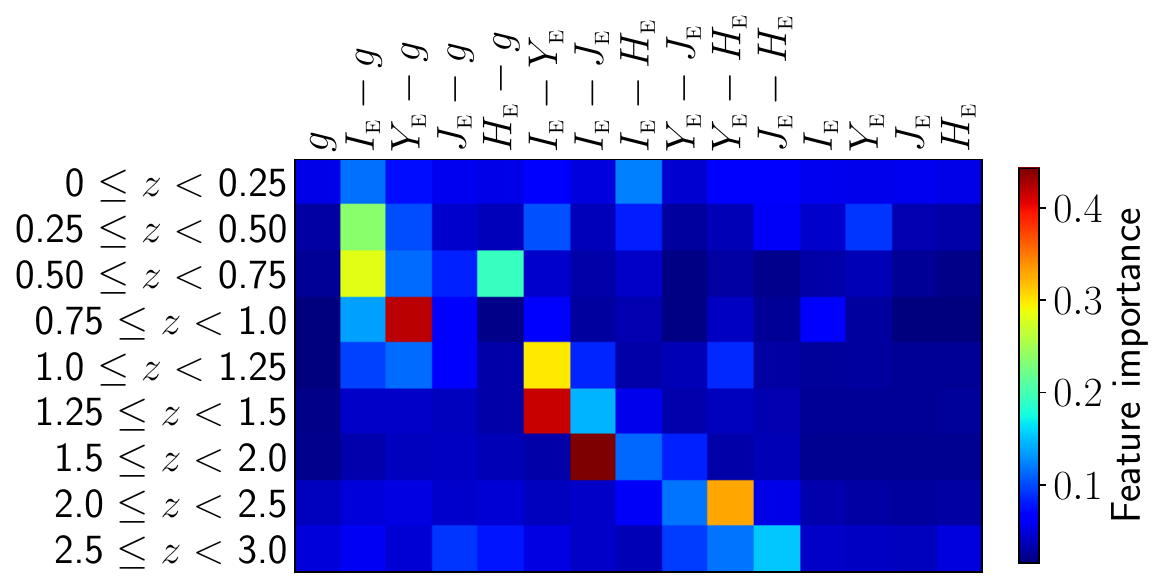}
\includegraphics[width=1\columnwidth]{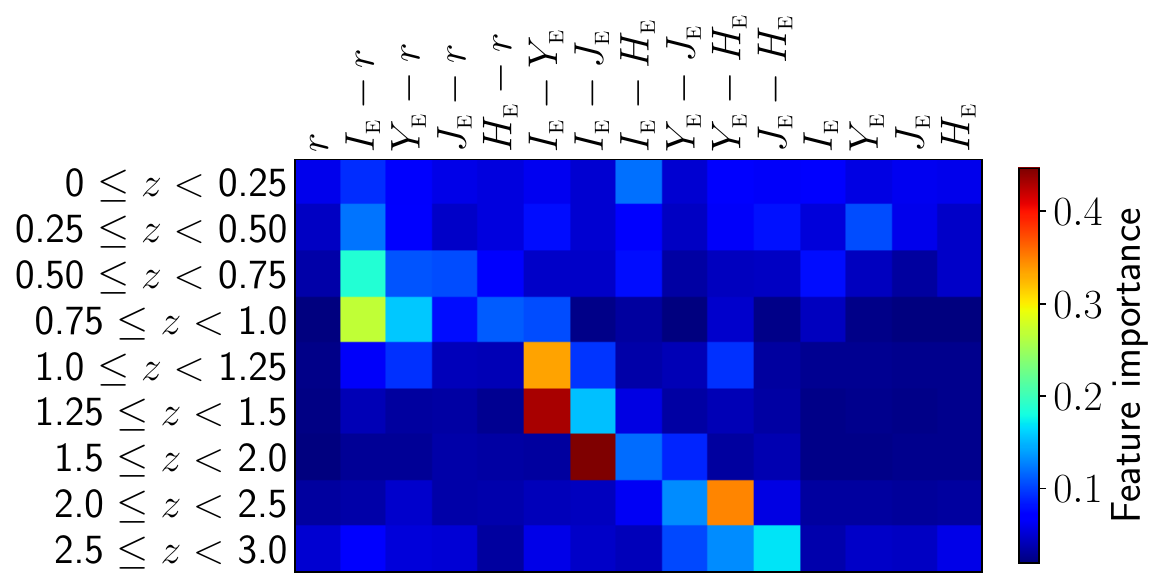}

\includegraphics[width=1\columnwidth]{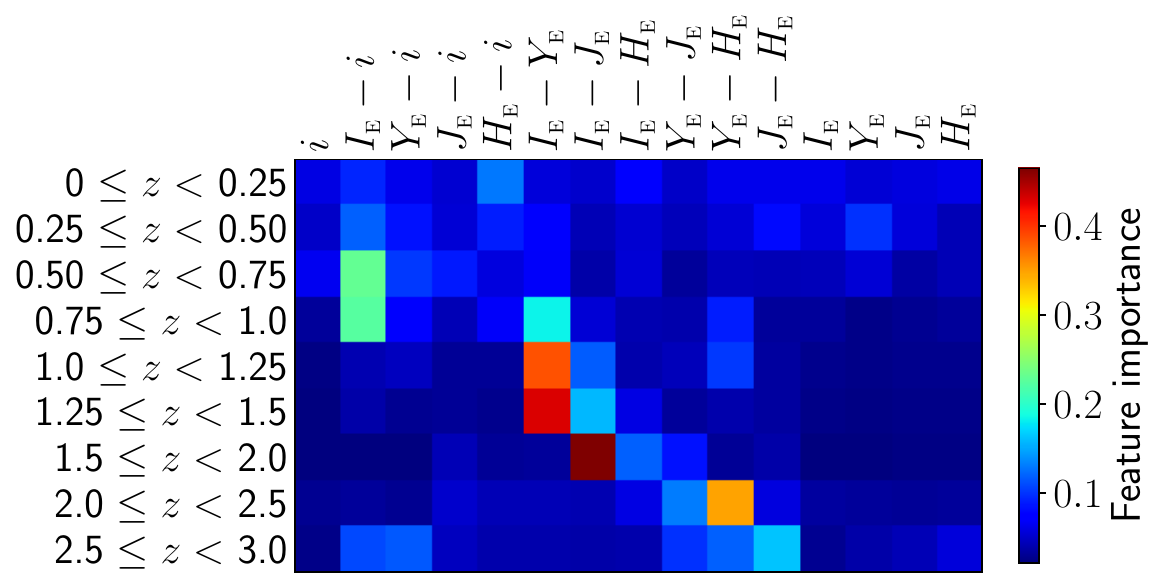}
\includegraphics[width=1\columnwidth]{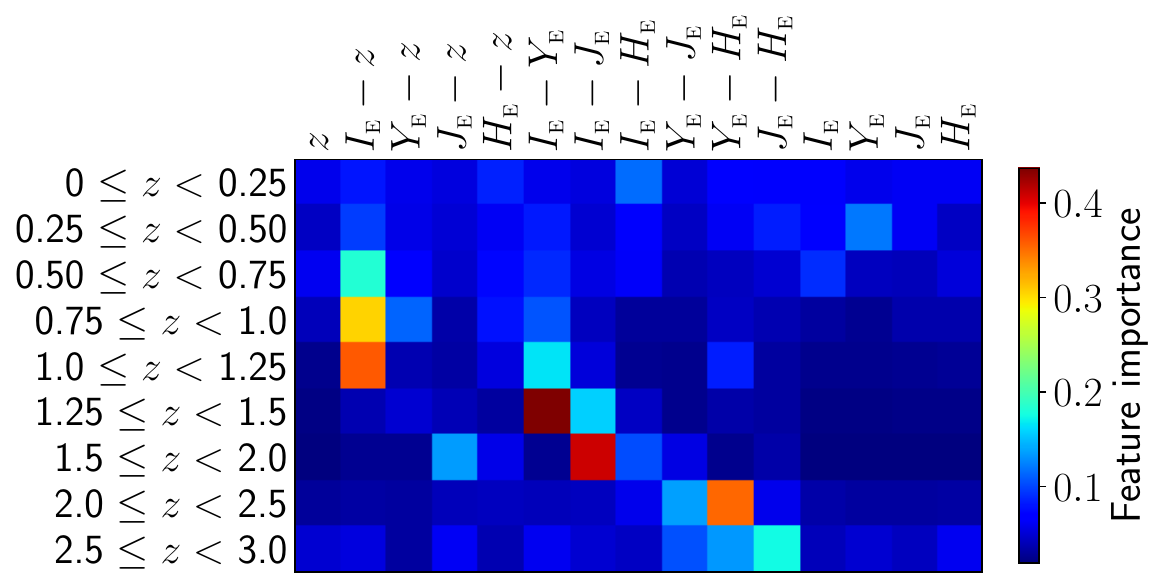}
\caption{Visual representation of the feature importances derived from \texttt{XGBoostClassifier} models trained to select quiescent galaxies
using only \Euclid\ photometry and colours (top left panel), or \Euclid\ photometry with the addition of one ground-based optical band, and the
relevant broadband colours. The optical bands are $u$ (top right), $g$ (middle left), $r$ (middle right), $i$ (bottom left), and $z$ (bottom right).
The \texttt{XGBoostClassifier} models were trained without foreknowledge of the redshifts.}
\label{fig:fimportance_images}
\end{figure*}

\begin{figure}
\centering
\includegraphics[width=1\columnwidth]{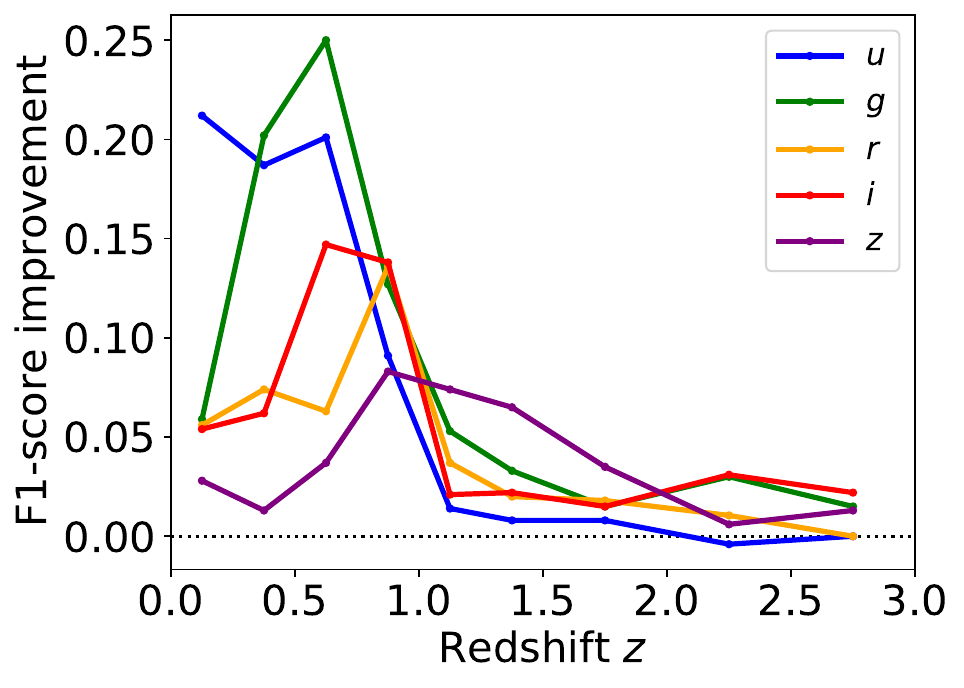}
\caption{Effect on the F1-score due to the including one additional band with the \Euclid\ photometry set. The \texttt{ARIADNE} pipeline
was run in {fast mode} using the \texttt{LightGBMClassifier} base learner, or \texttt{XGBoostClassifier} at $2.5 < z < 3$, without 
foreknowledge of galaxy redshifts.}
\label{fig:f1_increase}
\end{figure}

\begin{figure}
\centering
\includegraphics[width=1\columnwidth]{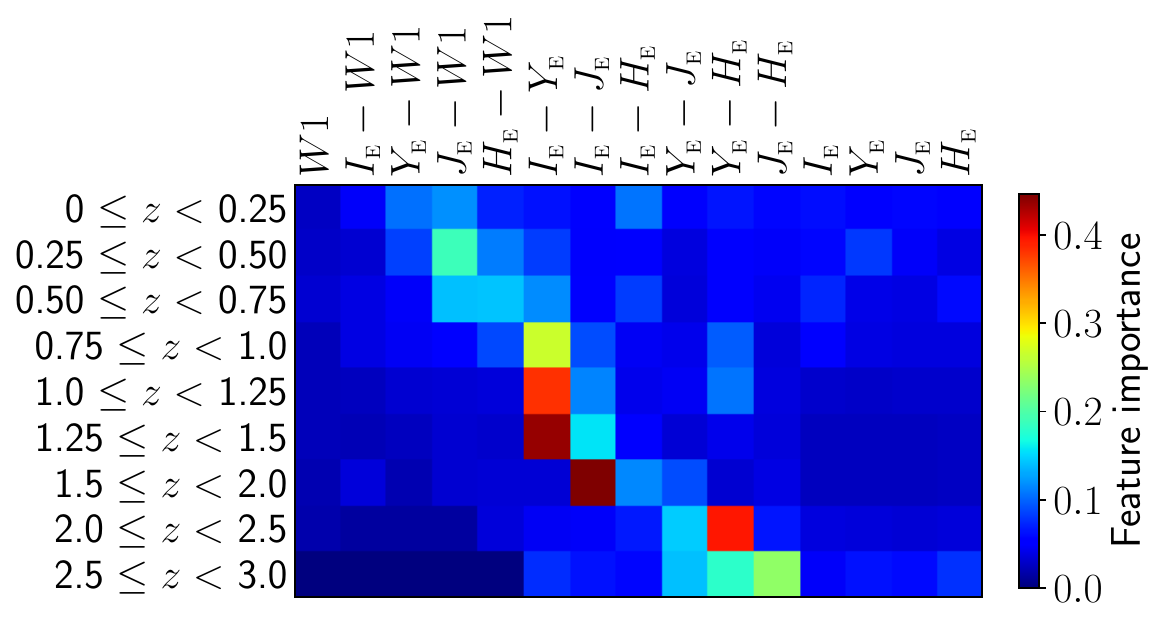}
\includegraphics[width=1\columnwidth]{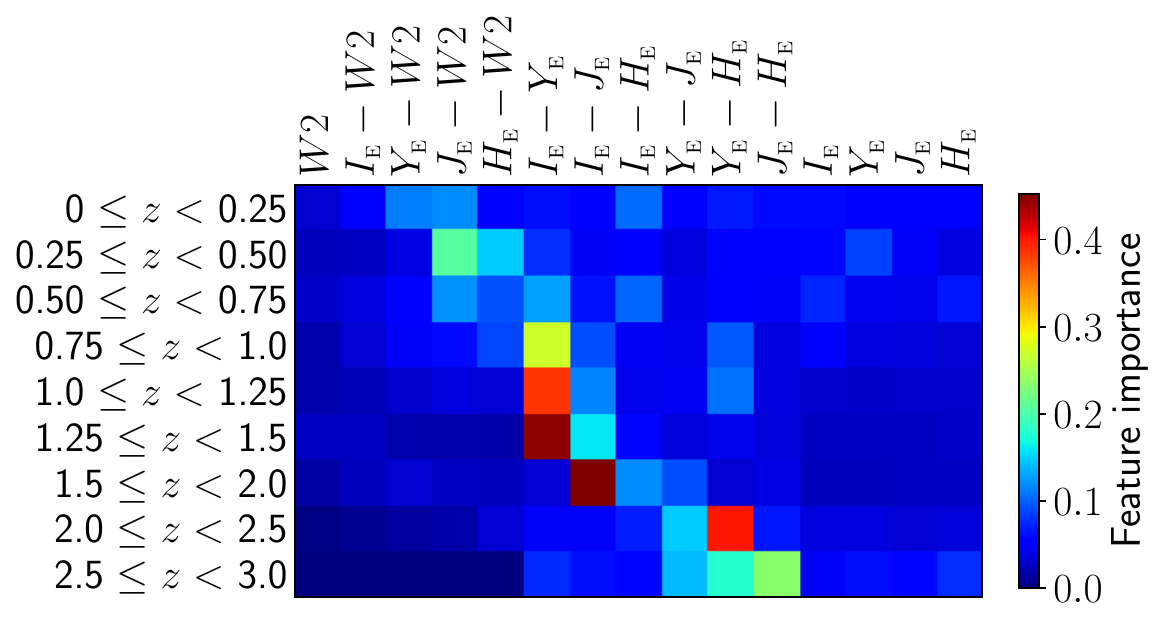}
\includegraphics[width=1\columnwidth]{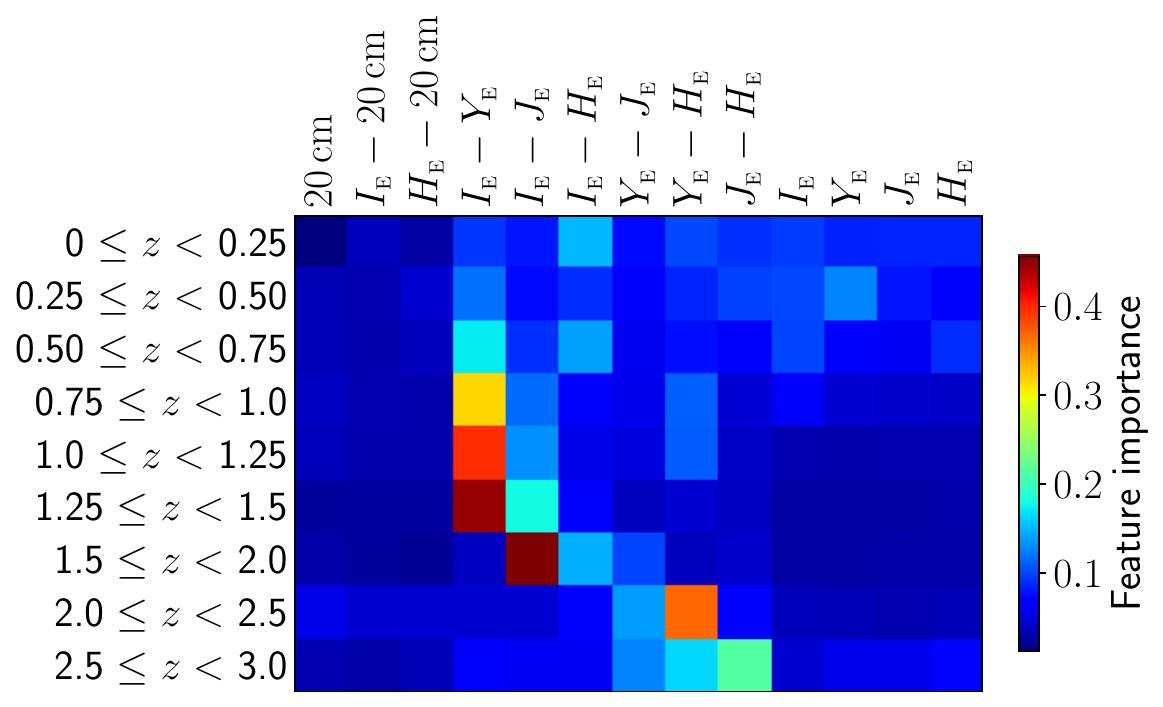}
\caption{Similar to Fig. \ref{fig:fimportance_images}, but showing the feature importances when the $W$1 (top), $W$2 (middle) or 20 cm (bottom) 
bands are included with the \Euclid\ photometry.}
\label{fig:fimportance_images_long_wavelength}
\end{figure}

Heretofore, we have approached the question of the usefulness of different features (colours, magnitudes, etc.) in terms of whether they are 
informative and whether their inclusion significantly improves the quality of our classification models (see Sect.\,\ref{sec:feat_imp}). However, it is 
also desirable to reach a deeper understanding of the usefulness of each feature, and how their usefulness depends on the circumstances under which 
quiescent galaxies are to be selected. For instance, the colours that are most useful for selecting quiescent galaxies in one redshift range may 
be less useful in another. 

Moreover, while it may be self-evident that the inclusion of photometry in various optical bands allows for better
 characterisation of galaxy SEDs, and thus more accurate classification of galaxies, the degree of improvement may not always justify the cost of 
 acquiring additional observations. Figure \ref{fig:bisigello_comparison_euclid} illustrates such an example, where the inclusion of $ugriz$ 
 photometry provides little or no significant improvement in the selection of quiescent galaxies at $z\ga1.25$, compared to when only the \Euclid\
 photometry is used (see Sect.\,\ref{sec:euclid_only} for further details). 

Our objective for this section, therefore, is to provide analyses that can inform the planning of future surveys regarding 
which filters or frequencies would be important to include, and to guide the construction of new selection methods, be they colour-colour, template-
fitting, or machine-learning-based. 
For this we employ feature importance analysis (see also Sect.\,\ref{sec:feat_imp}). The machine-learning models we use for these analyses 
are trained using the \texttt{XGBoostClassifier} learning algorithm and the Int Wide mock catalogue, because its feature importance values are more robust, 
as discussed in Appendix \ref{appendix_redundancy}. We also perform split-run tests using our pipeline in {fast mode} to examine the impact on the F1-score from the addition 
of ancillary bands to the \Euclid\ \IE, \YJH photometry. 

It is worth noting that these analysis methods provide information about machine-learning models and how they make use of the input data, rather 
than the data itself. As such, the information provided is limited by what the learning algorithms were able to learn from the data, and it is quite 
possible that additional relationships exist between features and the target that the learning algorithm was unable to find. Thus, while features found
to be important are highly likely to be useful for the selection of quiescent galaxies, features found to be unimportant (or less important) may
nevertheless hold hidden information that other selection methods (e.g. template fitting) may find useful when selecting quiescent galaxies.

\subsubsection{Which single optical band is most useful?}
\label{opt_incl}
In Fig. \ref{fig:fimportance_images} we show the effect of adding one optical band to the \Euclid\ photometry set, when selecting
quiescent galaxies in specific redshift ranges from the Int Wide catalogue using \texttt{XGBoostClassifier}, with no foreknowledge of 
redshifts. In addition, Fig.~\ref{fig:f1_increase} shows the improvement in F1-score due to the inclusion of each optical band. 

First, we examine feature importance as a function of redshift when only the \Euclid\ data are available 
(Fig. \ref{fig:fimportance_images}, top left). In this circumstance, there is no single, decisive broadband colour for the selection of 
quiescent galaxies at $z \la 0.5$. The lack of sensitivity to the 4000 \AA~break, or to emission bluewards thereof, results in models 
with relatively low F1-scores (see Fig.~\ref{fig:bisigello_comparison_euclid}). In this regime, our \texttt{XGBoostClassifier} models 
assign a fairly similar importance value to each feature. 

Above $z=0.4$, the 4000 \AA~break becomes redshifted into the \IE\ band, becoming potentially detectable via one or more of the \Euclid\ broadband 
colours. In the range $0.5 < z < 1.5$, the colour $I_{\scriptscriptstyle{\rm E}}-Y_{\scriptscriptstyle{\rm E}}$ is the most sensitive to the presence of the 4000 \AA~break, and our 
\texttt{XGBoostClassifier} models consider this feature to be the single most important for these redshifts. At $z > 1.5$, where the 
4000 \AA~break is now redshifted into the NIR bands, other colours become more important: $I_{\scriptscriptstyle{\rm E}}-J_{\scriptscriptstyle{\rm E}}$ at $1.5 < z < 2$, 
$Y_{\scriptscriptstyle{\rm E}}-H_{\scriptscriptstyle{\rm E}}$ at $2 < z < 2.5$, and $J_{\scriptscriptstyle{\rm E}}-H_{\scriptscriptstyle{\rm E}}$ at $2.5 < z < 3$. 

The addition of $u$-band photometry allows the 4000 \AA~break to be detected at $z \la 0.5$, resulting in significantly increased F1-scores
in this redshift range (see Fig.~\ref{fig:f1_increase}). In this case, $I_{\scriptscriptstyle{\rm E}}-u$ and/or $Y_{\scriptscriptstyle{\rm E}}-u$ become the most important in the redshift range
$0 < z < 0.75$ by a large margin. Features using the $u$-band photometry continue to have significant (or non-zero) importance up to $z \sim 2$,
before dropping to $\sim 0$ at $z > 2.5$.

In the case where models are trained using features derived from the $g$-band and \Euclid\ photometry, colours involving $g$ are found to be 
the most important within the range $0.25 < z < 1$. While the inclusion of $g$ also helps significantly at $z < 0.25$, the improvement is considerably
smaller compared to when the $u$ is used, because the 4000 \AA~break is bluewards of the $u$ band's wavelength range. While the importance values of features
using the $g$ band decrease substantially at $z > 1.25$, these features remain useful even in the $2.5 < z < 3$ bin.  

When the $r$, $i$, or $z$ band is used together with the \Euclid\ bands, the situation is similar to that described above for $g$, with the main
differences being the substantially lower feature importances at $0.25 < z < 0.5$ for colours that use one of the three bands, and the 
relatively high importance of $I_{\scriptscriptstyle{\rm E}}-z$ at $1 < z < 1.25$.

In summary, there is no single `ideal' optical band to include alongside the \Euclid\ bands when selecting quiescent galaxies, and a trade-off
should be made depending on whether low-redshift ($z \la 0.25$) or high-redshift ($z \ga 2$) galaxies are required. Of course, one may circumvent this choice 
if $ugriz$ photometry is available.

\begin{figure}
\centering
\includegraphics[width=1\columnwidth]{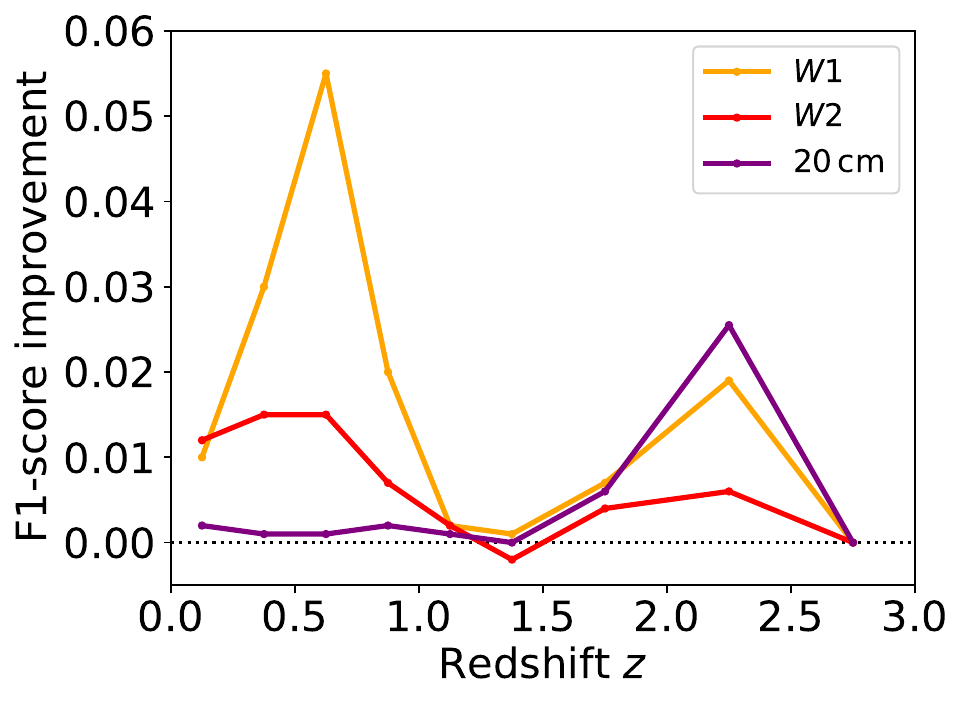}
\caption{Similar to Fig.~\ref{fig:f1_increase}, but showing the effect on the F1-score due to the including $W$1, $W$2, or the 20 cm radio band with 
the \Euclid\ photometry when training \texttt{LightGBMClassifier} ($z < 2.5$) or \texttt{XGBoostClassifier} ($2.5 < z < 3$) models.}
\label{fig:f1_increase_long_wavelength}
\end{figure}

\subsubsection{Importance of long wavelength bands}
We repeat the process described in Appendix~\ref{opt_incl}, this time considering the addition of the $W$1, $W$2, or 20 cm radio band and related 
broadband colours. The data for these three bands is very sparse, with detection fractions of 0.053, 0.024, and 0.0076 for $W$1, $W$2, and 
20 cm, respectively. 

The feature importance as a function of redshift is shown in Fig.~\ref{fig:fimportance_images_long_wavelength}, and the 
improvement from including the $W$1, $W$2, or 20 cm bands with the \Euclid\ photometry is shown in Fig.~\ref{fig:f1_increase_long_wavelength}. 
Despite being very sparse, each of the three bands provide a significant improvement in F1-score within the redshift ranges $0 < z < 1$ 
and $1.5 < z < 3$. We find little or no improvement evident in the redshift range $1 < z < 1.5$. 

The usefulness of the 20 cm band for discriminating quiescent galaxies from star-forming galaxies is intriguing and somewhat surprising. 
Clearly, there is a correlation (or correlations) between the galaxy class and the presence of radio continuum emission. 
We speculate that this may be due to a subset of star-forming galaxies being detected in radio continuum, and/or the presence of radio-loud 
massive elliptical galaxies whose X-ray emission is below the COSMOS2015 detection threshold.

\subsection{Selection of quiescent galaxies from SPRITZ: Euclid Deep Survey}
\label{sec:spritz}
To complement the results presented in Sect.\,\ref{sec:euclid_deep}, we also tested the selection of quiescent galaxies using the simulated
Euclid Deep Survey from SPRITZ \citep{Bisigello2021}. Compared to the Euclid Deep Survey catalogues, the SPRITZ catalogue has the advantage of being 
complete down to the expected depth for the actual survey Deep Survey and the expected ancillary ground-based data. 
For the \Euclid\ and $ugriz$ bands, we adopt identical photometric uncertainties and detection limits to those used for the SED Deep catalogue given 
in Sect.\,\ref{sec:mocks}. In the case of \Spitzer IRAC bands, we adopt the following four cases based on \citet{Moneti2022}:
{\bf Case 1}: no IRAC photometry;
{\bf Case 2}: IRAC photometry $3\,\sigma$ depths of 24.55, 24.39, 22.61, and 22.17\,mag in channel 1, 2, 3 and 4, respectively;
{\bf Case 3}: IRAC photometry $3\,\sigma$ depths of 25.55, 25.39, 23.61, and 23.17\,mag in channel 1, 2, 3 and 4, respectively;
and {\bf Case 4}: IRAC photometry $3\,\sigma$ depths of 26.55, 26.39, 23.61, and 23.17\,mag in channel 1, 2, 3 and 4, respectively.

\begin{figure}
\centering
\includegraphics[width=1\columnwidth]{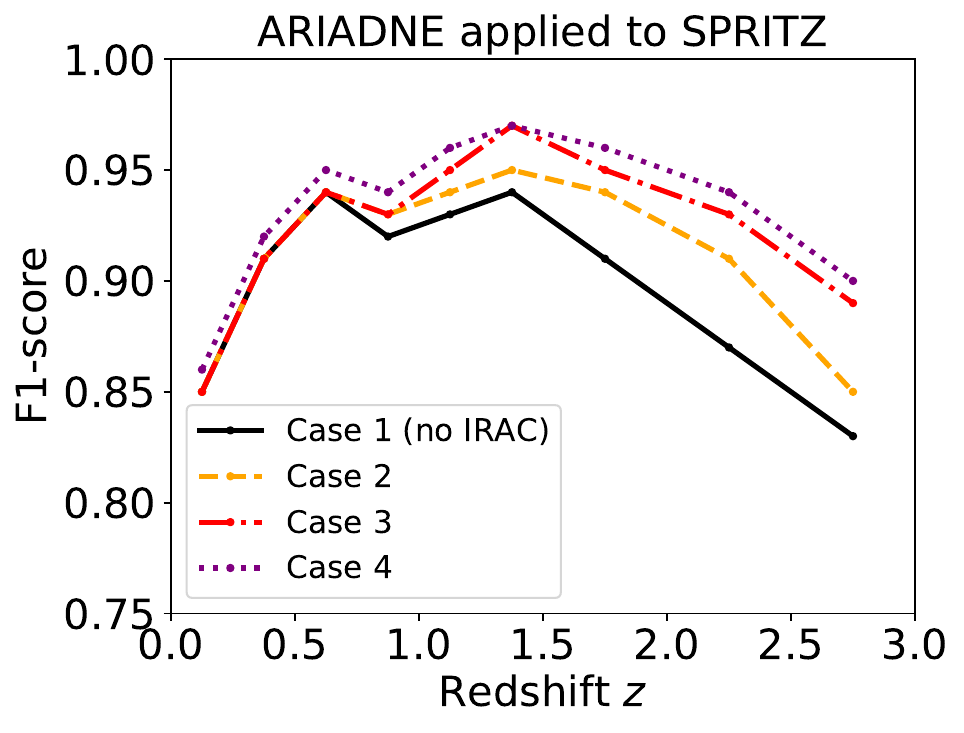}
\caption{Results from applying the \texttt{ARIADNE} pipeline to the selection of quiescent galaxies from the SPRITZ Euclid Deep Survey simulated catalogue. 
Four cases described in Appendix~\ref{sec:spritz} are shown.}
\label{fig:f1_spritz}
\end{figure}

The \texttt{ARIADNE} pipeline was used in its default configuration, where five base-learners are employed. Galaxies not detected in one or more
of the \Euclid\ bands were removed from the dataset, as were galaxies containing an active galactic nucleus. The results are shown in Table \ref{tab:spritz}
and Fig. \ref{fig:f1_spritz}.

\begin{table*}
        \centering
        \caption{Selection of quiescent galaxies in the redshift range $0 \ge z \ge 3$, from the SPRITZ Euclid Deep Survey simulation.}
        \resizebox{\textwidth}{!}{%
        \begin{tabular}{cccccccccccccc} 
                \hline
                                & & \multicolumn{3}{c}{$\overbrace{\rule{10em}{0pt}}^{\text{\small SPRITZ Case 1}}$} & \multicolumn{3}{c}{$\overbrace{\rule{10em}{0pt}}^{\text{\small SPRITZ Case 2}}$} & \multicolumn{3}{c}{$\overbrace{\rule{10em}{0pt}}^{\text{\small SPRITZ Case 3}}$} & \multicolumn{3}{c}{$\overbrace{\rule{10em}{0pt}}^{\text{\small SPRITZ Case 4}}$} \\
                                Redshift range & Redshift binning & $P$ & $R$ & F1-score & $P$ & $R$ & F1-score & $P$ & $R$ & F1-score & $P$ & $R$ & F1-score \\
                (1) & (2) & (3) & (4) & (5) & (6) & (7) & (8) & (9) & (10) & (11) & (12) & (13) & (14) \\
                \hline
                $0-3$      & No & 0.92 & 0.86 & 0.89 & 0.92 & 0.88 & 0.90 & 0.93 & 0.90 & 0.91 & 0.93 & 0.91 & 0.92 \\
                \hline
                $0-0.25$   & Yes & 0.84 & 0.85 & 0.85 & 0.85 & 0.85 & 0.85 & 0.85 & 0.85 & 0.85 & 0.85 & 0.86 & 0.86 \\
                $0.25-0.5$ & Yes & 0.92 & 0.90 & 0.91 & 0.92 & 0.91 & 0.91 & 0.92 & 0.91 & 0.91 & 0.92 & 0.91 & 0.92 \\
                $0.5-0.75$ & Yes & 0.96 & 0.92 & 0.94 & 0.96 & 0.92 & 0.94 & 0.96 & 0.93 & 0.94 & 0.96 & 0.93 & 0.95 \\
                $0.75-1$   & Yes & 0.93 & 0.91 & 0.92 & 0.94 & 0.91 & 0.93 & 0.94 & 0.92 & 0.93 & 0.95 & 0.92 & 0.94 \\
                $1-1.25$   & Yes & 0.95 & 0.92 & 0.93 & 0.96 & 0.92 & 0.94 & 0.97 & 0.93 & 0.95 & 0.97 & 0.94 & 0.96 \\
                $1.25-1.5$ & Yes & 0.93 & 0.95 & 0.94 & 0.96 & 0.95 & 0.95 & 0.97 & 0.96 & 0.97 & 0.98 & 0.97 & 0.97 \\
                $1.5-2$    & Yes & 0.90 & 0.93 & 0.91 & 0.94 & 0.94 & 0.94 & 0.95 & 0.95 & 0.95 & 0.96 & 0.96 & 0.96 \\
                $2-2.5$    & Yes & 0.91 & 0.84 & 0.87 & 0.93 & 0.89 & 0.91 & 0.95 & 0.91 & 0.93 & 0.95 & 0.93 & 0.94 \\
                $2.5-3$    & Yes & 0.82 & 0.83 & 0.83 & 0.86 & 0.84 & 0.85 & 0.89 & 0.89 & 0.89 & 0.91 & 0.89 & 0.90 \\
                \hline
        \end{tabular}%
}
        \label{tab:spritz}
	\tablefoot{The columns are as follows:\\
	      (1) redshift range in which the test was conducted;  \protect\\
          (2) information on whether photometric redshifts were used to restrict the dataset to source in the specified redshift range (`Yes' or `No'); \protect\\
          (3) precision $P$ for quiescent galaxy selection for SPRITZ Case 1;\protect\\
          (4) recall $R$ for quiescent galaxy selection for SPRITZ Case 1;\protect\\
          (5) the F1-score for quiescent galaxy selection for SPRITZ Case 1;\protect\\
          (6) precision $P$ for quiescent galaxy selection for SPRITZ Case 2;;\protect\\
          (7) recall $R$ for quiescent galaxy selection for SPRITZ Case 2;\protect\\
          (8) the F1-score for quiescent galaxy selection for SPRITZ Case 2;;\protect\\
          (9) precision $P$ for quiescent galaxy selection for SPRITZ Case 3;;\protect\\
          (10) recall $R$ for quiescent galaxy selection for SPRITZ Case 3;;\protect\\
          (11) the F1-score for quiescent galaxy selection for SPRITZ Case 3;;\protect\\
          (12) precision $P$ for quiescent galaxy selection for SPRITZ Case 4;;\protect\\
          (13) recall $R$ for quiescent galaxy selection for SPRITZ Case 4;;\protect\\
          (14) the F1-score for quiescent galaxy selection for SPRITZ Case 4.}
\end{table*}

\section{Number of detections in each catalogue and band}

\begin{table*}[!ht]
        \caption{Number of $3\,\sigma$ detections in each of the bands, for each mock catalogue.}
        \centering
        \setlength\tabcolsep{2.5pt}
        \resizebox{\textwidth}{!}{%
        \begin{tabular}{cccccccccccccc}
        \hline
        Catalogue & \IE\ & \YE\ & \JE\ & \HE\ & $u$ & $g$ & $r$ & $i$ & $z$ & $W$1 & $W$2 & 20 cm & Quiescent \\
        \hline
        Int Wide  & 315755 & 212019 & 231039 & 250077 & 140782 & 226514 & 198564 & 194912 & 204649 & 10476 & 4704 & 1536 & 21998 (7.0\,\%)\\
        Int Deep  & 517890 & 486394 & 491588 & 500299 & 499565 & 504416 & 490457 & 493018 & 499140 & 10476 & 4704 & 1698 & 30990 (6.0\,\%)\\
        SED Wide  & 3249101 & 2056800 & 2270138 & 2455887 & 1498518 & 2330338 & 2091921 & 2040916 & 2031115 & 131703 & 65904 & -- & 213837 (6.6\,\%)\\
        SED Deep  & 5121526 & 4763050 & 4890667 & 4963038 & 3971472 & 4796448 & 4828112 & 4802416 & 4766807 & 134656 & 69493 & -- & 303761 (5.9\,\%)\\
        \hline
        \end{tabular}%
}
        \label{tab:detections}
	\tablefoot{Also shown are the numbers of quiescent galaxies. The SED catalogues do not contain 20 cm radio band.}
\end{table*}

\end{appendix}


\label{lastpage}
\end{document}